\documentclass[10pt,twocolumn,letterpaper]{article}

\usepackage{cvpr}              
\usepackage{tikz}
\usepackage{amsmath}

\usepackage{filecontents}
\usepackage{caption}
\usepackage{subcaption}
\usepackage{times}
\usepackage{epsfig}
\usepackage{graphicx}
\usepackage{amsmath}
\usepackage{amssymb}
\usepackage{url}
\usepackage{booktabs}
\usepackage{subcaption}
\usepackage{graphicx}
\usepackage{wrapfig}
\usepackage{caption}
\usepackage{xcolor}
\usepackage[accsupp]{axessibility}  
\usepackage{amsthm}
\usepackage{makecell}

\usepackage{multirow}

\usepackage{bm}
\newcommand{\ourmodel}{PoisonedFL}

\newcommand{\myparatight}[1]{\smallskip\noindent{\bf {#1}:}~}

\definecolor{cvprblue}{rgb}{0.21,0.49,0.74}
\usepackage[pagebackref,breaklinks,colorlinks,allcolors=cvprblue]{hyperref}

\def\paperID{9493} 
\def\confName{CVPR}
\def\confYear{2025}

\title{Model Poisoning Attacks to Federated Learning via Multi-Round Consistency}

\author{Yueqi Xie\\
HKUST\\
{\tt\small yxieay@connect.ust.hk}
\and
Minghong Fang\\
University of Louisville\\
{\tt\small minghong.fang@louisville.edu}
\and
Neil Zhenqiang Gong\\
Duke University\\
{\tt\small neil.gong@duke.edu}
}

\begin{document}
\maketitle
\begin{abstract}
Model poisoning attacks are critical security threats to Federated Learning (FL). Existing model poisoning attacks suffer from two key limitations: 1) they achieve suboptimal effectiveness when defenses are deployed, and/or 2) they require knowledge of the model updates or local training data on genuine clients. In this work, we make a key observation that their suboptimal effectiveness arises from only leveraging model-update consistency among malicious clients within individual training rounds, making the attack effect self-cancel across training rounds. In light of this observation, we propose PoisonedFL, which enforces multi-round consistency among the malicious clients' model updates while not requiring any knowledge about the genuine clients. Our empirical evaluation on five benchmark datasets shows that PoisonedFL breaks eight state-of-the-art defenses and outperforms seven existing model poisoning attacks. Our study shows that FL systems are considerably less robust than previously thought, underlining the urgency for the development of new defense mechanisms. Our source code is available at \url{https://github.com/xyq7/PoisonedFL/}.
\vspace{-4mm}
\end{abstract}    
\section{Introduction}

Federated Learning (FL) enables clients--such as smartphones, self-driving cars, and IoT devices--to collaboratively train a model without sharing their raw local training data with a central server~\cite{konevcny2016federated,mcmahan2017communication}. 
Typically, FL involves multiple training rounds, each of which consists of three key steps: 1) the server sends the current \emph{global model} to the clients or a subset of them; 2) a client trains a \emph{local model} based on the current global model and its local training data, and sends the \emph{model update} (i.e., the difference between local model and global model) to the server; and 3) the server aggregates the clients' model updates according to an \emph{aggregation rule} and adds the \emph{aggregated model update} to the global model.  
 FL has wide applications across diverse domains, including healthcare~\cite{melloddy,kairouz2021advances,rieke2020future} and finance~\cite{webank,yang2019federated}.

However, due to the distributed nature, FL is fundamentally vulnerable to \emph{model poisoning attacks}~\cite{baruch2019little,cao2022mpaf,fang2020local,shejwalkar2021manipulating}. In particular, an attacker can control some \emph{malicious clients}  and send carefully crafted \emph{malicious model updates} to the server to poison the global model. The malicious clients could be compromised  \emph{genuine clients}  or  injected \emph{fake clients}. The poisoned global model misclassifies many inputs indiscriminately, i.e., it has a large testing error rate (also known as untargeted model poisoning attacks). 

\begin{figure}[t]
    
    \begin{subfigure}{0.495\linewidth}
        \centering
        \includegraphics[width=1\linewidth]{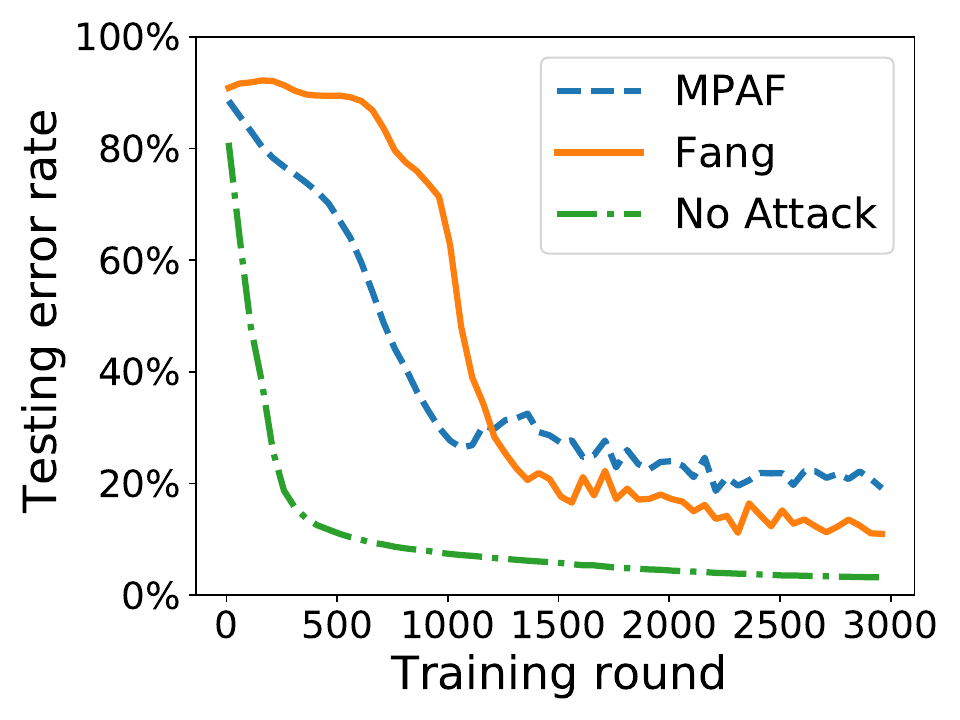}
        \subcaption{Testing error rate}
        \label{fig:moti_acc}
    \end{subfigure}
    \begin{subfigure}{0.495\linewidth}
        \centering
        \includegraphics[width=1\linewidth]{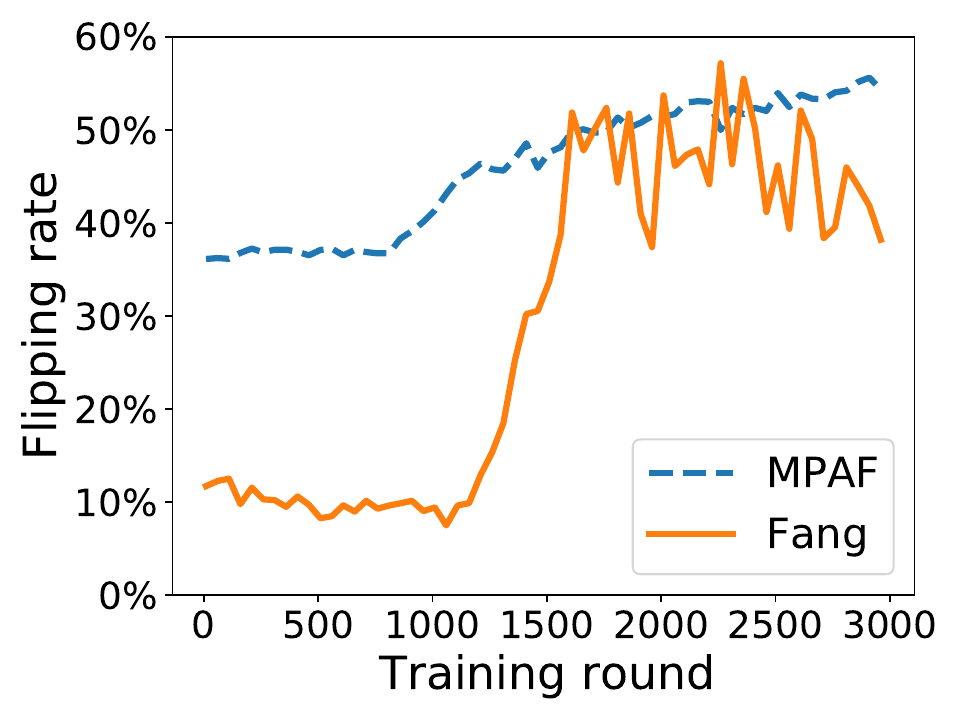}
        \subcaption{Flipping rate}
        \label{fig:moti_flip}
    \end{subfigure}
    \caption{Illustration of (a) limited attack effectiveness  and (b)  inconsistent malicious model updates of existing attacks.}
    \label{fig:motivation}
    \vspace{-5mm}
\end{figure}

Existing model poisoning attacks suffer from two key limitations. First, they~\cite{baruch2019little,fang2020local,shejwalkar2021manipulating,cao2022mpaf} only leverage \emph{in-round consistency}, 
i.e.,  they send the same or similar malicious model updates in a single training round to amplify the attack effect, but neglect \emph{multi-round consistency}.
Fig.~\ref{fig:motivation} reveals that many parameters' attack directions ($+1$/$-1$) reverse from the previous round (a large flipping rate), leading to self-cancellation and diminished impact after extensive training (small testing error rate). 
Detailed setups and discussion on this observation is deferred to Appendix~\ref{sec:motivation}.
Second, most existing attacks~\cite{baruch2019little,fang2020local,shejwalkar2021manipulating} require knowledge of genuine clients' model updates or local training data. They often assume \emph{compromising a significant number of genuine clients} to obtain such knowledge, which is notably impractical.  
These attacks with suboptimal effectiveness and/or impractical assumptions give a false sense of FL security.

\myparatight{Our work}
In this work, we propose \ourmodel, a novel model poisoning attack to FL, which overcomes the above limitations of existing attacks.  
Specifically, to enhance attack effectiveness, \ourmodel{} leverages two key components: 1) multi-round consistency, and 2) dynamic attack magnitude adjustment. 
The first component ensures that even if the attack effect is weakened in individual training rounds by defenses, the cumulative attack effect over multiple training rounds remains undeterred.
The second component aims to prevent the malicious model updates from being entirely filtered out in individual training rounds by defenses. Moreover, \ourmodel{} does not require genuine clients' information, and is agnostic to the defenses deployed by the server, thus can be performed on \emph{injected fake clients} with least information of the FL system.

For the first component, 
our key observation is that a model, when updated substantially in a random direction,  exhibits significantly degraded accuracy (see Fig.~\ref{fig:norm-performance} in Appendix). Based on this observation,
we formulate an optimization problem that 
maximizes the magnitude of the total aggregated model updates in a random \emph{update direction},  which is characterized by a random and fixed \emph{sign vector} $\bm{s}$. With such total aggregated model updates, the final global model is substantially moved along the random update direction, leading to a large testing error rate.
To achieve this, we propose a simple attack strategy in each round, where the malicious model update is a dimension-wise product of the fixed sign vector $\bm{s}$ and a dynamic non-negative \emph{magnitude vector}, a dimension of which indicates the magnitude of the corresponding dimension of the malicious model update.
Note that our objective stands substantially distinct from those of existing attacks, as it enforces multi-round consistency without self-cancellation. 



For the second component, we propose to dynamically adjust the magnitude vector to have a large magnitude while avoiding the malicious model update being filtered out by the unknown defense. On one hand, based on the global models and malicious model updates in the previous training rounds, we estimate the model updates of genuine clients to determine an \emph{unit magnitude vector}. The magnitude vector is the product of the unit magnitude vector and a \emph{scaling factor}.  On the other hand, we set the scaling factor based on whether the attack achieves the objective in the past rounds. In particular, our attack objective is that the total aggregated model updates have a random update direction $\bm{s}$. We leverage \emph{statistical hypothesis testing} to check whether our attack in the previous rounds achieves this objective. If not, it means the malicious model updates were filtered out by the defense in previous rounds and thus we decrease the scaling factor to be more stealthy.



We evaluate our \ourmodel{} on five benchmark datasets. Our results show that,  under the least knowledge scenario -- that is, no knowledge about the defense nor genuine local training data/models -- \ourmodel{} can break eight state-of-the-art FL defenses~\cite{blanchard2017machine,cao2020fltrust,cao2022flcert,nguyen2022flame,sun2019can,yin2018byzantine,zhang2022fldetector} in different FL settings. 
We also explore new defenses tailored to \ourmodel{}. However, we can still adapt \ourmodel{} to counter such defense with a minor impact on attack effectiveness. Our results highlight the need for new defense mechanisms. 


To summarize, our key contributions are  as follows:
\begin{itemize}
    \item We uncover a fundamental limitation of existing model poisoning attacks, i.e., their inconsistent malicious model updates across multiple rounds self-cancel the attack effect, leading to suboptimal effectiveness. 
     \item We propose \ourmodel{} that leverages consistent malicious model updates, does not require knowledge about genuine local training data or local models, and is agnostic to the defense deployed by the server.
    \item We perform extensive experiments to demonstrate that \ourmodel{} breaks state-of-the-art FL defenses and outperforms state-of-the-art attacks. 
    \item We explore new defenses tailored to \ourmodel{} and show their insufficiency by adapting \ourmodel{} to counter them. 
\end{itemize}  

\section{Background and Related Work}


\subsection{Federated Learning (FL)}
Suppose we have $n$ clients and each client has a local training dataset $\mathcal{D}_i$, $i=1,2,\cdots,n$. 
FL iteratively trains the global model $\boldsymbol{w}$. In the $t$-th training round, the server first distributes the current global model $\boldsymbol{w}^{t-1}$ to all clients or a subset of them.  For simplicity, we assume all clients are selected, but we will also evaluate a subset of them are selected in each training round in our experiments.  
Each client $i$ then trains a local model $\boldsymbol{w}_i^{t}$ based on the current global model and its local training data. For instance, client $i$ initializes $\boldsymbol{w}_i^{t}$ as $\boldsymbol{w}^{t-1}$ and iteratively updates $\boldsymbol{w}_i^{t}$ via minimizing  $\mathcal{L}{(\boldsymbol{w},\mathcal{D}_i)}$ using Stochastic Gradient Descent (SGD). After training the local model $\boldsymbol{w}_i^{t}$, client $i$ sends the model update $\bm{g}^t_i=\boldsymbol{w}^{t}_i-\boldsymbol{w}^{t-1}$ to the server. Finally, the server aggregates the received  model updates and updates the global model as follows:
\begin{equation}
\boldsymbol{w}^{t}=\boldsymbol{w}^{t-1} + \bm{g}^t,
\label{eq_FedAVG}
\end{equation}
where $\bm{g}^t={{AR}}(\{\bm{g}^t_i\}_{i \in[1, n]})$ is the aggregated model update, ${AR}$ denotes the aggregation rule used by the server, and $\{\bm{g}^t_i\}_{i \in[1, n]}$ is the set of clients' model updates.  For instance, when ${AR}$ is FedAvg~\cite{mcmahan2017communication}, the aggregated model update $\bm{g}^t$ is the average of the clients' model updates.

\subsection{Poisoning Attacks to FL}
In model poisoning attacks~\cite{bagdasaryan2020backdoor,baruch2019little,fang2020local,shejwalkar2021manipulating}, malicious clients corrupt the global model via sending carefully crafted malicious model updates to the server. 
This work focuses on untargeted attacks~\cite{baruch2019little,fang2020local,shejwalkar2021manipulating}, where a poisoned global model 
indiscriminately misclassifies a large number of clean inputs, leading to a large testing error rate and eventually denial-of-service attacks. 

\myparatight{Requiring genuine clients' information} One category of model poisoning attacks~\cite{baruch2019little,fang2020local,shejwalkar2021manipulating} requires that an attacker has access to genuine clients' model updates or local training data. To obtain such knowledge, these attacks often assume an attacker can compromise a large fraction of genuine clients. 
For instance, \emph{a little is enough (LIE) attack}~\cite{baruch2019little} manipulates malicious model updates with small perturbation to the genuine model updates, while Fang et al.~\cite{fang2020local} propose to craft malicious model updates towards the reverse direction of genuine model updates.
\myparatight{Not requiring genuine clients' information} The other category of model poisoning attacks~\cite{cao2022mpaf} do not rely on any genuine clients' information. Therefore, they only assume that an attacker can inject fake clients into an FL system, which is more realistic than compromising genuine clients. For instance, in MPAF~\cite{cao2022mpaf}, the attacker crafts malicious model updates on fake clients towards the initial model.


\myparatight{Limitations of existing attacks}
Existing attacks suffer from the following limitations.
{First, the first category of attacks requires an attacker to compromise a large fraction of genuine clients, which is not affordable in production FL~\cite{shejwalkar2022back}. }
Second, both categories of attacks suffer from self-cancellation of attack effect across training rounds. As a result, they achieve suboptimal attack effectiveness when defenses are deployed. 




\subsection{Defenses against Poisoning Attacks}
Defenses against poisoning attacks to FL can be roughly categorized
into \emph{Byzantine-robust aggregation rules}, \emph{provably robust defenses}, and \emph{malicious clients detection}.

\myparatight{Byzantine-robust aggregation rules} Instead of using the average of the clients' model updates to update the global model, Byzantine-robust aggregation rules~\cite{blanchard2017machine,cao2020fltrust, guerraoui2018hidden,nguyen2022flame,sun2019can,yin2018byzantine,fang2024byzantine,fang2025we} typically employ statistical analysis to filter out or clip outlier model updates before aggregating them.   
For instance, 
 Multi-Krum~\cite{blanchard2017machine} 
selects model update with vector-wise analysis, while  Median (or Trimmed Mean)~\cite{yin2018byzantine} uses dimension-wise median (or trimmed-mean) of the clients' local models.
Norm Bound~\cite{sun2019can} clips the magnitudes of all model updates.
FLTrust~\cite{cao2020fltrust} assumes the server uses a clean and small dataset to bootstrap trust.
Under some assumptions about the learning problem and  clients' local training data distribution, these Byzantine-robust aggregation rules often can theoretically guarantee that the learnt global model does not change much when the number of malicious clients is bounded. 
However, they are still vulnerable to our attack because these assumptions often do not hold and a bounded change in the parameters of the global model does not guarantee a small change of error rate.

\myparatight{Provably robust defenses} Provably robust defenses~\cite{cao2021provably,cao2022flcert,xie2021crfl} provide theoretical guarantees of a lower bound of testing accuracy given a number of malicious clients.    
For instance, FLCert~\cite{cao2022flcert} trains multiple global models using a subset of clients and predicts with a majority vote. 
A major limitation of these provably robust defenses is that they can only tolerate a small number of malicious clients, i.e., the lower bound of testing accuracy reduces to 0 when the number of malicious clients grows (e.g., 20\%). As a result, they are still vulnerable to our attack since an attacker can easily inject some fake clients.

\myparatight{Malicious clients detection} These defenses~\cite{li2020learning,zhang2022fldetector,yueqifedredefense} aim to detect malicious clients during or after the training process. 
For instance, FLDetector~\cite{zhang2022fldetector} detects malicious clients by utilizing the inconsistency of a client's model updates across multiple training rounds.
However, in our attack, we enforce the malicious model updates of a malicious client to be consistent across training rounds, making FLDetector ineffective.

\section{Threat Model}
\label{sec:threat_model}

\myparatight{Attacker's goal}
Following previous works ~\cite{baruch2019little,fang2020local,shejwalkar2021manipulating},
we consider the goal is to poison the training process of FL such that the final learnt global model makes incorrect predictions for clean inputs indiscriminately. In other words, the poisoned global model has a large testing error rate. Such a global model would eventually  be abandoned, leading to denial-of-service attacks~\cite{barreno2006can}.    

\vspace{-1mm}

\myparatight{Attacker's capability}
 We assume that the attacker can inject fake clients into an FL system and control them as malicious clients. We note that this is a practical threat model. Multiple platforms~\cite{simulatefake,NoxPlayer,bluestacks} can be used to automatically inject fake devices/clients in a distributed system. For instance, when the FL system learns a global model on Android smartphones (clients), e.g., Google's next-word prediction on Android devices~\cite{gboard}, an attacker can use a laptop to simulate fake smartphones via Android emulators~\cite{cao2022mpaf}.



\myparatight{Attacker's background knowledge} 
We consider the \emph{minimal} background knowledge scenario, which represents the most realistic threat model. Specifically, the attacker has access to the global models during the training since the server sends them to all clients. However, the attacker has no access to genuine local training data nor local models on any genuine client. Moreover, we assume the attacker does not know the defense deployed by the server.
We note that our threat model is the same as previous fake client based model poisoning attacks~\cite{cao2022mpaf}. 

\section{\ourmodel}
\label{sec:our_framework}

\begin{figure}[t]
    
    \begin{subfigure}{0.33\linewidth}
        \centering
        \includegraphics[width=1\linewidth]{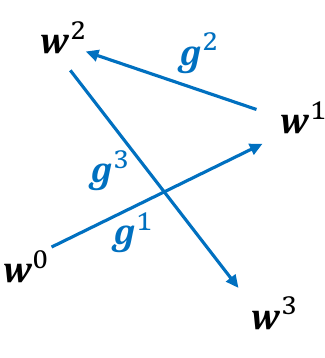}
        \subcaption{Existing attacks}
        \label{fig:method-existing}
    \end{subfigure}
    \begin{subfigure}{0.66\linewidth}
        \centering
        \includegraphics[width=0.98\linewidth]{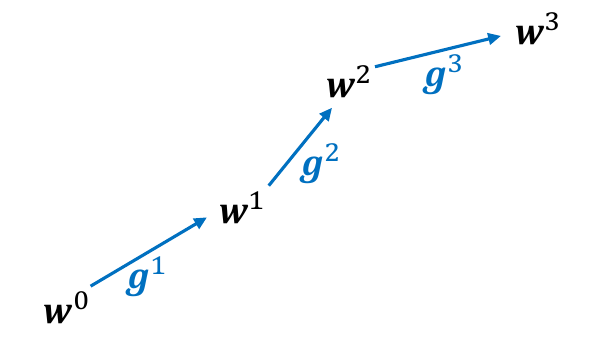}
        \subcaption{\ourmodel{}}
        \label{fig:method-fedfaker}
    \end{subfigure}
    \vspace{-6mm}
    \caption{Illustration of the global-model evolution in three training rounds under existing attacks and \ourmodel{}. The attack effect self-cancels in existing attacks, while \ourmodel{} consistently moves the global model along the same direction.} 
    \label{fig:method}
    \vspace{-5mm}
\end{figure}

\subsection{Overview}

Our \ourmodel{} aims to craft malicious model updates on the fake clients such that the final global model is substantially moved along a random update direction, leading to a large testing error rate.   Towards this goal, our malicious model updates on the fake clients are a dimension-wise product between a sign vector (characterizing the update direction of each dimension) and a magnitude vector (characterizing the update size of each dimension). The sign vector is randomly picked at the beginning and keeps the same across training rounds to enforce consistency; while the magnitude vector dynamically changes to simultaneously have large magnitudes and avoid being filtered out by the unknown defense that may be deployed by the server.
In particular, we  set the magnitude vector in a training round by considering whether our attack is successful in the previous training rounds to avoid being filtered out by a defense.  Thus,  the aggregated model updates across training rounds are likely to have consistent sign vectors (i.e., update direction) and add up instead of cancelling out with each other. As a result,  the final global model is substantially moved along the random update direction, leading to a large testing error rate. 

Fig.~\ref{fig:method} illustrates how the global model evolves in existing attacks and our attack in three training rounds, where $\bm{w}^0$, $\bm{w}^1$, $\bm{w}^2$, and $\bm{w}^3$ are the global models, while $\bm{g}^1$, $\bm{g}^2$, and $\bm{g}^3$ are the aggregated model updates in the three training rounds.  The aggregated model updates self-cancel across training rounds in existing attacks due to inconsistent update directions (i.e., sign vectors), while they add up towards the same, random update direction in our attack.  

\subsection{Formulating an Optimization Problem }
\label{ssec:general}
\myparatight{All-round optimization problem} Suppose we have $n$ genuine clients and $m$ fake clients. The genuine clients are numbered as $1, 2, \cdots, n$, while the fake clients are numbered as $n+1,\cdots, n+m$.  We denote by $T$ the total number of training rounds and $\bm{w}^T$ the final global model. Recall that our attack goal is to poison the training process such that the final global model $\bm{w}^T$  has a large testing error rate. We observe that $\bm{w}^T=\bm{w}^0 +\sum_{t=1}^T \bm{g}^t$, where $\bm{w}^0$ is the initial global model and $\bm{g}^t=AR(\{\bm{g}_i^t\}_{i\in [1,n+m]})$ is the aggregated model update in training round $t$. We denote by $\text{sign}(\sum_{t=1}^T \bm{g}^t)$ the \emph{sign vector} (i.e., update direction) of the \emph{total aggregated model update} $\sum_{t=1}^T \bm{g}^t$ in the $T$ rounds. 

If the {total aggregated model update} $\sum_{t=1}^T \bm{g}^t$ has a large magnitude and a random update direction, then the final global model $\bm{w}^T$ is substantially moved along the random update direction  and would have a large testing error rate no matter what the initial global model  $\bm{w}^0$ is. For instance, Fig.~\ref{fig:norm-performance} in Appendix shows an example where a global model becomes nearly random guessing when moved along a random update direction substantially. Based on this observation, our goal is to craft the malicious model updates $\{\bm{g}_i^t\}_{i\in [n+1,n+m],\ t\in [1,T]}$ on the fake clients in the $T$ rounds such that the total aggregated model update $\sum_{t=1}^T \bm{g}^t$ has a large magnitude along a random update direction. Formally, we formulate the following \emph{all-round} optimization problem:
\begin{equation}
\begin{gathered}
\max _{\bm{g}_i^t, i\in[n+1, n+m], t \in[1, T]}\left\|\sum_{t=1}^T \bm{g}^t \right\| \\
s.t.\ \text{sign}(\sum_{t=1}^T \bm{g}^t) =\bm{s},
 \label{eua:global}
\end{gathered}
\end{equation}
where $\left\|\cdot \right\|$ is $\ell_2$-norm and $\bm{s}$ is a random sign vector that characterizes a random update direction. Each dimension of $\bm{s}$ is picked as either +1 or -1 uniformly at random. However, the all-round optimization problem is hard to solve. In particular, when solving the malicious model updates in training round $t$, the model updates in the future training rounds  are not available yet. Therefore, we transform our {all-round} optimization problem into a \emph{per-round} one.

\myparatight{Per-round optimization problem} Our key idea is to enforce the aggregated model update $\bm{g}^t$ in each training round $t$ to have the sign vector  $\bm{s}$. In particular, for each training round $t$, we reformulate a per-round optimization problem as follows:
\begin{align}
\max _{\bm{g}_i^t, i\in[n+1, n+m]}  \left\| \bm{g}^t \right\|,\ s.t.\  \text{sign}(\bm{g}^t) = \bm{s}. 
 \label{eua:per-round}
\end{align}
Since the aggregated model updates across rounds have the same sign vector $\bm{s}$, their magnitudes add up in the total aggregated model update, leading to a bad global model.

\subsection{Solving the Optimization Problem}
\label{ssec:solve_problem}

It is still challenging to solve the per-round optimization problem in Equation~\ref{eua:per-round}. This is because  $\bm{g}^t=AR(\{\bm{g}_i^t\}_{i\in [1,n+m]})$, but the attacker does not have access to the aggregation rule $AR$ nor the model updates of the genuine clients. We address the challenge via proposing a method to approximately solve the optimization problem. Our key idea is to enforce that the malicious model updates on the fake clients have the same sign vector $\bm{s}$, so the aggregated model updates are  likely to have the sign vector $\bm{s}$ (see Fig.~\ref{fig:matched-fraction} in Appendix for an example in our experiments).  Specifically, we assume that the malicious model update $\bm{g}_{i}^t$ of fake client $i$ in training round $t$ has the following form:
\begin{equation}
\bm{g}_{i}^t =\bm{k}^t \odot \bm{s}, \forall i\in [n+1, n+m], \forall t\in [1,T], 
\end{equation}
where $\bm{k}^t$ is the magnitude vector of the malicious model updates in training round $t$. 

Recall that the objective function of our per-round optimization problem is to maximize the magnitude of the aggregated model update $\left\| \bm{g}^t \right\|$. One naive way to achieve this objective is to select a large magnitude vector $\bm{k}^t$. However, such a magnitude vector $\bm{k}^t$ may be filtered out by the server when a defense (e.g.,  Multi-Krum~\cite{blanchard2017machine},  FLTrust~\cite{cao2020fltrust}, or FLAME~\cite{nguyen2022flame}) is deployed, leading to suboptimal attack effectiveness as  our experiments in Section~\ref{sec:exp} show. Therefore, our goal is to maximize the magnitude vector $\bm{k}^t$ while avoiding being filtered out by an unknown defense that may be deployed by the server. To achieve this goal, our key idea is to pick $\bm{k}^t$ that has a large magnitude but is also similar to the model updates on the genuine clients. In particular, we explicitly decompose the magnitude vector $\bm{k}^t$ into the following form:
\begin{equation}
    \boldsymbol{k}^t = \lambda^t \cdot \boldsymbol{v}^t,
\end{equation}
where $\lambda^t$ is a scaling factor that governs the overall magnitude of $\boldsymbol{k}^t$, and $\boldsymbol{v}^t$ is a unit magnitude vector used to control the relative magnitudes of different dimensions.  $\boldsymbol{v}^t$ has an $\ell_2$-norm of 1. Such decomposition enables us to explicitly solve   $\boldsymbol{v}^t$ and  $\lambda^t$, which we discuss next.  


\myparatight{Dynamically adjusting the unit magnitude vector $\boldsymbol{v}^t$} A straightforward approach to pick the unit magnitude vector $\boldsymbol{v}^t$ is to set each dimension to the same value, which means that we treat all dimensions equally.  
However, as our experiments in Section~\ref{sec:exp} show, this strategy achieves suboptimal attack effectiveness. This is because the dimensions of the model updates on the genuine clients are unlikely to have the same magnitude. As a result, the  model updates of the fake clients are dissimilar to those of the genuine clients, making them be filtered out when a defense is deployed on the server. It is challenging to  construct an unit magnitude vector similar to those of the model updates on the genuine clients, since the attacker does not have access to the model updates on the genuine clients in our threat model. 

To address this challenge, we observe that the global model difference $\bm{w}^{t-1}-\bm{w}^{t-2}$ is the aggregated model update $\bm{g}^{t-1}$ in the previous training round $t-1$. Moreover, $\bm{g}^{t-1}$ is an aggregation of the model updates on the genuine clients and those on the fake clients. The attacker knows $\bm{w}^{t-1}$, $\bm{w}^{t-2}$, and the model updates $\bm{g}_i^{t-1}=\bm{k}^{t-1} \odot \bm{s}$ on the fake clients in the previous training round. Therefore, the attacker can estimate the unit magnitude vector of the genuine clients in the previous training round and use it as $\boldsymbol{v}^t$ in the current training round. In particular, we can normalize the model update $\bm{k}^{t-1} \odot \bm{s}$ on the fake clients in the previous training round to have the same magnitude as $\bm{g}^{t-1}$ and then subtract it from $\bm{g}^{t-1}$ to estimate the model updates on the genuine clients. Formally, we have the following:
\begin{equation}
\label{unitvector}
    \boldsymbol{v}^t = \frac{\left|\bm{g}^{t-1} -  \frac{\left\|\bm{g}^{t-1} \right\|}{\left\| \bm{k}^{t-1} \odot \bm{s}\right\|} \cdot \bm{k}^{t-1} \odot \bm{s} \right|  }{\left\|\bm{g}^{t-1} -  \frac{\left\|\bm{g}^{t-1} \right\|}{\left\|\bm{k}^{t-1} \odot \bm{s}\right\|} \cdot \bm{k}^{t-1} \odot \bm{s} \right\|},
\end{equation}
where $\bm{g}^{t-1}=\boldsymbol{w}^{t-1} -\boldsymbol{w}^{t-2}$ is the aggregated model update in the previous training round,  $\frac{\left\|\boldsymbol{w}^{t-1} -\boldsymbol{w}^{t-2} \right\|}{\left\| \bm{k}^{t-1} \odot \bm{s}\right\|}$ is used to normalize $\bm{k}^{t-1} \odot \bm{s}$ to have the same magnitude as $\bm{g}^{t-1}$ so they are on the same scale for subtraction, the absolute value $|\cdot|$ is used in the numerator because $\boldsymbol{v}^t$ is a magnitude vector, and the denominator is used to normalize $\boldsymbol{v}^t$ to have $\ell_2$-norm of 1.  

\myparatight{Dynamically adjusting the scaling factor $\lambda^t$} A naive approach to pick $\lambda^t$ is to set it as a large number. However, as our experiments in Appendix~\ref{asec:variants} show, this approach may be filtered out by a defense. To address the challenge, we dynamically set $\lambda^t$ based on the global model difference (i.e., aggregated model update) in the previous training round. Our intuition is that, if the aggregated model update  in the previous training round has a large (or small) magnitude, the server may also expect an aggregated model update with a large (or small) magnitude in the current training round. Therefore, we set $\lambda^t$ to be proportional to the magnitude of the aggregated model update (i.e., $\left\|\boldsymbol{w}^{t-1} -\boldsymbol{w}^{t-2}\right\|$) in the previous training round. Formally, we have the following:
\begin{equation}
\lambda^t = c^t  \cdot  \left\|\boldsymbol{w}^{t-1} -\boldsymbol{w}^{t-2}\right\|,
\label{scalingfactor}
\end{equation}
where $c^t$ is a factor.

\myparatight{Dynamically adjusting $c^t$ based on hypothesis testing} We adjust $c^t$ in training round $t$ based on whether the attack is successful in the past $e$ training rounds. 
Specifically, we formulate the following two hypotheses:
\begin{itemize}
    \item {\bf Null hypothesis $H_0$.} The attack is not successful in the past $e$ training rounds. 
    \item {\bf Alternative hypothesis $H_1$.} The attack is successful in the past $e$ training rounds. 
\end{itemize}
If the hypothesis testing favors $H_1$ with a p-value $p$,  $c^t$ does not change, i.e., we set $c^t=c^{t-1}$; otherwise the malicious model updates may have been filtered out by the defense in the past training rounds and thus we decrease $c^t$ by a factor, i.e., $c^t=\beta c^{t-1}$, where $\beta < 1$. 
Specifically, the attacker can calculate the total aggregated model update in the past $e$ training rounds, i.e., $\bm{w}^{t-1} - \bm{w}^{t-e}$. Moreover, the attacker can calculate the number (denoted as $X$) of dimensions of $\bm{w}^{t-1} - \bm{w}^{t-e}$ whose signs match with those in $\bm{s}$. 
Under the null hypothesis $H_0$,  the malicious model updates were filtered out by the defense and thus the signs of the total aggregated model update $\bm{w}^{t-1} - \bm{w}^{t-e}$ are not correlated with   $\bm{s}$. Since $\bm{s}$ is picked uniformly at random, $X$ is a random variable and follows a binomial distribution $Bin(d, 0.5)$, where $d$ is the number of dimensions of the global model. Therefore, if we have $\text{Pr}(x\geq X)\leq p$, then we reject $H_0$ and favor $H_1$,  where $x\sim Bin(d, 0.5)$ and $p$ is the p-value. We use $p=0.01$ in experiments.


\begin{table*}[t]
\centering
\caption{Testing error rate (\%) of the global model under different attacks and defenses for different datasets. 
We assume Fang, Opt. Fang, LIE, Min-Max, and Min-Sum have access to the model updates on \emph{all} genuine clients, and Fang and Opt. Fang further have access to the aggregation rule, which give advantages to these attacks. Random, MPAF, and \ourmodel{} do not have access to these knowledge.  
In each row, we highlight in \textbf{bold} the highest testing error rate and those no better than random guessing.
}
\vspace{-2mm}
\resizebox{0.82\linewidth}{!}{
\begin{tabular}{{c|c|c|cccccccc}}
\toprule
\midrule
\multirow{1}{*}{Dataset}                 & \multirow{1}{*}{Defense} &    \multirow{1}{*}{No Attack}                             & 
Fang    &  Opt. Fang     & LIE  & Min-Max   & Min-Sum   &Random &MPAF   & \ourmodel{}   \\ \midrule
\multirow{9}{*}{MNIST}  
& FedAvg & 2.11 & 13.66 & \textbf{90.08} & 2.28 & \textbf{97.89} & 2.73 & \textbf{90.02} & \textbf{90.04} & \textbf{90.02} \\
& Multi-Krum & 2.13 & 5.98 & 6.80 & 2.34 & 6.23 & 5.02 & 2.55 & 2.80 &  \textbf{75.28}\\
& Median & 4.72 & 8.43 & 7.49 & 4.27 & 8.52 & 6.28 & 3.66 & 8.71 & \textbf{86.49} \\
& TrMean & 2.14 & 5.66 & 6.79 & 2.82 & 5.66 & 3.92 & 3.02 & 11.27 & \textbf{88.73} \\
& Norm Bound & 2.33 & 5.46 & 6.02 & 2.46 & 6.87 & 2.94 & 2.67 & 31.02 & \textbf{90.26} \\
& FLTrust& 3.43 & 4.00&	5.04&	3.41&	12.56&	8.89&	3.44&	3.43&	\textbf{88.65} \\
& FLAME  &2.86&	2.66	&2.45	&2.61	&2.60	&4.26&	2.84	&2.72&	\textbf{88.59}\\
& FLCert  &3.34& 4.61&4.61&2.83&4.57&4.18&3.35&6.46&\textbf{88.06}\\
& FLDetector & 4.72 & 4.72 & 4.72 &6.73 & \textbf{95.21} & \textbf{99.65} & 4.72 & 4.72 & \textbf{90.02} \\
\midrule
\multirow{9}{*}{FashionMNIST} 
& FedAvg & 14.96 & 33.46 & \textbf{90.00} & 16.87 & \textbf{99.04} & 17.33 & \textbf{90.00} & \textbf{90.00} & \textbf{90.00} \\
& Multi-Krum & 15.58 & 22.06 & 22.79 & 15.36 & 17.35 & 16.27 & 16.09 & 15.52 & \textbf{78.06} \\
& Median & 20.61 & 27.82 & 28.05 & 20.38 & 23.30 & 20.95 & 21.23 & 23.19 & \textbf{87.75} \\
& TrMean & 15.36 & 27.96 & 28.29 & 19.50 & 23.68 & 20.41 & 18.90 & 26.57 & \textbf{85.36} \\
& Norm Bound & 15.57 & 28.76 & 30.31 & 16.03 & 27.68 & 18.37 & 16.45 & 43.67 & \textbf{86.33} \\
& FLTrust& 16.73&17.52	&18.99&	12.50&	21.32&	20.23	&15.78	&16.83	&\textbf{88.41} \\
& FLAME  & 15.94&	18.02&	15.51	&15.45	&17.75	&19.10	&16.14&	16.58&\textbf{71.91}\\
& FLCert  & 17.49&18.78&19.89&16.76&19.28&18.28&17.25&24.20&\textbf{71.92}\\
& FLDetector & 20.61 & 20.61 & 20.61 & 20.73 & 22.44 & \textbf{90.00} & 20.61 & 20.61 & \textbf{90.00} \\
\midrule
\multirow{9}{*}{Purchase} 
& FedAvg & 8.67 & \textbf{97.01} & \textbf{99.32} & 8.45 & 34.23 & 25.02 & \textbf{99.21} & \textbf{99.20} & \textbf{99.34} \\
& Multi-Krum & 11.09 & 16.78 & 17.04 & 11.87 & 14.56 & 12.77 & 12.07 & 12.11 & \textbf{73.59} \\
& Median & 16.73 & 50.22 & 49.56 & 16.07 & 38.80 & 38.21 & 16.55 & 37.34 & \textbf{70.50} \\
& TrMean & 8.35 & \textbf{70.02} & 65.23 & 8.78 & 20.79 & 20.97 & 9.91 & 35.05 &  53.86\\
& Norm Bound & 8.26 & 20.75 & 20.80 & 8.97 & 18.43 & 18.97 & 9.04 & 26.02 &  \textbf{70.72}\\
& FLTrust&21.45	&56.05	&20.24	&24.76&	65.16&	57.69&	25.54&	23.45	&\textbf{67.27}\\
& FLAME  & 12.25	&12.36	&11.89&	12.27&	12.02	&22.46	&12.04	&11.95&	\textbf{28.09}  \\
& FLCert  & 18.32 & 38.89 &40.05 &19.62&29.03&38.47&25.67 &34.93 &\textbf{43.56} \\
& FLDetector  & 16.73 & 16.73 & 16.73 &37.08& \textbf{99.97} & 38.84 & 16.73 & 16.73 & \textbf{99.38}\\
 \midrule
\multirow{9}{*}{CIFAR-10} 
& FedAvg & 34.16 & 72.10 & \textbf{90.00} & 34.90 & \textbf{95.03} & 50.90 & \textbf{90.00} & \textbf{90.00} & \textbf{90.01} \\
& Multi-Krum & 34.75 & 44.36  & 45.02 & 34.63 & 44.76 &  40.62& 34.98 & 34.77 &  \textbf{90.00}\\
& Median & 34.47 & 52.01 & 50.93 & 34.32 & 54.55 & 46.00 & 37.43 & 61.04 & \textbf{90.01} \\
& TrMean & 34.68 & 50.03 & 53.45 & 35.22 & 49.04 & 49.03 & 34.78 & 67.73 & \textbf{90.00} \\
& Norm Bound & 34.99 & 58.00 & 57.92 & 36.02 & 62.90 & 43.22 & 34.67 & \textbf{90.02} & \textbf{90.02} \\
& FLTrust&34.83& 33.40&	33.91	&34.21	&34.25&	46.51	&34.58&	34.03	&\textbf{88.81}\\
& FLAME  & 37.52&	37.94&	36.79&	35.43&	37.31	&43.26&	36.44&	37.00&	\textbf{80.12} \\ 
& FLCert &37.92&37.79&37.20&40.03&37.70&38.24&39.05&61.63&\textbf{90.00}  \\
& FLDetector  & 34.47 & 34.47& 34.47 &37.75& \textbf{90.01}& \textbf{90.02} & 34.47 & 34.47  & \textbf{90.00}\\
\midrule
\multirow{9}{*}{FEMNIST} 
& FedAvg & 25.67 & 40.04 & 95.02 & 25.56 & 37.99 & 17.58 & \textbf{95.17} & \textbf{95.17} & \textbf{95.17} \\
& Multi-Krum & 25.63 & 29.53 & 32.25 & 26.27 & 32.31 & 30.28 & 26.21 & 25.97 & \textbf{95.18} \\
& Median & 29.90 & 39.02 & 39.94 & 26.66 & 40.09 & 38.04 & 30.03 & 38.26 & \textbf{91.54} \\
& TrMean & 25.55 & 40.42 & 41.31 & 27.01 & 40.02 & 33.54 & 26.77 & 41.47 & \textbf{75.23} \\
& Norm Bound & 26.17 & 30.54 & \textbf{95.54} & 26.52 & 35.22 & 28.65 & 26.82 & 87.31 & \textbf{95.54} \\
& FLTrust&20.13&	20.74 & 23.88 & 22.51& 95.09& 95.11 & 22.42& 20.29 & \textbf{95.17}\\
& FLAME  & 27.23	&26.60	&26.64	&24.41	&31.48	&30.14	&26.72	&26.34&	\textbf{94.91} \\ 
& FLCert  & 25.12 &30.41&30.32&23.45&30.09&28.97&24.69 & 31.75&\textbf{92.46}\\
& FLDetector  & 30.02  & 29.90 & 29.90  & 35.95& 94.88 & \textbf{94.97} & 29.90 & 29.90  & 91.25\\
\midrule
\bottomrule
\end{tabular}
}
\label{tab:overall}
\end{table*}
\section{Evaluation}
\label{sec:exp}

\subsection{Experimental Setup}
\label{sec:setup}


\subsubsection{Datasets}
We use five datasets from different domains, including MNIST~\cite{lecun2010mnist}, FashionMNIST~\cite{xiao2017/online}, Purchase~\cite{purchase}, CIFAR-10~\cite{krizhevsky2009learning}, and FEMNIST~\cite{caldas2018leaf}.
Detailed introduction and default non-IID degrees are deferred to Appendix~\ref{app:dataset}. 

\subsubsection{Evaluation Metric}

We use the testing error rate of the global model as the evaluation metric, which is the fraction of testing inputs that are misclassified by the global model. 
A larger testing error rate indicates a  better attack. 

\subsubsection{Defenses and Compared Attacks}

We evaluate eight state-of-the-art defenses, including six Byzantine-robust aggregation rules (Multi-Krum~\cite{blanchard2017machine}, Median~\cite{yin2018byzantine}, Trimmed Mean~\cite{yin2018byzantine}, Norm Bound~\cite{sun2019can}, FLTrust~\cite{cao2020fltrust}, and FLAME~\cite{nguyen2022flame}), one provably robust defense (FLCert~\cite{cao2022flcert}) and one malicious clients detection method (FLDetector~\cite{zhang2022fldetector}). We also consider the non-robust FedAvg~\cite{mcmahan2017communication} as a baseline. 
We compare our  \ourmodel{} with seven attacks, including five attacks (Fang~\cite{fang2020local}, Opt. Fang~\cite{shejwalkar2021manipulating}, LIE~\cite{baruch2019little}, Min-Max~\cite{shejwalkar2021manipulating}, and Min-Sum~\cite{shejwalkar2021manipulating}) that require genuine clients' information and two attacks (Random~\cite{cao2022mpaf} and MPAF~\cite{cao2022mpaf}) that do not require such information. 
Detailed introduction is deferred to Appendix~\ref{app:defense} and \ref{app:attack}.

\begin{figure*}[t]
\includegraphics[width=1\linewidth]{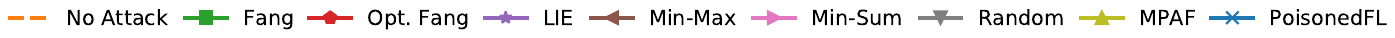}
        \begin{subfigure}{0.121\linewidth}
        \centering
        \includegraphics[width=1\linewidth]{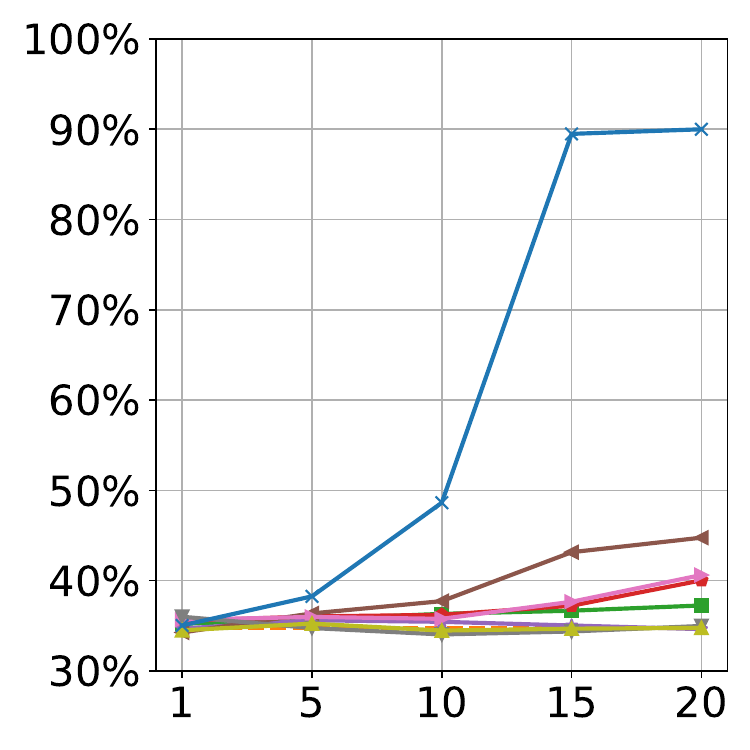}
        \caption{Multi-Krum}
    \end{subfigure}
    \begin{subfigure}{0.121\linewidth}
        \centering
        \includegraphics[width=1\linewidth]{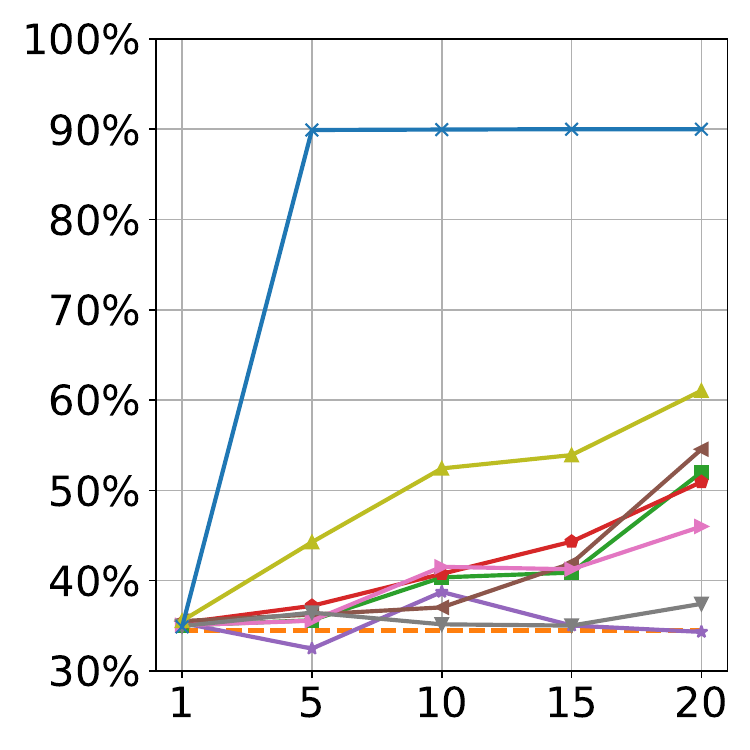}
        \caption{Median}
    \end{subfigure}
    \begin{subfigure}{0.121\linewidth}
        \centering
        \includegraphics[width=1\linewidth]{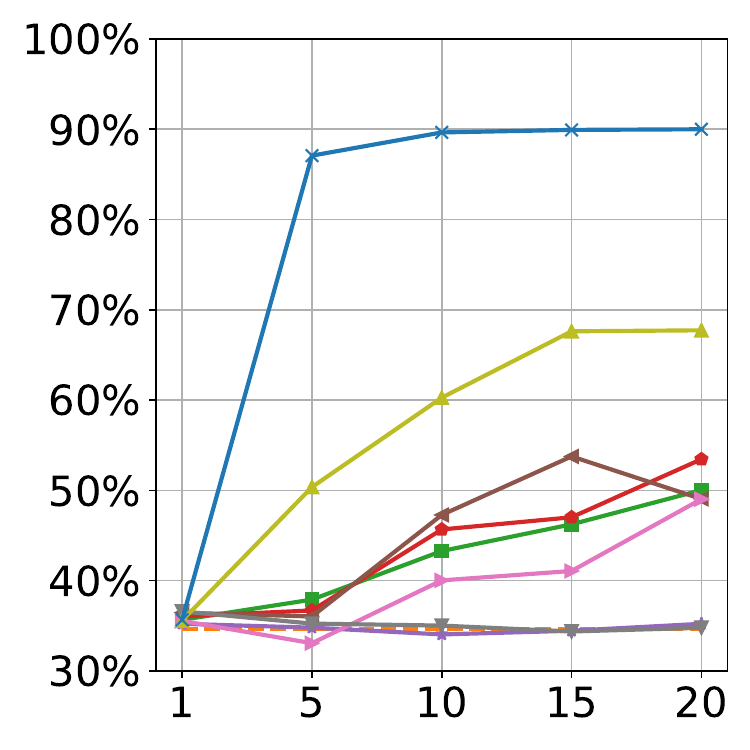}
        \caption{TrMean}
    \end{subfigure}
    \begin{subfigure}{0.121\linewidth}
        \centering
        \includegraphics[width=1\linewidth]{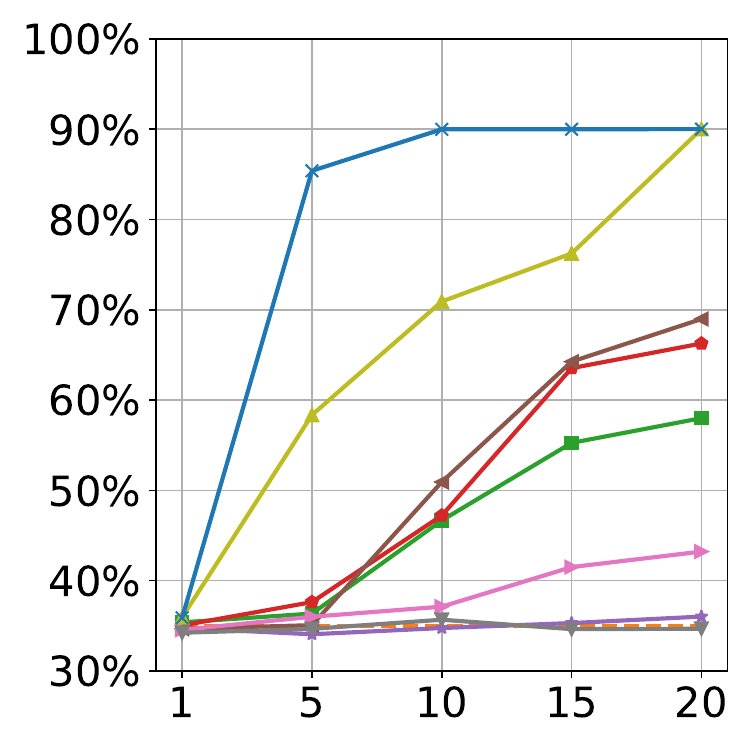}
        \caption{Norm Bound}
    \end{subfigure}
                \begin{subfigure}{0.121\linewidth}
        \centering
        \includegraphics[width=1\linewidth]{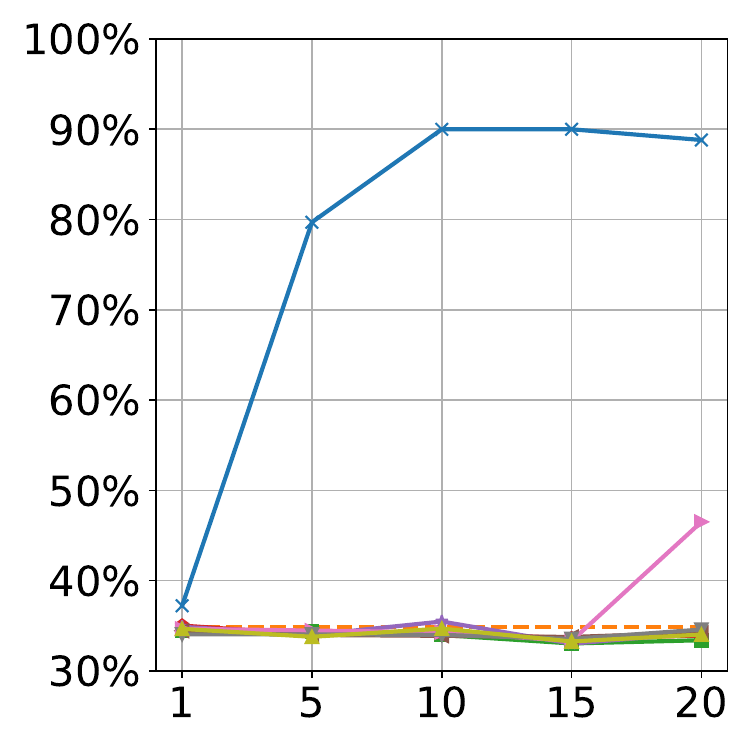}
        \caption{FLTrust}
         \end{subfigure}
         \begin{subfigure}{0.121\linewidth}
        \centering
        \includegraphics[width=1\linewidth]{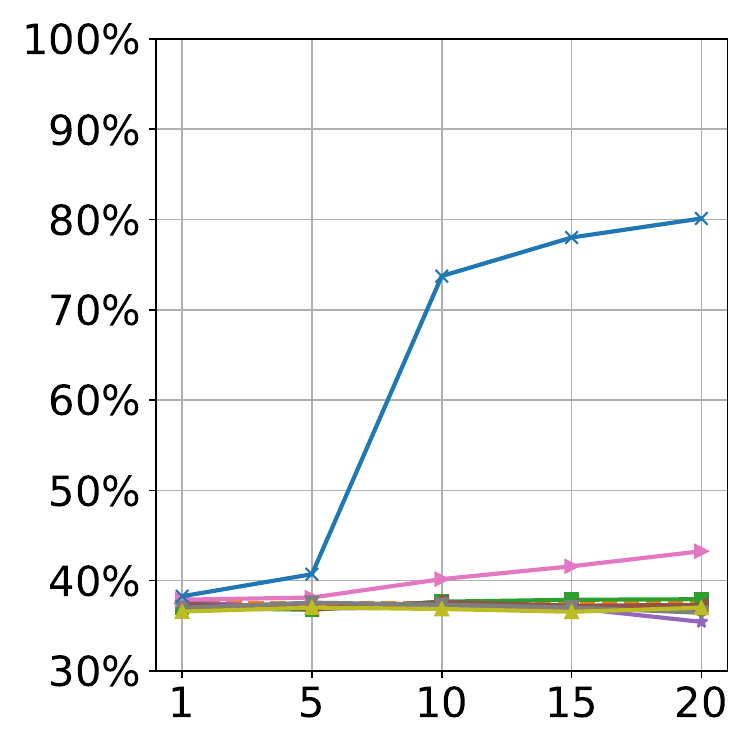}
        \caption{FLAME}
    \end{subfigure}
            \begin{subfigure}{0.121\linewidth}
        \centering
        \includegraphics[width=1\linewidth]{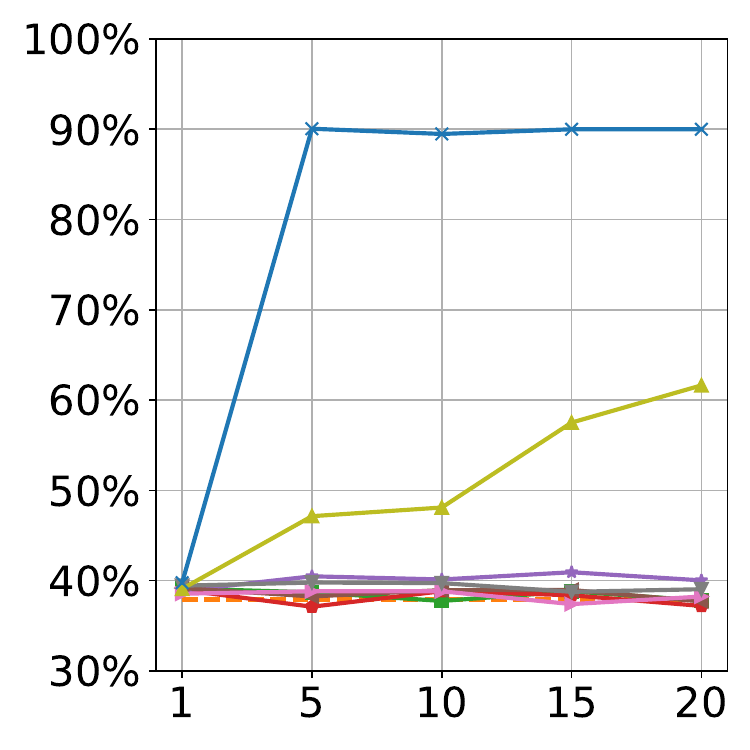}
        \caption{FLCert}
    \end{subfigure}
        \begin{subfigure}{0.121\linewidth}
        \centering
        \includegraphics[width=1\linewidth]{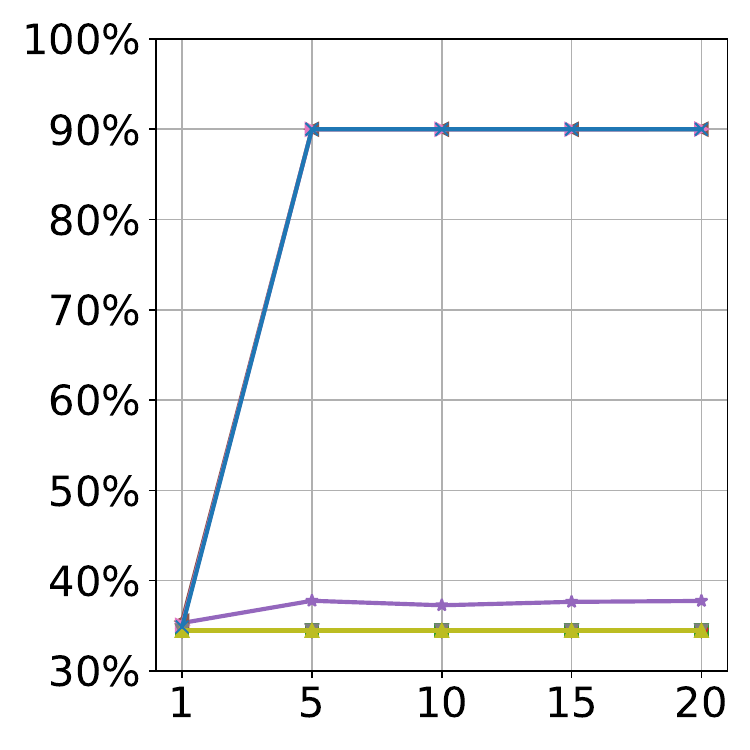}
        \caption{FLDetector}
    \end{subfigure}
      \caption{Testing error rate of the global model as a function of the fraction of fake clients (\%) under  different scenarios.}
    \label{fig:fraction}
    \vspace{-3mm}
\end{figure*}

\begin{figure*}[t]
\includegraphics[width=1\linewidth]{Figures/legend2.pdf}
    \begin{subfigure}{0.121\linewidth}
        \centering
        \includegraphics[width=1\linewidth]{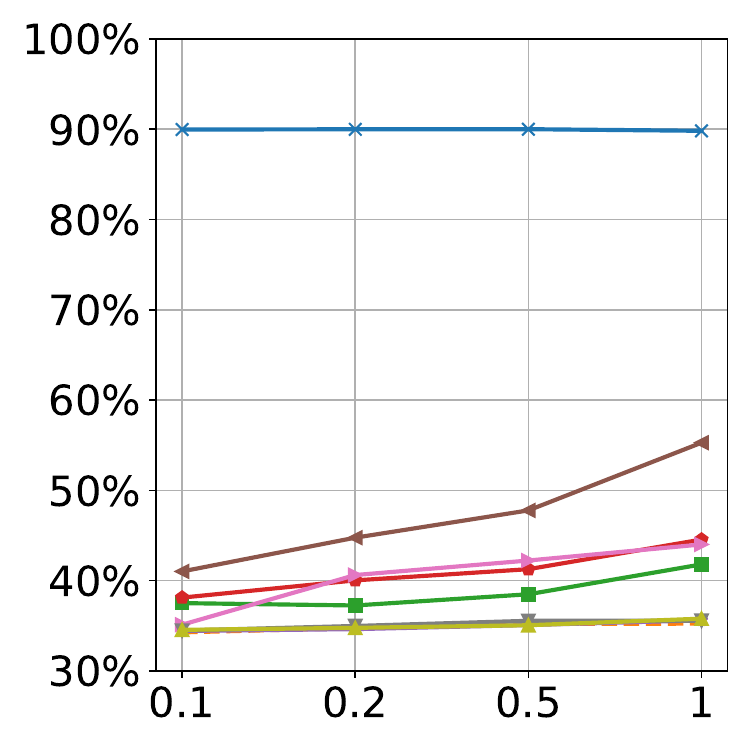}
        \subcaption{Multi-Krum}
    \end{subfigure}
    \begin{subfigure}{0.121\linewidth}
        \centering
        \includegraphics[width=1\linewidth]{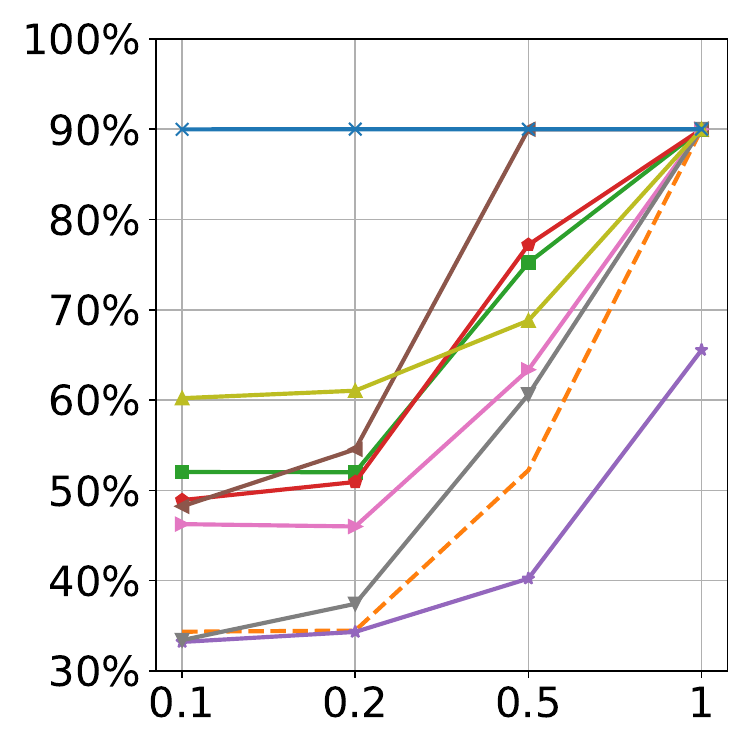}
        \subcaption{Median}
    \end{subfigure}
    \begin{subfigure}{0.121\linewidth}
        \centering
        \includegraphics[width=1\linewidth]{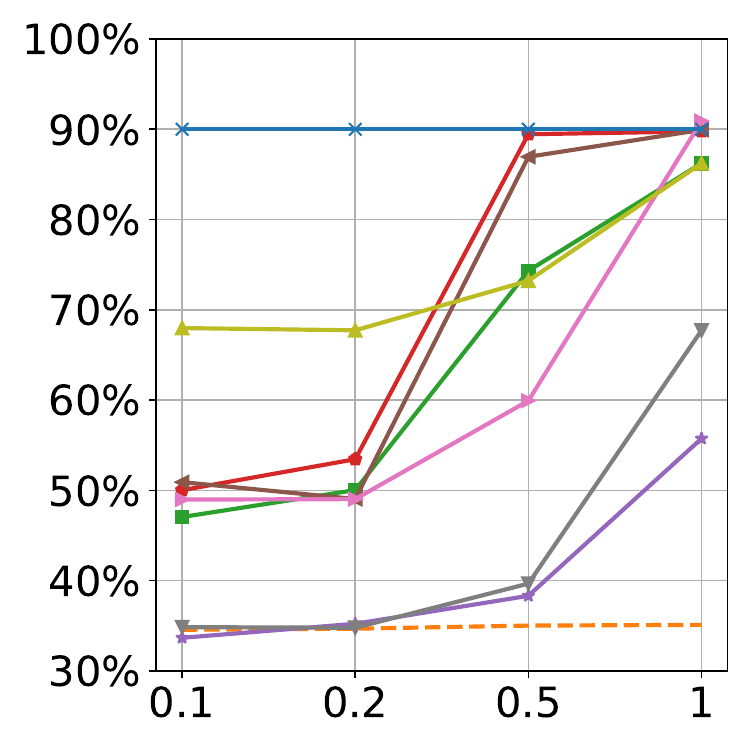}
        \subcaption{TrMean}
    \end{subfigure}
    \begin{subfigure}{0.121\linewidth}
        \centering
        \includegraphics[width=1\linewidth]{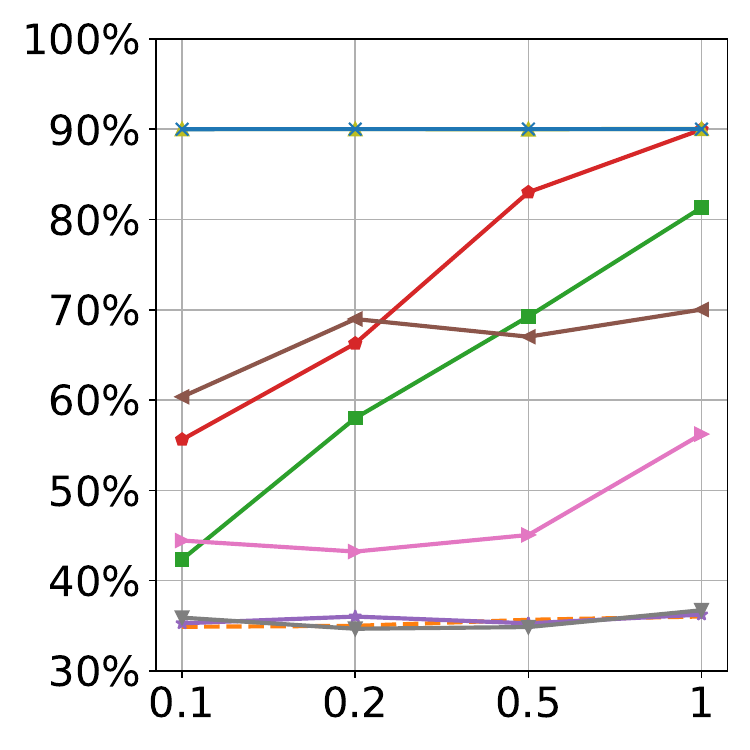}
        \subcaption{Norm Bound}
    \end{subfigure}
        \begin{subfigure}{0.121\linewidth}
        \centering
        \includegraphics[width=1\linewidth]{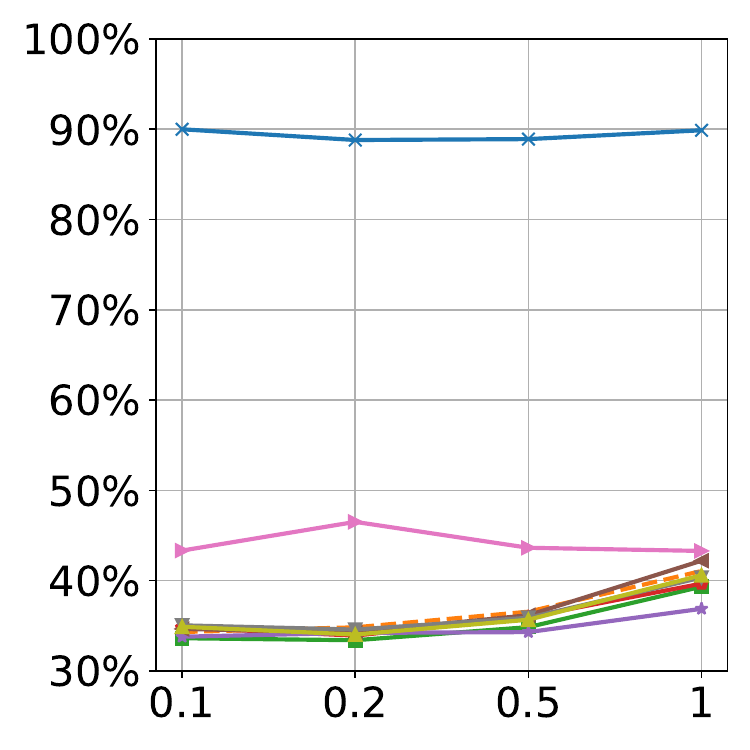}
        \subcaption{FLTrust}
    \end{subfigure}
    \begin{subfigure}{0.121\linewidth}
        \centering
        \includegraphics[width=1\linewidth]{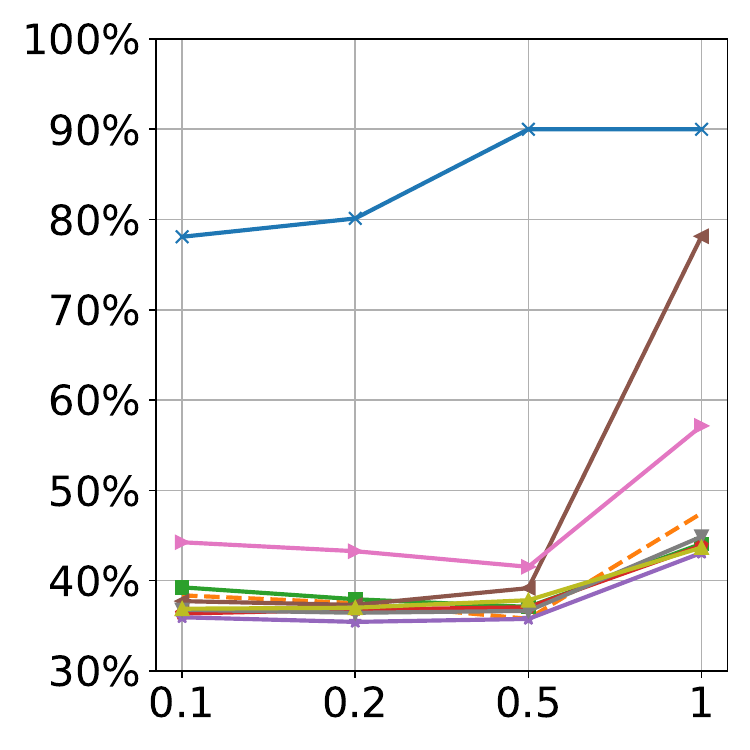}
        \caption{FLAME}
    \end{subfigure}
            \begin{subfigure}{0.121\linewidth}
        \centering
        \includegraphics[width=1\linewidth]{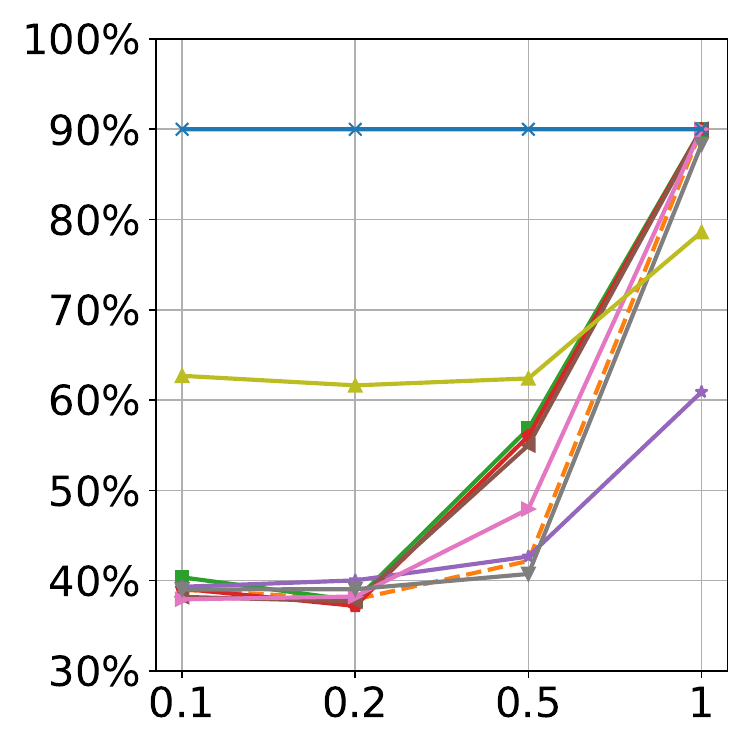}
        \caption{FLCert}
    \end{subfigure}
        \begin{subfigure}{0.121\linewidth}
        \centering
        \includegraphics[width=1\linewidth]{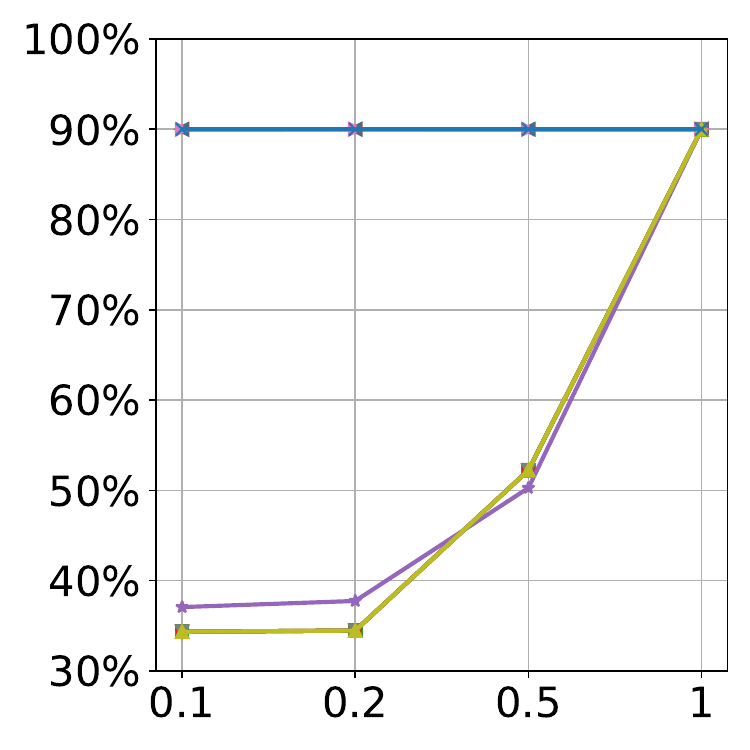}
        \subcaption{FLDetector}
    \end{subfigure}
    \caption{Testing error rate of the global model as a function of the degree of non-IID under different defenses and attacks.}
    \label{fig:noniid}
    \vspace{-3mm}
\end{figure*}

\begin{figure*}[t!]
\includegraphics[width=1\linewidth]{Figures/legend2.pdf}
    \begin{subfigure}{0.121\linewidth}
        \centering
        \includegraphics[width=1\linewidth]{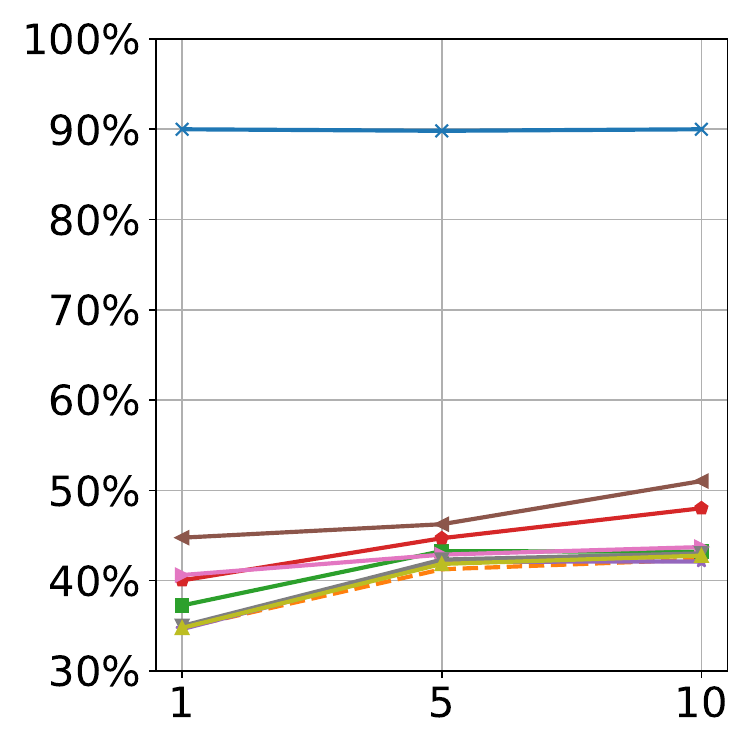}
        \subcaption{Multi-Krum}
    \end{subfigure}
    \begin{subfigure}{0.121\linewidth}
        \centering
        \includegraphics[width=1\linewidth]{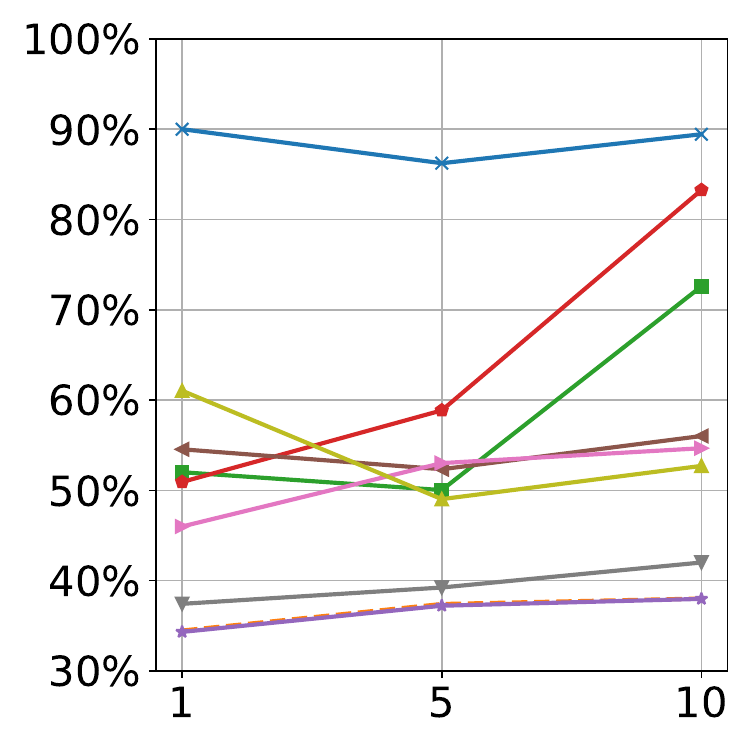}
        \subcaption{Median}
    \end{subfigure}
    \begin{subfigure}{0.121\linewidth}
        \centering
        \includegraphics[width=1\linewidth]{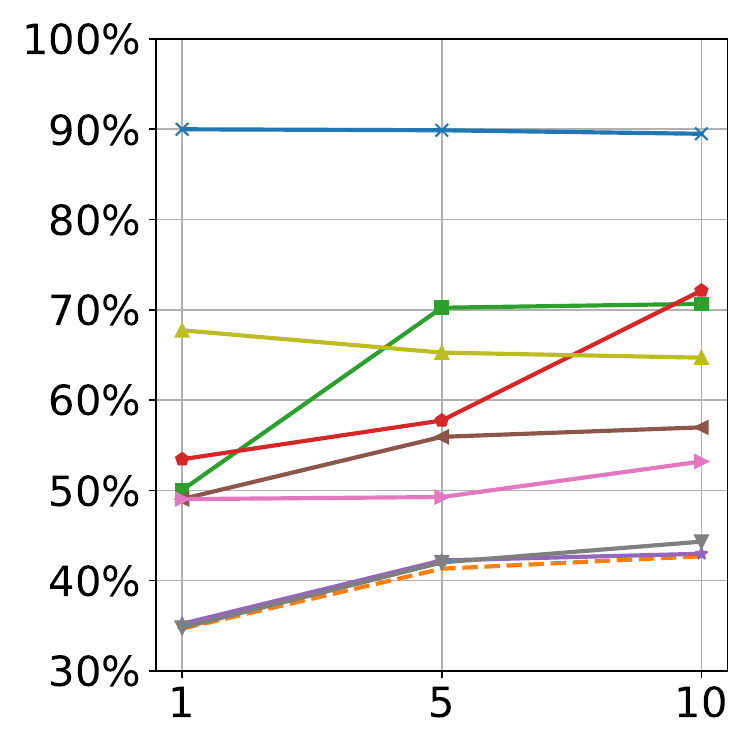}
        \subcaption{TrMean}
    \end{subfigure}
    \begin{subfigure}{0.121\linewidth}
        \centering
        \includegraphics[width=1\linewidth]{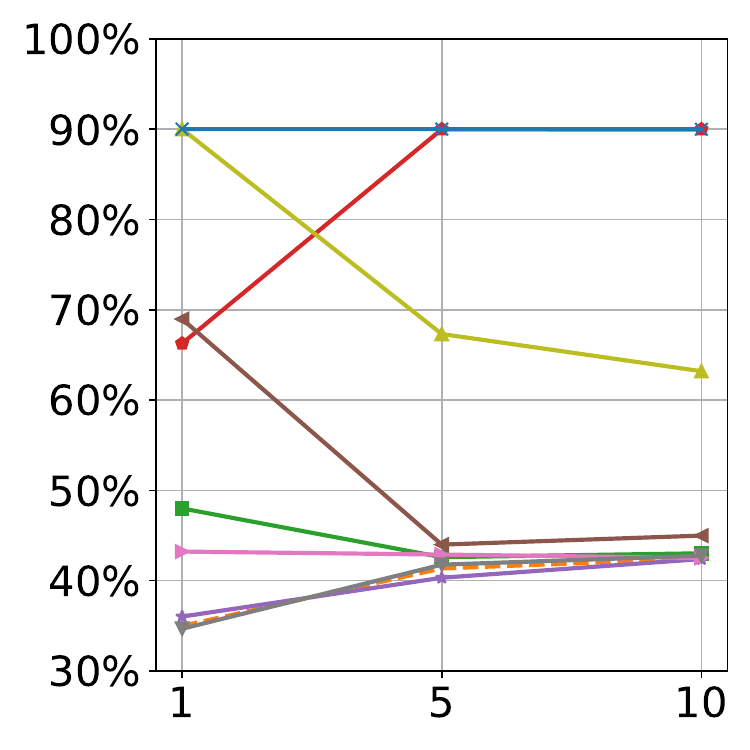}
        \subcaption{Norm Bound}
    \end{subfigure}
        \begin{subfigure}{0.121\linewidth}
        \centering
        \includegraphics[width=1\linewidth]{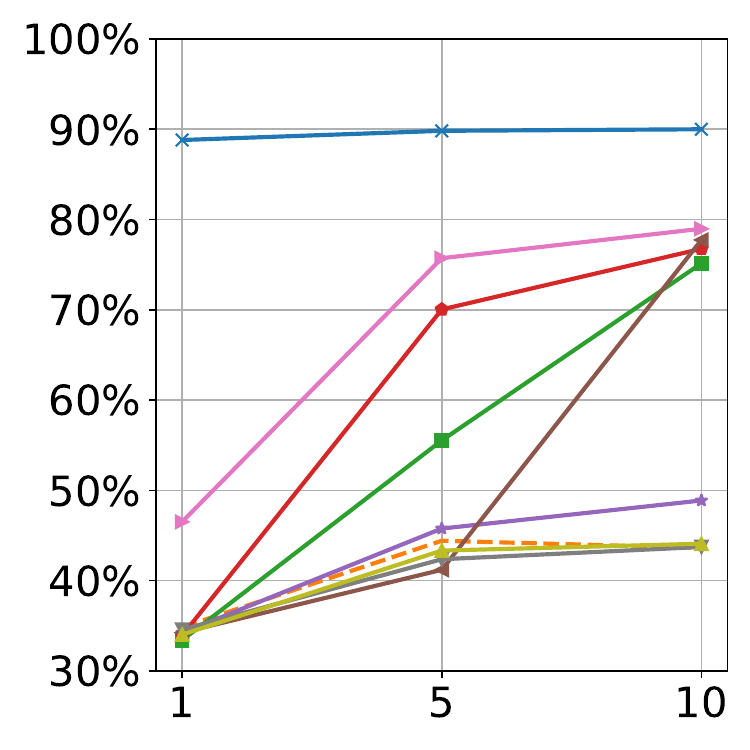}
        \subcaption{FLTrust}
        \end{subfigure}
        \begin{subfigure}{0.121\linewidth}
        \centering
        \includegraphics[width=1\linewidth]{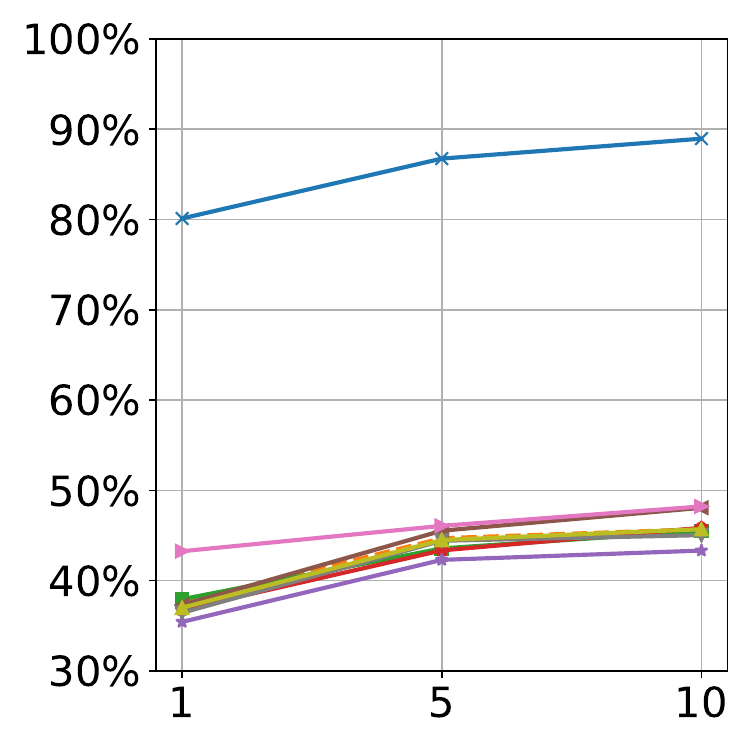}
        \caption{FLAME}
    \end{subfigure}
                \begin{subfigure}{0.121\linewidth}
        \centering
        \includegraphics[width=1\linewidth]{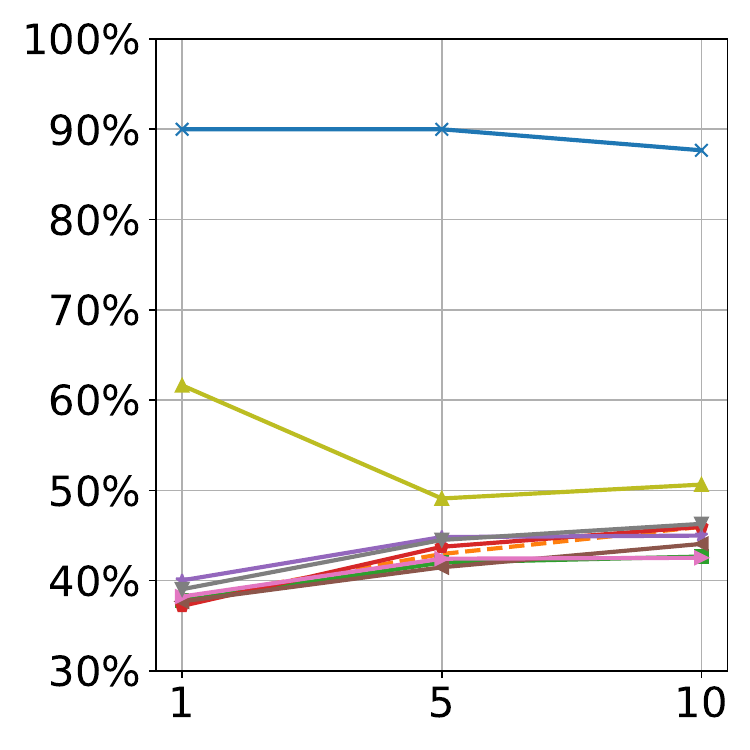}
        \caption{FLCert}
    \end{subfigure}
    \begin{subfigure}{0.121\linewidth}
        \centering
        \includegraphics[width=1\linewidth]{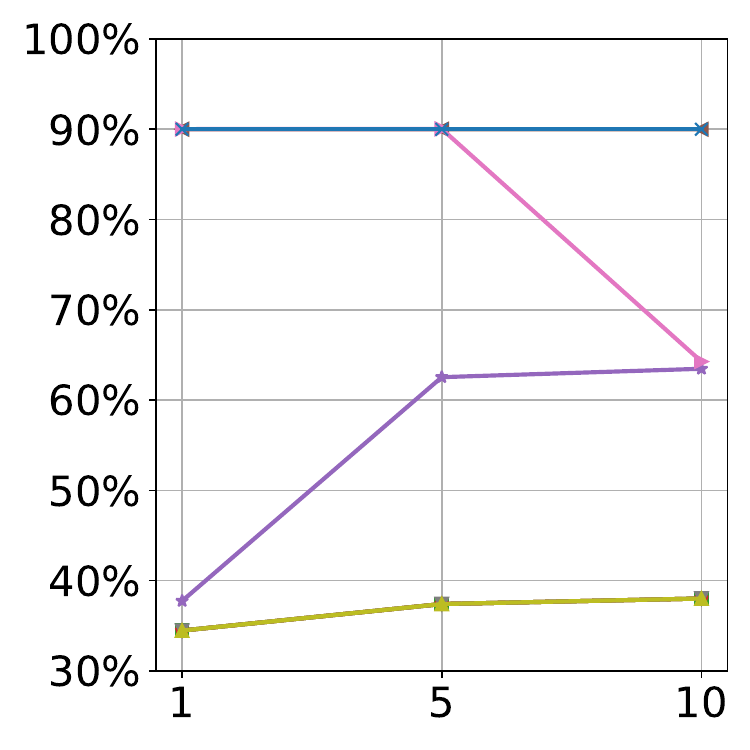}
        \subcaption{FLDetector}
    \end{subfigure}
    \caption{Testing error rate of the global model as a function of local training epochs under different defenses and attacks.}
    \label{fig:local}
     \vspace{-3mm}
\end{figure*}

\subsubsection{FL Settings}\label{sec:paramtersetting}
By default, we assume $1,200$ genuine clients and 20\% of fake clients, i.e., the number of fake clients is 20\% of the number of genuine clients. 
In each training round, the server selects a random subset of 10\% of clients to participate in training.
A  genuine client trains its local model for 1 local epoch in a training round using SGD. 
Due to limited space, more FL settings for the five datasets are detailed in Appendix~\ref{app:setting}. 
We also study the impact of different FL settings (fraction of fake clients, non-IID degree, local training epochs, and participation rate) in Section~\ref{sec:settingimpact}.
Our  \ourmodel{} has parameters $c^t$, $e$, and $\beta$. We set $c^0=8$, $e=50$, and $\beta=0.7$, and we do not decrease $c^t$ to be smaller than $0.5$ since a too small $c^t$ is ineffective.  
We conduct our experiments using 16 GeForce RTX 3090 GPUs.






\subsection{Main Results}

Table~\ref{tab:overall} shows the testing error rates of the learnt global models under different attacks and defenses for the five datasets. ``No Attack'' means no fake clients are injected. 

\myparatight{\ourmodel{}  breaks state-of-the-art FL defenses}
 \ourmodel{} substantially increases the testing error rates of the global models learnt by FedAvg and  state-of-the-art defenses, compared to No Attack.  Specifically, \ourmodel{} increases the testing error rate by 15.84 to 90.67 across the five datasets and nine FL methods. Moreover, in 29 out of the 45 cases,  \ourmodel{} makes the global model nearly random guessing or worse. Our results show that existing FL defenses are not robust even if an attacker has no knowledge about the defense nor the genuine clients' local training data or models. 
We detail \textbf{why \ourmodel{} breaks FL defenses} in Appendix~\ref{app:analysis}.

\myparatight{\ourmodel{} outperforms existing attacks} 
For the non-robust FedAvg,  both  \ourmodel{} and some existing attacks can reduce the global models to random guessing or worse. However, for the five datasets and eight FL defenses, our \ourmodel{} achieves a much larger testing error rate than existing attacks in 32 out of the 40 cases with the gap ranging from 2.11 to 84.33. In six of the remaining cases, both our \ourmodel{} and some existing attacks make the learnt global models nearly random guessing or worse. Our \ourmodel{} achieves a smaller testing error rate than some existing attacks only in  two cases, where  \ourmodel{} still achieves large  testing error rates even in these rare cases.

Among the three attacks Random, MPAF, and \ourmodel{}, which do not require knowledge about the local training data/models on the genuine clients nor the defense, Random is ineffective in most cases when a defense is deployed; while MPAF is effective to  some extent at  attacking some defenses, e.g., TrMean and Norm  Bound, but ineffective at attacking other defenses, e.g., Multi-Krum, FLTrust, and FLDetector.  \ourmodel{} substantially outperforms Random and MPAF when a defense is deployed in all cases except (CIFAR10, Norm Bound), for which both  MPAF and \ourmodel{} make the global models random guessing. For instance, on MNIST dataset, \ourmodel{} achieves 59.24 to 85.87 higher testing error rates than MPAF at attacking a defense.  
We detail \textbf{why \ourmodel{} outperforms existing attacks} in Appendix~\ref{app:analysis}.

\subsection{Impact of FL Settings}
\label{sec:settingimpact}
We study the impact of different FL settings, including the fraction of fake clients, degree of non-IID, the number of local training epochs, and participation rate, on the effectiveness of existing attacks and our attack.
Unless otherwise mentioned, we show experimental results on CIFAR-10 dataset for simplicity, but we have similar observations on other datasets. Moreover, when studying the impact of one parameter, we use the default settings for other parameters discussed in Section~\ref{sec:paramtersetting}.

\myparatight{Impact of the fraction of fake clients} 
Fig.~\ref{fig:fraction} shows the testing error rate of the global model as the fraction of fake clients increases from 1\% to 20\%. 
First, we observe that  our  \ourmodel{} is more effective, i.e., the testing error rate increases, as more  fake clients are injected. This is because the aggregated model update in each training round is more likely to have the sign vector $\bm{s}$ when more fake clients send the malicious model updates with the sign vector $\bm{s}$ to the server. 
Second, using only 5\% of fake clients, our \ourmodel{} can already break all FL defenses except Multi-Krum and FLAME. 
However, existing attacks cannot reach such effectiveness even if 20\% of fake clients are injected, except MPAF under Norm Bound defense.  


\myparatight{Impact of the degree of non-IID} 
In our experiments, we use a parameter $q$ to control the degree of non-IID among the genuine clients' local training data, where $0.1 \le q \le 1$~\cite{fang2020local}.  
A larger $q$ indicates a larger degree of non-IID, where $q=0.1$ means the genuine clients' local training data is IID and $q=1$ indicates highly non-IID (i.e., each class of data is distributed among a subset of clients). 
Fig.~\ref{fig:noniid} shows the testing error rate of the global model as $q$ increases from 0.1 to 1 under different attacks and defenses. 
We observe that \ourmodel{} consistently achieves high testing error rates (i.e., random guessing) across different settings of $q$ under different defenses. However, it is harder for existing attacks to compromise the defenses when $q$ is smaller, e.g., $q=0.1$. 
This is because, in such cases, the  genuine model updates are less diverse, leaving less room for existing attacks to craft malicious model updates. We note that existing attacks are substantially less effective than \ourmodel{} under Multi-Krum and FLTrust even if $q=1.0$. 

\myparatight{Impact of the local training epochs} 
Fig.~\ref{fig:local} shows the testing error rate of the global model under different defenses and attacks as the number of local training epochs ranges from 1 to 10.
We observe that our \ourmodel{} can effectively break the defenses (i.e., reducing the global models to be nearly random guessing) in all considered scenarios. Moreover, our \ourmodel{} consistently achieves testing error rates no smaller than existing attacks. We do not observe any particular pattern with respect to how the effectiveness of an existing attack varies as the number of local training epochs changes. For instance, MPAF achieves a larger testing error rate as the number of local training epochs increases under Multi-Krum, but its testing error rate decreases under defenses like TrMean and Norm Bound.

Due to limited space, we show the impact of  \textbf{participation rate} and \textbf{ablation study} on variants of \ourmodel{} in Appendix \ref{asec:pr}, \ref{asec:variants}, and \ref{app:sign}.

\section{Discussion and Limitations}
\myparatight{Countermeasures and adaptive attacks} If the attack is known, we propose \textbf{tailored defenses}. Specifically, we study a new defense that normalizes the total aggregated model update since \ourmodel{} aims to increase its magnitude.  We also explore \textbf{GMM-Sign} and \textbf{GMM-Magnitude} to detect fake clients based on how \ourmodel{} crafts the malicious model updates. We show that the normalization-based defense has limited effectiveness at mitigating \ourmodel{} and we can still adapt \ourmodel{} to break the detection based defenses. Due to limited space, the detailed methods and evaluation are deferred to Appendix~\ref{app:counter}.

Further discussion on the potential application of \textbf{synthetic data based attacks} and \textbf{limitations in cross-silo FL} is specified in Appendix~\ref{app:sythetic} and \ref{app:limitation}, respectively.

\section{Conclusion}
In this work, we uncover the fundamental self-cancellation issue of existing model poisoning attacks to FL. Based on our discovery, we propose \ourmodel{}, a new model poisoning attack to FL. 
We show that, by maintaining multi-round consistency, an attacker can substantially increase the testing error rate of the learnt global model, even if the attacker does not have access to the genuine local training data, genuine local models, or deployed defense.  
\ourmodel{} highlights the severe vulnerability of FL and the urgency for developing new defense mechanisms.

{
    \small
    \bibliographystyle{ieeenat_fullname}
    \bibliography{example_paper}  

\begin{thebibliography}{40}
\providecommand{\natexlab}[1]{#1}
\providecommand{\url}[1]{\texttt{#1}}
\expandafter\ifx\csname urlstyle\endcsname\relax
  \providecommand{\doi}[1]{doi: #1}\else
  \providecommand{\doi}{doi: \begingroup \urlstyle{rm}\Url}\fi

\bibitem[Nox()]{NoxPlayer}
Noxplayer, the perfect android emulator to play mobile games on pc.
\newblock https://www.bignox.com/.

\bibitem[blu()]{bluestacks}
The world's first cloud-based android gaming platform.
\newblock https://www.bluestacks.com/.

\bibitem[gbo()]{gboard}
Federated learning: Collaborative machine learning without centralized training data.
\newblock \url{ https://ai.googleblog.com/2017/04/federated-learning-collaborative.html}.

\bibitem[mel()]{melloddy}
Machine learning ledger orchestration for drug discovery (melloddy).
\newblock https://www.melloddy.eu/.

\bibitem[sim()]{simulatefake}
Android-x86 run android on your pc.
\newblock https://www.android-x86.org/.

\bibitem[web()]{webank}
Utilization of fate in risk management of credit in small and micro enterprises.
\newblock https://www.fedai.org/cases/utilization-of-fate-in-risk-management-of-credit-in-small-and-micro-enterprises/.

\bibitem[pur(Last accessed April, 2021)]{purchase}
Acquire valued shoppers challenge at kaggle.
\newblock \url{https://www.kaggle.com/c/acquire-valued-shoppers-challenge/data}, Last accessed April, 2021.

\bibitem[Bagdasaryan et~al.(2020)Bagdasaryan, Veit, Hua, Estrin, and Shmatikov]{bagdasaryan2020backdoor}
Eugene Bagdasaryan, Andreas Veit, Yiqing Hua, Deborah Estrin, and Vitaly Shmatikov.
\newblock How to backdoor federated learning.
\newblock In \emph{AISTATS}, 2020.

\bibitem[Barreno et~al.(2006)Barreno, Nelson, Sears, Joseph, and Tygar]{barreno2006can}
Marco Barreno, Blaine Nelson, Russell Sears, Anthony~D Joseph, and J~Doug Tygar.
\newblock Can machine learning be secure?
\newblock In \emph{ASIACCS}, 2006.

\bibitem[Baruch et~al.(2019)Baruch, Baruch, and Goldberg]{baruch2019little}
Gilad Baruch, Moran Baruch, and Yoav Goldberg.
\newblock A little is enough: Circumventing defenses for distributed learning.
\newblock In \emph{NeurIPS}, 2019.

\bibitem[Blanchard et~al.(2017)Blanchard, El~Mhamdi, Guerraoui, and Stainer]{blanchard2017machine}
Peva Blanchard, El~Mahdi El~Mhamdi, Rachid Guerraoui, and Julien Stainer.
\newblock Machine learning with adversaries: Byzantine tolerant gradient descent.
\newblock In \emph{NeurIPS}, 2017.

\bibitem[Caldas et~al.(2018)Caldas, Duddu, Wu, Li, Kone{\v{c}}n{\`y}, McMahan, Smith, and Talwalkar]{caldas2018leaf}
Sebastian Caldas, Sai Meher~Karthik Duddu, Peter Wu, Tian Li, Jakub Kone{\v{c}}n{\`y}, H~Brendan McMahan, Virginia Smith, and Ameet Talwalkar.
\newblock Leaf: A benchmark for federated settings.
\newblock \emph{arXiv preprint arXiv:1812.01097}, 2018.

\bibitem[Cao and Gong(2022)]{cao2022mpaf}
Xiaoyu Cao and Neil~Zhenqiang Gong.
\newblock Mpaf: Model poisoning attacks to federated learning based on fake clients.
\newblock In \emph{CVPR Workshops}, 2022.

\bibitem[Cao et~al.(2021{\natexlab{a}})Cao, Fang, Liu, and Gong]{cao2020fltrust}
Xiaoyu Cao, Minghong Fang, Jia Liu, and Neil~Zhenqiang Gong.
\newblock Fltrust: Byzantine-robust federated learning via trust bootstrapping.
\newblock In \emph{NDSS}, 2021{\natexlab{a}}.

\bibitem[Cao et~al.(2021{\natexlab{b}})Cao, Jia, and Gong]{cao2021provably}
Xiaoyu Cao, Jinyuan Jia, and Neil~Zhenqiang Gong.
\newblock Provably secure federated learning against malicious clients.
\newblock In \emph{AAAI}, 2021{\natexlab{b}}.

\bibitem[Cao et~al.(2022)Cao, Zhang, Jia, and Gong]{cao2022flcert}
Xiaoyu Cao, Zaixi Zhang, Jinyuan Jia, and Neil~Zhenqiang Gong.
\newblock Flcert: Provably secure federated learning against poisoning attacks.
\newblock In \emph{IEEE Transactions on Information Forensics and Security}, 2022.

\bibitem[Fang et~al.(2020)Fang, Cao, Jia, and Gong]{fang2020local}
Minghong Fang, Xiaoyu Cao, Jinyuan Jia, and Neil Gong.
\newblock Local model poisoning attacks to byzantine-robust federated learning.
\newblock In \emph{USENIX Security Symposium}, 2020.

\bibitem[Fang et~al.(2024)Fang, Zhang, Hairi, Khanduri, Liu, Lu, Liu, and Gong]{fang2024byzantine}
Minghong Fang, Zifan Zhang, Hairi, Prashant Khanduri, Jia Liu, Songtao Lu, Yuchen Liu, and Neil Gong.
\newblock Byzantine-robust decentralized federated learning.
\newblock In \emph{CCS}, 2024.

\bibitem[Fang et~al.(2025)Fang, Nabavirazavi, Liu, Sun, Iyengar, and Yang]{fang2025we}
Minghong Fang, Seyedsina Nabavirazavi, Zhuqing Liu, Wei Sun, Sundararaja~Sitharama Iyengar, and Haibo Yang.
\newblock Do we really need to design new byzantine-robust aggregation rules?
\newblock In \emph{NDSS}, 2025.

\bibitem[Fredrikson et~al.(2015)Fredrikson, Jha, and Ristenpart]{fredrikson2015model}
Matt Fredrikson, Somesh Jha, and Thomas Ristenpart.
\newblock Model inversion attacks that exploit confidence information and basic countermeasures.
\newblock In \emph{CCS}, 2015.

\bibitem[Kairouz et~al.(2021)Kairouz, McMahan, Avent, Bellet, Bennis, Bhagoji, Bonawitz, Charles, Cormode, Cummings, et~al.]{kairouz2021advances}
Peter Kairouz, H~Brendan McMahan, Brendan Avent, Aur{\'e}lien Bellet, Mehdi Bennis, Arjun~Nitin Bhagoji, Kallista Bonawitz, Zachary Charles, Graham Cormode, Rachel Cummings, et~al.
\newblock Advances and open problems in federated learning.
\newblock In \emph{Foundations and Trends{\textregistered} in Machine Learning}, 2021.

\bibitem[Kone{\v{c}}n{\`y} et~al.(2016)Kone{\v{c}}n{\`y}, McMahan, Yu, Richt{\'a}rik, Suresh, and Bacon]{konevcny2016federated}
Jakub Kone{\v{c}}n{\`y}, H~Brendan McMahan, Felix~X Yu, Peter Richt{\'a}rik, Ananda~Theertha Suresh, and Dave Bacon.
\newblock Federated learning: Strategies for improving communication efficiency.
\newblock \emph{arXiv preprint arXiv:1610.05492}, 2016.

\bibitem[Krizhevsky et~al.(2009)Krizhevsky, Hinton, et~al.]{krizhevsky2009learning}
Alex Krizhevsky, Geoffrey Hinton, et~al.
\newblock Learning multiple layers of features from tiny images.
\newblock 2009.

\bibitem[LeCun et~al.(1998)LeCun, Cortes, and Burges]{lecun2010mnist}
Yann LeCun, Corinna Cortes, and CJ Burges.
\newblock Mnist handwritten digit database.
\newblock \emph{Available: http://yann. lecun. com/exdb/mnist}, 1998.

\bibitem[Li et~al.(2020)Li, Cheng, Wang, Liu, and Chen]{li2020learning}
Suyi Li, Yong Cheng, Wei Wang, Yang Liu, and Tianjian Chen.
\newblock Learning to detect malicious clients for robust federated learning.
\newblock \emph{arXiv preprint arXiv:2002.00211}, 2020.

\bibitem[McMahan et~al.(2017)McMahan, Moore, Ramage, Hampson, and y~Arcas]{mcmahan2017communication}
H.~Brendan McMahan, Eider Moore, Daniel Ramage, Seth Hampson, and Blaise~Ag{\"u}era y Arcas.
\newblock Communication-efficient learning of deep networks from decentralized data.
\newblock In \emph{AISTATS}, 2017.

\bibitem[Mhamdi et~al.(2018)Mhamdi, Guerraoui, and Rouault]{guerraoui2018hidden}
El~Mahdi~El Mhamdi, Rachid Guerraoui, and S{\'{e}}bastien Rouault.
\newblock The hidden vulnerability of distributed learning in byzantium.
\newblock In \emph{ICML}, 2018.

\bibitem[Nguyen et~al.(2022)Nguyen, Rieger, De~Viti, Chen, Brandenburg, Yalame, M{\"o}llering, Fereidooni, Marchal, Miettinen, et~al.]{nguyen2022flame}
Thien~Duc Nguyen, Phillip Rieger, Roberta De~Viti, Huili Chen, Bj{\"o}rn~B Brandenburg, Hossein Yalame, Helen M{\"o}llering, Hossein Fereidooni, Samuel Marchal, Markus Miettinen, et~al.
\newblock Flame: Taming backdoors in federated learning.
\newblock In \emph{USENIX Security Symposium}, 2022.

\bibitem[Permuter et~al.(2006)Permuter, Francos, and Jermyn]{permuter2006study}
Haim Permuter, Joseph Francos, and Ian Jermyn.
\newblock A study of gaussian mixture models of color and texture features for image classification and segmentation.
\newblock In \emph{Pattern recognition}, 2006.

\bibitem[Pi et~al.(2023)Pi, Zhang, Xie, Gao, Wang, Kim, and Chen]{pi2023dynafed}
Renjie Pi, Weizhong Zhang, Yueqi Xie, Jiahui Gao, Xiaoyu Wang, Sunghun Kim, and Qifeng Chen.
\newblock Dynafed: Tackling client data heterogeneity with global dynamics.
\newblock In \emph{CVPR}, 2023.

\bibitem[Rieke et~al.(2020)Rieke, Hancox, Li, Milletari, Roth, Albarqouni, Bakas, Galtier, Landman, Maier-Hein, et~al.]{rieke2020future}
Nicola Rieke, Jonny Hancox, Wenqi Li, Fausto Milletari, Holger~R Roth, Shadi Albarqouni, Spyridon Bakas, Mathieu~N Galtier, Bennett~A Landman, Klaus Maier-Hein, et~al.
\newblock The future of digital health with federated learning.
\newblock In \emph{NPJ digital medicine}, 2020.

\bibitem[Shejwalkar and Houmansadr(2021)]{shejwalkar2021manipulating}
Virat Shejwalkar and Amir Houmansadr.
\newblock Manipulating the byzantine: Optimizing model poisoning attacks and defenses for federated learning.
\newblock In \emph{NDSS}, 2021.

\bibitem[Shejwalkar et~al.(2022)Shejwalkar, Houmansadr, Kairouz, and Ramage]{shejwalkar2022back}
Virat Shejwalkar, Amir Houmansadr, Peter Kairouz, and Daniel Ramage.
\newblock Back to the drawing board: A critical evaluation of poisoning attacks on production federated learning.
\newblock In \emph{IEEE Symposium on Security and Privacy}, 2022.

\bibitem[Sun et~al.(2019)Sun, Kairouz, Suresh, and McMahan]{sun2019can}
Ziteng Sun, Peter Kairouz, Ananda~Theertha Suresh, and H~Brendan McMahan.
\newblock Can you really backdoor federated learning?
\newblock \emph{arXiv preprint arXiv:1911.07963}, 2019.

\bibitem[Xiao et~al.(2017)Xiao, Rasul, and Vollgraf]{xiao2017/online}
Han Xiao, Kashif Rasul, and Roland Vollgraf.
\newblock Fashion-mnist: a novel image dataset for benchmarking machine learning algorithms.
\newblock \emph{arXiv preprint arXiv:1708.07747}, 2017.

\bibitem[Xie et~al.(2021)Xie, Chen, Chen, and Li]{xie2021crfl}
Chulin Xie, Minghao Chen, Pin-Yu Chen, and Bo Li.
\newblock Crfl: Certifiably robust federated learning against backdoor attacks.
\newblock In \emph{ICML}, 2021.

\bibitem[Xie et~al.()Xie, Fang, and Gong]{yueqifedredefense}
Yueqi Xie, Minghong Fang, and Neil~Zhenqiang Gong.
\newblock Fedredefense: Defending against model poisoning attacks for federated learning using model update reconstruction error.
\newblock In \emph{Forty-first International Conference on Machine Learning}.

\bibitem[Yang et~al.(2019)Yang, Liu, Chen, and Tong]{yang2019federated}
Qiang Yang, Yang Liu, Tianjian Chen, and Yongxin Tong.
\newblock Federated machine learning: Concept and applications.
\newblock In \emph{ACM Transactions on Intelligent Systems and Technology}, 2019.

\bibitem[Yin et~al.(2018)Yin, Chen, Kannan, and Bartlett]{yin2018byzantine}
Dong Yin, Yudong Chen, Ramchandran Kannan, and Peter Bartlett.
\newblock Byzantine-robust distributed learning: Towards optimal statistical rates.
\newblock In \emph{ICML}, 2018.

\bibitem[Zhang et~al.(2022)Zhang, Cao, Jia, and Gong]{zhang2022fldetector}
Zaixi Zhang, Xiaoyu Cao, Jinyuan Jia, and Neil~Zhenqiang Gong.
\newblock Fldetector: Defending federated learning against model poisoning attacks via detecting malicious clients.
\newblock In \emph{KDD}, 2022.

\end{thebibliography}
}
\newpage
\appendix
\maketitlesupplementary

\section{Motivation}

\label{sec:motivation}

We first introduce our key observation about why existing model poisoning attacks achieve suboptimal attack effectiveness, which motivates the design of our \ourmodel{} in the next section. We show experimental results on the MNIST dataset with the default parameter settings described in Section~\ref{sec:setup}. We assume the server uses Trimmed Mean~\cite{yin2018byzantine} as the aggregation rule. Moreover, we assume the ratio of fake clients to genuine clients is $20\%$. We consider  MPAF~\cite{cao2022mpaf}, an attack not requiring genuine clients' information, and Fang~\cite{fang2020local}, a representative  attack that requires genuine clients' information. 
To give advantages to Fang, we assume the attacker knows the local models of \emph{all} participating genuine clients in each training round when crafting malicious model updates on fake clients. 

In each training round, these attacks craft a malicious model update for each fake client selected by the server to participate in training. A malicious model update $\bm{g}_i^t$ of fake client $i$ in training round $t$  can be decomposed into a dimension-wise product of a  \emph{sign vector} $\bm{s}_i^t$ and a \emph{magnitude vector} $|\bm{g}_i^t|$, where $|\cdot|$ means dimension-wise absolute value of a vector. Each dimension of $\bm{s}_i^t$ is either +1 or -1, while all dimensions of $|\bm{g}_i^t|$ are non-negative. A dimension with a +1 (or -1) sign aims to increase (or decrease) the corresponding dimension/parameter of the global model. For a fake client, we say a dimension is flipped in training round $t$  if its sign in the malicious model update is flipped, compared to that in the previous training round. A flipped dimension in a training round $t$ may cancel the attack effect of the previous training round $t-1$. This is because the malicious model updates aim to increase the dimension of the global model in one training round but decrease it in the other. For each training round $t$, we define \emph{flipping rate} as the fraction of flipped dimensions of a malicious model update averaged over the fake clients. 

Fig.~\ref{fig:moti_acc} shows the testing error rate of the global model as a function of training round when no attack, MPAF, or Fang is used; while Fig.~\ref{fig:moti_flip} shows the flipping rate as a function of training round when MPAF or Fang is used to craft malicious model updates on the fake clients. We observe that the flipping rate of MPAF is always larger than 35\% and that of Fang is larger than 40\% in the last around 1,500 training rounds, leading to self-cancellation of attack effect in many training rounds. For instance, in Fang attack, the flipping rate is around 10\% in  nearly the first 1,000 training rounds, i.e., the malicious model updates maintain some degree of ``consistency''. As a result, the global model under Fang attack has a large testing error rate in those training rounds as shown in Fig.~\ref{fig:moti_acc}. However, the flipping rate  rapidly increases as the training proceeds, eventually leading to self-cancellation of attack effect and an accurate global model.

\begin{figure}[t]
\centering
\includegraphics[width=0.8\linewidth]{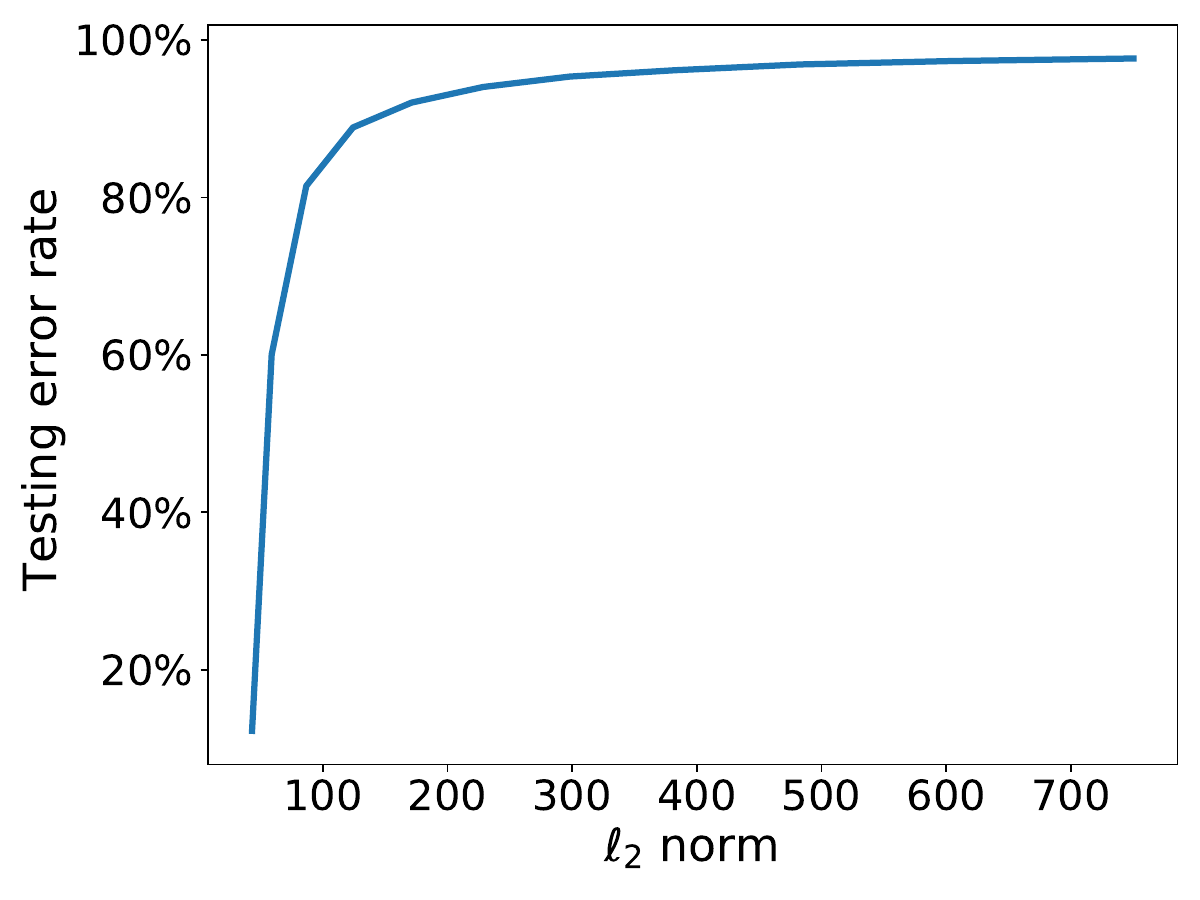}
    \caption{The testing error rate of a global model as a function of the magnitude (measured by $\ell_2$ norm) of the noise  added to it along a random update direction. We assume the initial global model is accurate and trained under no attack. The dataset is Purchase and FL defense is Median.}
    \label{fig:norm-performance}
\end{figure}

\section{Additional Details of Experimental Setup}
\begin{table}[t]
    \centering
        \caption{Model architecture for MNIST and FashionMNIST.}
        \resizebox{0.5\linewidth}{!}{
    \begin{tabular}{c|c}
    \toprule
\midrule
         Layer Type&Size  \\
         \midrule
         Convolution + ReLU & 3x3x30 \\
         \midrule
         Max Pooling & 2x2 \\
         \midrule
         Convolution + ReLU & 3x3x50 \\
         \midrule
         Max Pooling & 2x2 \\
         \midrule
         Fully Connected + ReLU & 100\\
         \midrule
         Softmax & 10\\
         \midrule
         \bottomrule
    \end{tabular}}
    \label{tab:model_mnist_fashion}
\end{table}

\begin{table}[t]
    \centering
        \caption{Model architecture for CIFAR-10.}
        \resizebox{0.5\linewidth}{!}{
    \begin{tabular}{c|c}
    \toprule
\midrule
         Layer Type&Size  \\
         \midrule
         Convolution + ReLU & 3x3x32 \\
         \midrule
         Max Pooling & 2x2 \\
         \midrule
         Convolution + ReLU & 3x3x64 \\
         \midrule
         Max Pooling & 2x2 \\
         \midrule
         Fully Connected + ReLU & 512\\
         \midrule
         Softmax & 10\\
         \midrule
         \bottomrule
    \end{tabular}}
    \label{tab:model_cifar}
\end{table}

\begin{table}[t]
    \centering
    \caption{Model architecture for FEMNIST.}
    \resizebox{0.5\linewidth}{!}{
    \begin{tabular}{c|c}
    \toprule
\midrule
         Layer Type&Size  \\
         \midrule
         Convolution + ReLU & 3x3x30 \\
         \midrule
         Max Pooling & 2x2 \\
         \midrule
         Convolution + ReLU & 3x3x50 \\
         \midrule
         Max Pooling & 2x2 \\
         \midrule
         Fully Connected + ReLU & 200\\
         \midrule
         Softmax & 62\\
         \midrule
         \bottomrule
    \end{tabular}}
    \label{tab:model_femnsit}
\end{table}

\subsection{Datasets}
\label{app:dataset}
    We use the following five datasets from different domains.
    
    \myparatight{MNIST~\cite{lecun2010mnist}} 
    MNIST is a 10-class handwritten digits classification dataset.
    It comprises 60,000 training examples and 10,000 testing examples. 
    In FL, the training data is typically not independently and identically distributed (non-IID) among clients. 
    Following~\cite{cao2022mpaf,fang2020local}, we distribute the training examples among clients based on a non-IID degree of $q = 0.5$ by default.
    We train a convolutional neural network (CNN) for MNIST dataset. The architecture of CNN is shown in Table~\ref{tab:model_mnist_fashion} in Appendix.

    \myparatight{FashionMNIST~\cite{xiao2017/online}} 
    FashionMNIST is a benchmark dataset of Zalando's article images, which contains 60,000 training examples and 10,000 testing examples. 
    For FashionMNIST, the degree of non-IID is also set to $q=0.5$ by default.
    We use the same CNN architecture for FashionMNIST dataset as employed in MNIST dataset.  

    \myparatight{Purchase~\cite{purchase}} 
    Purchase is a 100-class customer purchase style prediction dataset. 
    Each input consists of 600 binary features.
    Following~\cite{cao2022mpaf}, we split the total 197,324 examples into 180,000 training examples  and 17,324 testing examples. 
    Following~\cite{cao2022mpaf}, we evenly distribute the training examples to clients and train a fully connected neural network.

     \myparatight{CIFAR-10~\cite{krizhevsky2009learning}} 
    CIFAR-10 is 10-class color image classification dataset. 
    It comprises 50,000 training examples and 10,000 testing examples. 
    To consider a different degree of non-IID, we set $q = 0.2$ for CIFAR-10 dataset. 
    We train a CNN for CIFAR-10, whose  architecture  is shown in Table~\ref{tab:model_cifar}.

    \myparatight{FEMNIST~\cite{caldas2018leaf}} 
    FEMNIST is a  62-class classification dataset. 
     The dataset is already distributed among 3,550 clients with a total of 805,263 examples.
     We randomly sample 1,200 clients.
     We train a CNN for FEMNIST, whose architecture is shown in Table~\ref{tab:model_femnsit}. 


\subsection{Defenses}
\label{app:defense}
We evaluate eight state-of-the-art defenses, including six Byzantine-robust aggregation rules (Multi-Krum~\cite{blanchard2017machine}, Median~\cite{yin2018byzantine}, Trimmed Mean~\cite{yin2018byzantine}, Norm Bound~\cite{sun2019can}, FLTrust~\cite{cao2020fltrust}, and FLAME~\cite{nguyen2022flame}), one provably robust defense (FLCert~\cite{cao2022flcert}) and one malicious clients detection method (FLDetector~\cite{zhang2022fldetector}). We also consider the non-robust FedAvg~\cite{mcmahan2017communication} as a baseline.

\myparatight{Multi-Krum~\cite{blanchard2017machine}}  
Multi-Krum  uses an iterative method to select a subset of clients' model updates. In each step, it selects the model update that has the smallest sum of Euclidean distance to its $n-2$ neighbors, where $n$ is the number of genuine clients. 
This selection process continues until $n$ clients are chosen. Finally, the server computes the average of the $n$ selected model updates as the aggregated model update.

\myparatight{Median~\cite{yin2018byzantine}} 
In Median aggregation rule, the server computes the coordinate-wise median of the clients' model updates as the aggregated model update.

 \myparatight{Trimmed Mean (TrMean)~\cite{yin2018byzantine}} 
For each dimension, the server first removes the largest $m$ values and smallest $m$ values, and then computes the average of the remaining $n-m$ values, where $m$ is the number of  fake clients. 

\myparatight{Norm Bound~\cite{sun2019can}}  
In Norm Bound aggregation rule, the server first clips each client's model update to have a predetermined norm and then computes the average of the clipped model updates as the aggregated model update.
We follow prior work~\cite{shejwalkar2022back} to set the predetermined norm  as the average norm of the genuine model updates in each training round.

\myparatight{FLTrust~\cite{cao2020fltrust}}
FLTrust assumes that the server has a small, clean  dataset to bootstrap trust. We assume the server's dataset includes training examples selected uniformly at random.  
The size of the server's dataset is  100 for MNIST, FashionMNIST, and CIFAR-10,  and 200 for  Purchase and FEMNIST.

 \myparatight{FLAME~\cite{nguyen2022flame}}  In each training round, FLAME considers the cosine similarity between clients' local models (i.e., a client's model update + current global model) to divide clients into clusters and filter out the clusters containing potentially malicious model updates.

\myparatight{FLCert~\cite{cao2022flcert}} 
 FLCert divides clients into multiple groups and trains a global model for each group using any existing FL algorithm (we use Median~\cite{yin2018byzantine}).   Given a testing input, FLCert predicts its label based on the majority vote among these global models. We divide the clients into 10 disjoint groups uniformly at random in our experiments.

\myparatight{FLDetector~\cite{zhang2022fldetector}}  
 FLDetector aims to detect malicious clients during training. 
 We first apply FLDetector with the full participation of clients for detection for 300 communication rounds. 
 Following~\cite{zhang2022fldetector}, if some clients are detected as malicious, we remove them from the system.
 Then we re-train a global model from scratch based on the remaining clients using Median~\cite{yin2018byzantine} aggregation rule with the default setting.


\subsection{Compared Attacks}
\label{app:attack}
We compare our  \ourmodel{} with seven attacks, including five attacks (Fang~\cite{fang2020local}, Opt. Fang~\cite{shejwalkar2021manipulating}, LIE~\cite{baruch2019little}, Min-Max~\cite{shejwalkar2021manipulating}, and Min-Sum~\cite{shejwalkar2021manipulating}) that require genuine clients' information and two attacks (Random~\cite{cao2022mpaf} and MPAF~\cite{cao2022mpaf}) that do not require such information. 
Note that when applying the first category of attacks to craft the malicious model updates on fake clients, we assume that the attacker has access to the model updates of \emph{all} genuine clients and the aggregation rule used by the server. In Section~\ref{sec:discussion}, we also study the scenarios where an attacker uses the global models to reconstruct synthetic data on the fake  clients and uses the local models trained on them to perform attacks. 
We do not make these assumptions for  Random, MPAF, and our \ourmodel{}. Therefore, we give advantages to the first category of attacks.

\myparatight{Fang~\cite{fang2020local}} 
In this attack, the attacker crafts malicious model updates for fake clients such that the aggregated model update after attack deviates substantially from the before-attack one. Fang has different versions for different aggregation rules. 
We apply their Krum attack for Multi-Krum defense and TrMean attack for others.

\myparatight{Opt. Fang~\cite{shejwalkar2021manipulating}} 
Following Fang, this attack also aims to maximize the deviation between the after-attack aggregated model update and before-attack one, but it uses a different way to solve the malicious model update. Specifically, a malicious model update is a variant of the average of all genuine clients' model updates.


\myparatight{A little is enough (LIE)~\cite{baruch2019little}} 
In LIE attack, fake clients craft their malicious model updates by adding a small amount of noise to the average of genuine model updates.
Specifically, for each dimension $j$, LIE calculates the mean $\mu_j$ and standard deviation $\sigma_j$ among the genuine model updates, and the dimension $j$ of the malicious model update is set as $\mu_j + 0.74 \cdot \sigma_j$.

    \myparatight{Min-Max~\cite{shejwalkar2021manipulating}} 
In this attack, a fake client crafts its malicious model update such that its distance to any genuine model update is no larger than the maximum distance between any two genuine model updates.

    \myparatight{Min-Sum~\cite{shejwalkar2021manipulating}} 
Each fake client crafts its malicious model update such that the sum of the distances between the malicious model update and all genuine model updates is not larger than the sum of distances between any genuine model update and other genuine model updates.

    \myparatight{Random~\cite{cao2022mpaf}} 
    In this attack, each fake client $i \in[n+1, n+m]$ sends a scaled random vector $\bm{g}_i^t = \lambda \cdot \boldsymbol{\epsilon} $ to the server, where $\boldsymbol{\epsilon} \sim \mathcal{N}(\mathbf{0}, \mathbf{I})$ and $\lambda$ is a scaling factor. 
    Following prior work~\cite{cao2022mpaf}, we set $\lambda=1e6$ in our experiments.

        \myparatight{MPAF~\cite{cao2022mpaf}}
    In MPAF attack, during training round $t$, fake client $i \in[n+1, n+m]$ crafts its malicious model update as $\bm{g}_i^t=\lambda \cdot (\boldsymbol{w}^{\prime}-\boldsymbol{w}^{t-1}$), where $\boldsymbol{w}^{\prime}$ is an attacker-chosen random model, $\boldsymbol{w}^{t-1}$ is the global model in the previous training round $t-1$, and $\lambda$ is a scaling factor. Following prior work~\cite{cao2022mpaf}, we set $\lambda=1e6$ in our experiments.

\subsection{FL Settings}
\label{app:setting}
We primarily establish the FL setting in a cross-device scenario, which is more feasible for model poisoning attacks, as discussed in Section~\ref{sec:discussion}. Therefore, we assume $1,200$ genuine clients and $20\%$ fake clients.
The learning rates for the five datasets are selected within the range of $0.01$ to $0.1$ for optimal training effectiveness. Specifically, the learning rates for MNIST, FashionMNIST, and FEMNIST are set to $0.01$, while CIFAR-10 is set to $0.03$, and Purchase is set to $0.1$.
The number of training rounds is configured to ensure full convergence under various scenarios: $6,000$ for MNIST, FashionMNIST, and FEMNIST, and $10,000$ for CIFAR-10 and FEMNIST.
The batch size is set to $32$ for MNIST, FashionMNIST, CIFAR-10, and FEMNIST, and to $128$ for Purchase, based on the number of local data samples.
\section{Additional Experimental Results}
\subsection{Analyzing the Effectiveness of \ourmodel{}}
\label{app:analysis}
\myparatight{Why \ourmodel{} breaks FL defenses} State-of-the-art defenses leverage filtering and clipping to reduce the impact of malicious model updates.  They can weaken the attack effect in individual training rounds.  However, since \ourmodel{} crafts malicious model updates that are consistent across training rounds and avoid being entirely filtered out by dynamically adjusting attack magnitudes, its cumulative attack effect still substantially moves the global model towards the designated random update direction $\bm{s}$, leading to a large testing error rate. For instance, Fig.~\ref{fig:matched-fraction} in Appendix shows that a large fraction of the dimensions of the total aggregated model updates have signs matching with $\bm{s}$  and Fig.~\ref{fig:norm} in Appendix shows that the magnitude of the total aggregated model updates becomes very large, after our attack effect accumulates over multiple training rounds, for all defenses.

Specifically, for Median and TrMean, even though the malicious model updates are not necessarily selected, their consistent nature  over multiple training rounds biases the aggregation of each dimension towards the designated random update direction. For Norm Bound, even if our attack effect is weakened in each training round, its cumulative effect over multiple training rounds remains pronounced with a significant magnitude for the total aggregated model update. For filtering-involved defenses based on Euclidean distance or cosine similarity, such as Multi-Krum, FLTrust, and FLAME, \ourmodel{} maintains effectiveness because the malicious model updates are not entirely filtered out in many training rounds due to dynamic magnitude adjustment. For instance, Table~\ref{tab:fnr} in Appendix shows that a large fraction of or all malicious model updates are not filtered out  by  FLAME and Multi-Krum, while Table~\ref{tab:trust} shows similar results for FLTrust. 
 FLCert is vulnerable to \ourmodel{} since the global models for ensembling are learnt by  existing aggregation rules, which are vulnerable to \ourmodel{}. The detection mechanism  FLDetector relies on inconsistent model updates of malicious clients. Since the malicious model updates are consistent in \ourmodel{}, FLDetector falsely classifies all malicious clients as benign across all datasets.

\myparatight{Why \ourmodel{} outperforms existing attacks} 
\ourmodel{} outperforms existing attacks because their crafted malicious model updates are inconsistent across training rounds, leading to self-cancellation of attack effect. In contrast, \ourmodel{} enforces the malicious model updates to be consistent (i.e., have the same sign vector) across training rounds, making the aggregated model updates add up in the global model across training rounds.  As a result, the final learnt global model has a very large magnitude, leading to a large testing error rate. For instance, Fig.~\ref{fig:norm} in Appendix shows the magnitude (measured by $\ell_2$ norm) of the total aggregated model updates as a function of training round for different attacks on the Purchase dataset. Our results show that the total  aggregated model updates under  existing attacks have  small magnitudes while those under our \ourmodel{} attack have much larger magnitudes, which is caused by the inconsistent malicious model updates in existing attacks vs. consistent malicious model updates in our attack. 

\begin{figure}[t]
\centering
\includegraphics[width=0.8\linewidth]{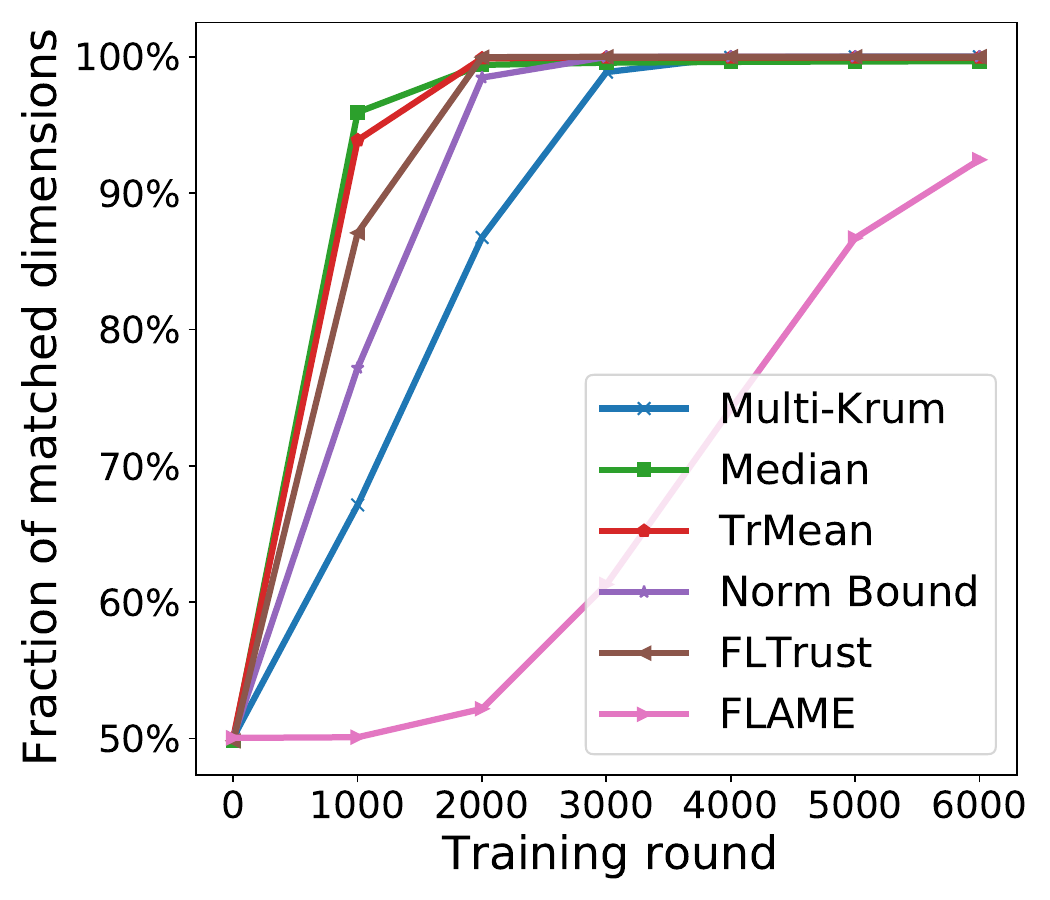}
    \caption{{Fraction of dimensions of the aggregated model update whose signs match with the sign vector $\bm{s}$ as a function of training round under \ourmodel{}. The dataset is Purchase.}}
    \label{fig:matched-fraction}
\end{figure}

\begin{figure*}[t]
\includegraphics[width=1\linewidth]{Figures/legend2.pdf}

    \begin{subfigure}{0.16\linewidth}
        \centering
        \includegraphics[width=1\linewidth]{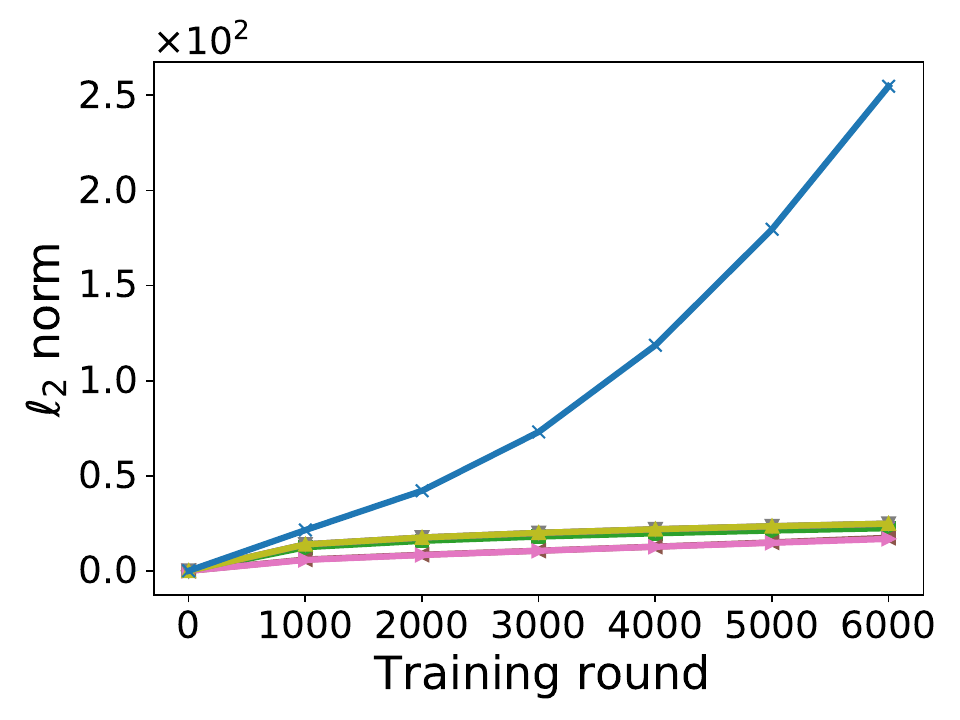}
        \caption{Multi-Krum}
    \end{subfigure}
    \begin{subfigure}{0.16\linewidth}
        \centering
        \includegraphics[width=1\linewidth]{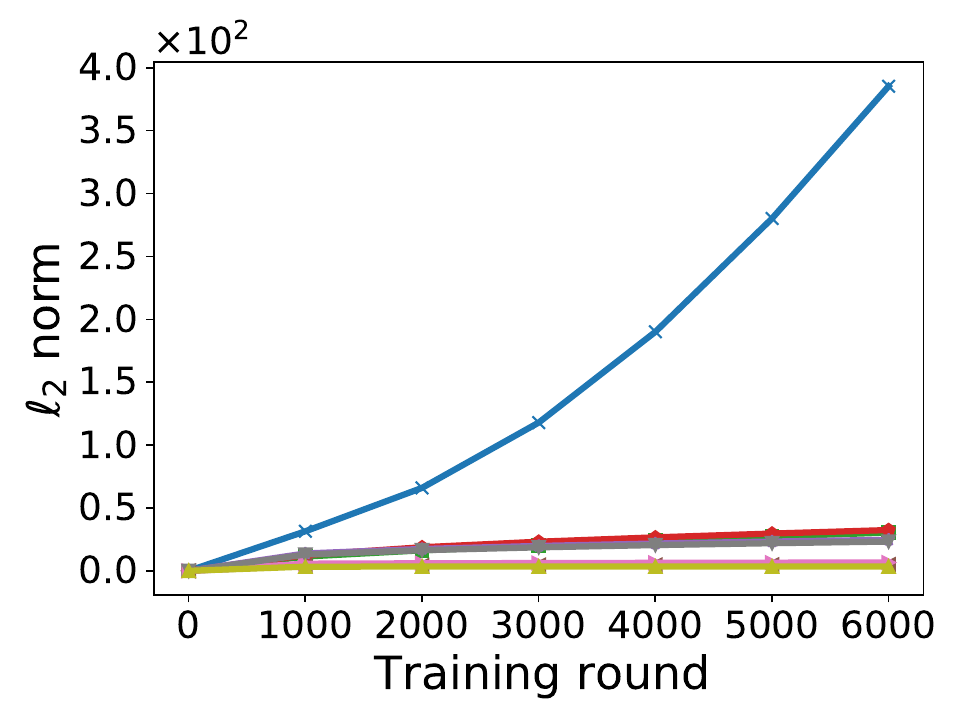}
        \caption{Median}
    \end{subfigure}
    \begin{subfigure}{0.16\linewidth}
        \centering
        \includegraphics[width=1\linewidth]{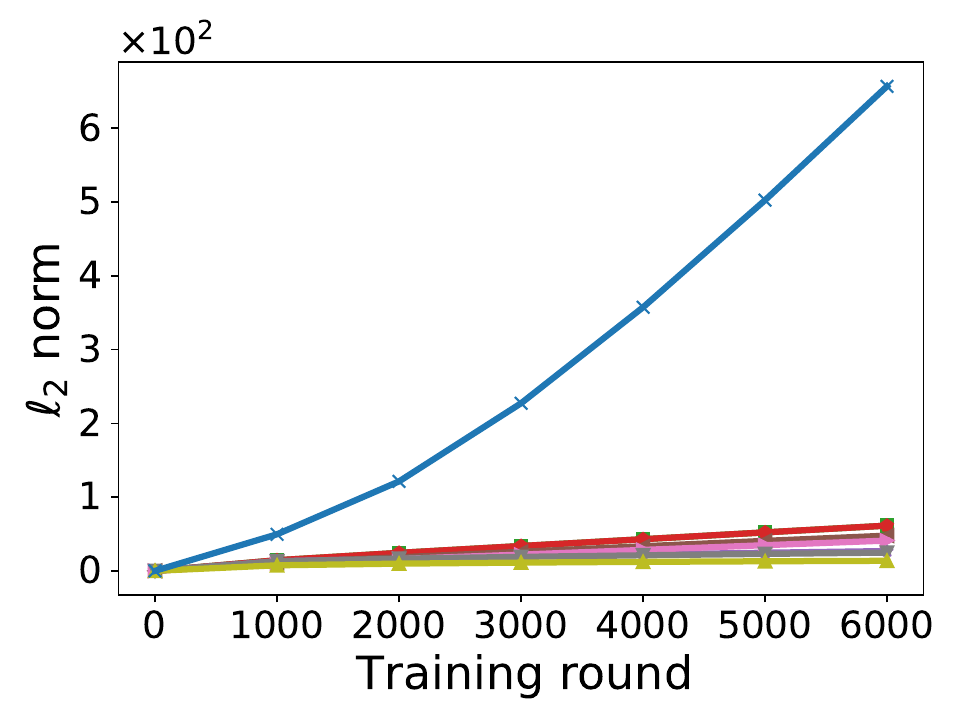}
        \caption{TrMean}
    \end{subfigure}
    \begin{subfigure}{0.16\linewidth}
        \centering
        \includegraphics[width=1\linewidth]{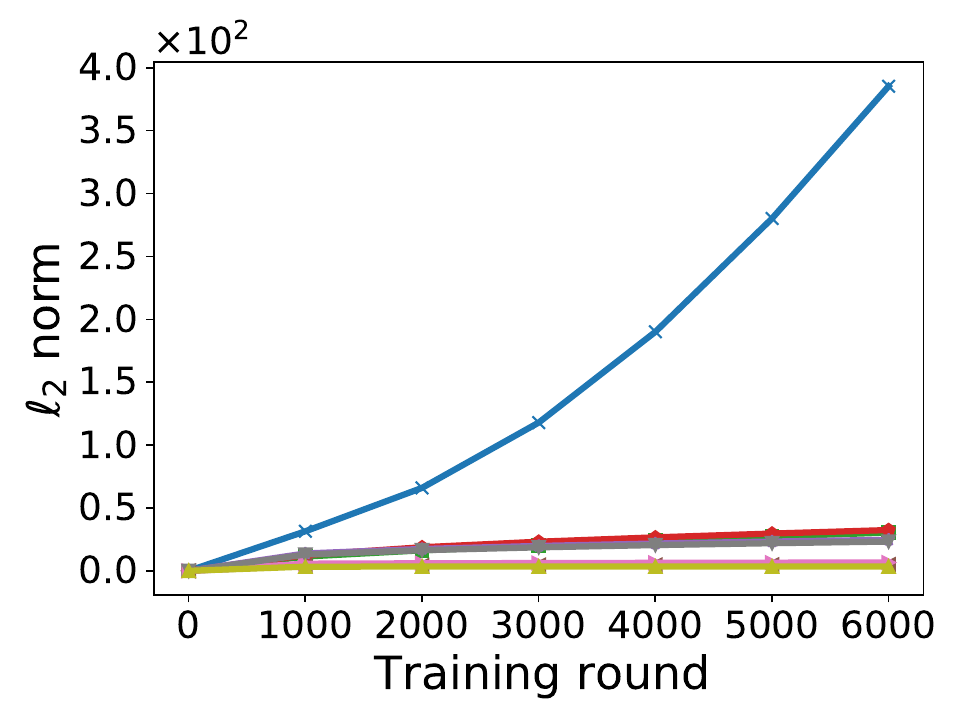}
        \caption {Norm Bound}
    \end{subfigure}
    \begin{subfigure}{0.16\linewidth}
        \centering
        \includegraphics[width=1\linewidth]{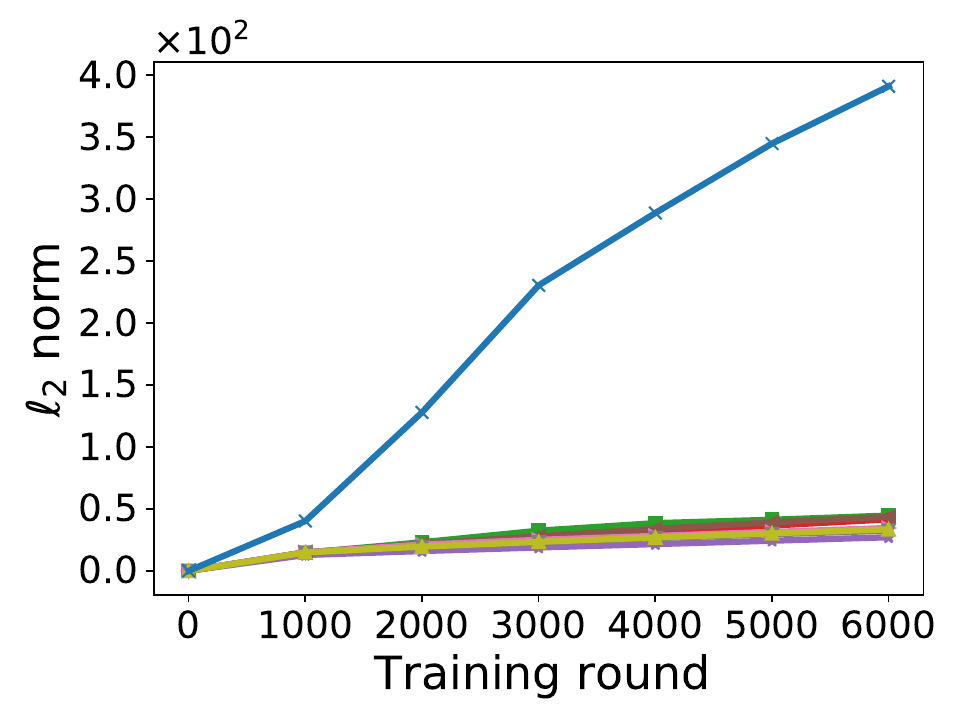}
        \caption{FLTrust}
    \end{subfigure}
    \begin{subfigure}{0.16\linewidth}
        \centering
        \includegraphics[width=1\linewidth]{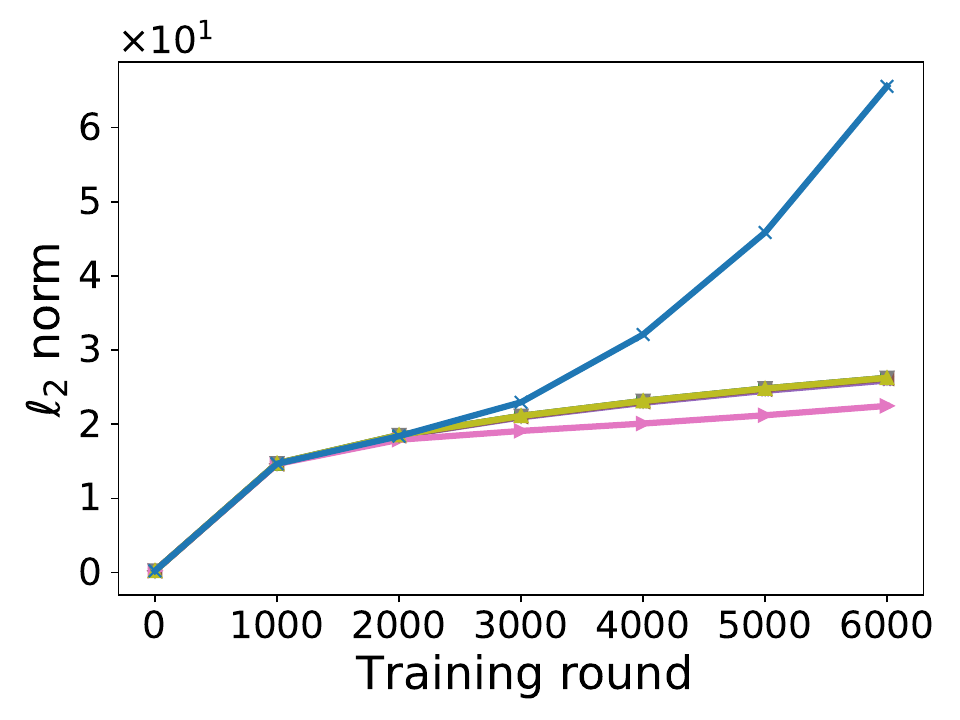}
        \caption{FLAME}
    \end{subfigure}
    \caption{{$\ell_2$ norm of the total aggregated model updates $\left\|\sum_{t^\prime = 1}^{t} \bm{g}^{t^\prime} \right\|$ (i.e., $\left\|\boldsymbol{w}^t - \boldsymbol{w}^0\right\|$) as a function of training round $t$ for different attacks on the Purchase dataset.}}
    \label{fig:norm}
\end{figure*}

\begin{table}[t]
    \centering
        \caption{The fraction (mean $\pm$ standard deviation, \%) of malicious model updates in \ourmodel{}  that are not filtered out by FLAME and Multi-Krum (i.e., false negative rate) across rounds. The dataset is Purchase.}
        \resizebox{0.7\linewidth}{!}{
    \begin{tabular}{c|cc}
    \toprule
\midrule
         &Multi-Krum&FLAME\\
         \midrule
         False Negative Rate& 32.38$\pm$2.04 & 44.25$\pm$8.18 \\
         \midrule
         \bottomrule
    \end{tabular}}
    \label{tab:fnr}
\end{table}

\begin{table}[t]
    \centering
        \caption{The normalized trust score (mean $\pm$ standard deviation) of the malicious/benign model updates in FLTrust against \ourmodel{} across rounds. The positive normalized trust scores indicate that the malicious model updates are not entirely filtered out.  The dataset is Purchase.}
        \resizebox{0.8\linewidth}{!}{
    \begin{tabular}{c|cc}
    \toprule
\midrule
          & Malicious & Benign \\
         \midrule
        Normalized Trust Score & 0.008$\pm$0.010 & 0.007$\pm$0.002 \\
         \midrule
         \bottomrule
    \end{tabular}}
    \label{tab:trust}
\end{table}



\begin{figure*}[t]
\includegraphics[width=1\linewidth]{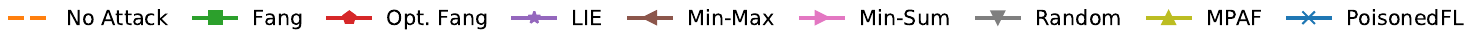}
    \begin{subfigure}{0.121\linewidth}
        \centering
        \includegraphics[width=1\linewidth]{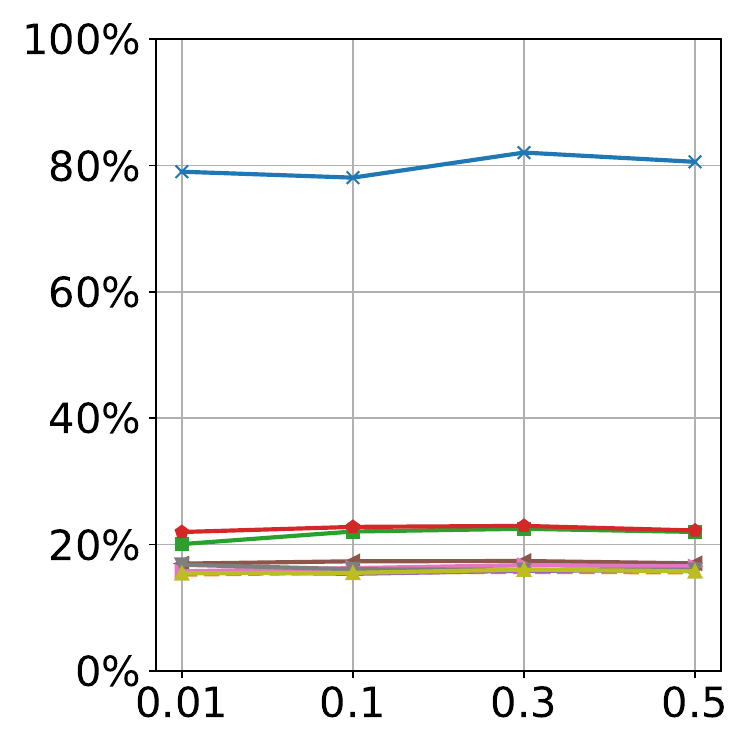}
        \subcaption{Multi-Krum}
    \end{subfigure}
    \begin{subfigure}{0.121\linewidth}
        \centering
        \includegraphics[width=1\linewidth]{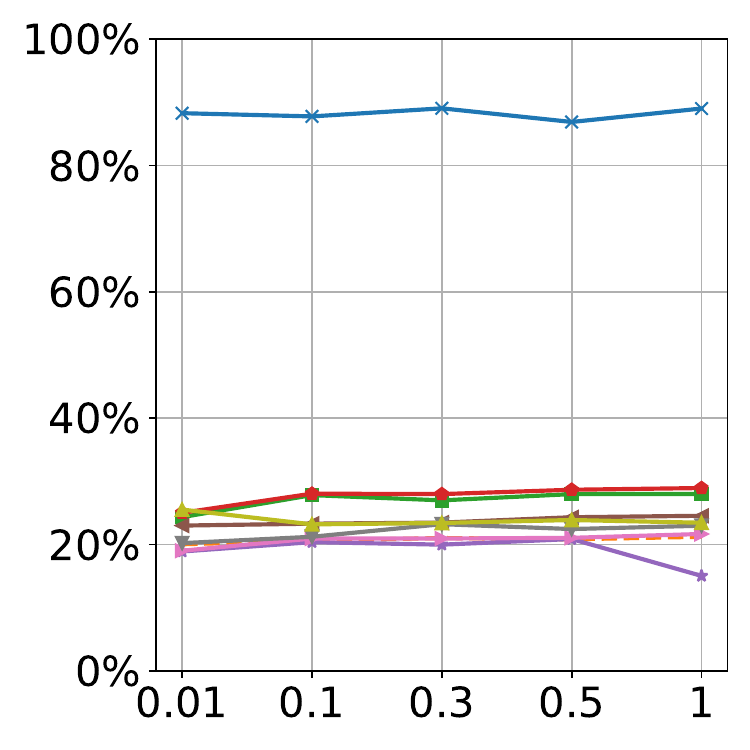}
        \subcaption{Median}
    \end{subfigure}
    \begin{subfigure}{0.121\linewidth}
        \centering
        \includegraphics[width=1\linewidth]{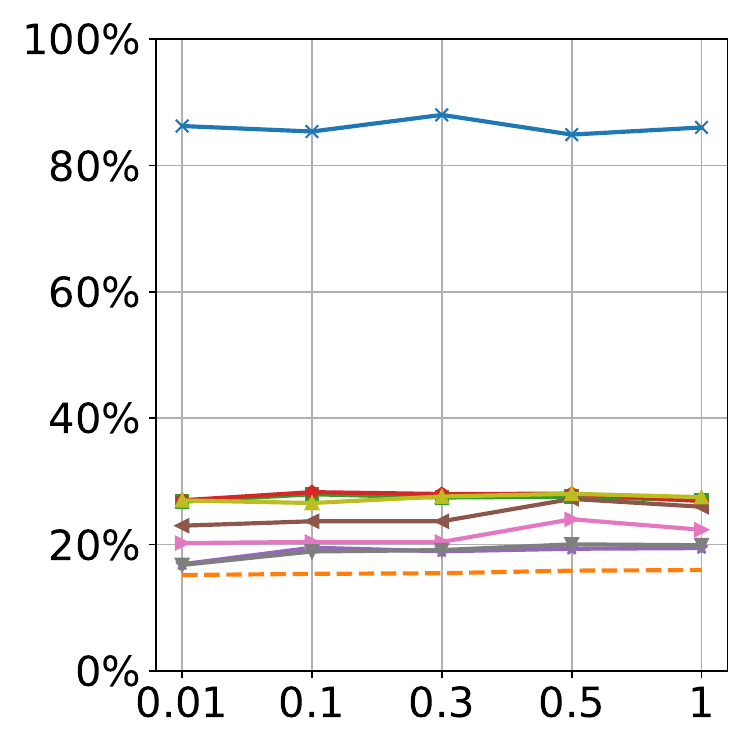}
        \subcaption{TrMean}
    \end{subfigure}
    \begin{subfigure}{0.121\linewidth}
        \centering
        \includegraphics[width=1\linewidth]{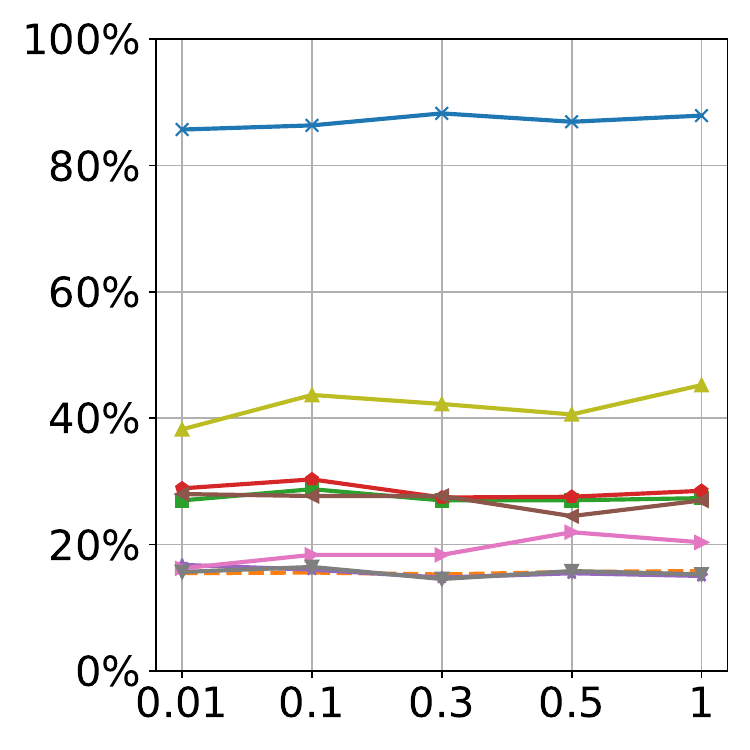}
        \subcaption{Norm Bound}
    \end{subfigure}
        \begin{subfigure}{0.121\linewidth}
        \centering
        \includegraphics[width=1\linewidth]{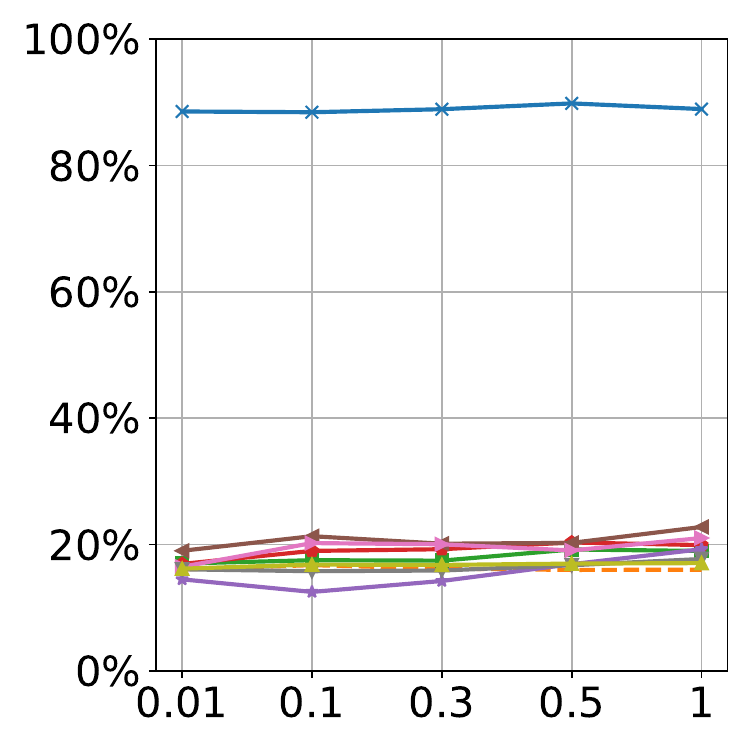}
        \subcaption{FLTrust}
    \end{subfigure}
    \begin{subfigure}{0.121\linewidth}
        \centering
        \includegraphics[width=1\linewidth]{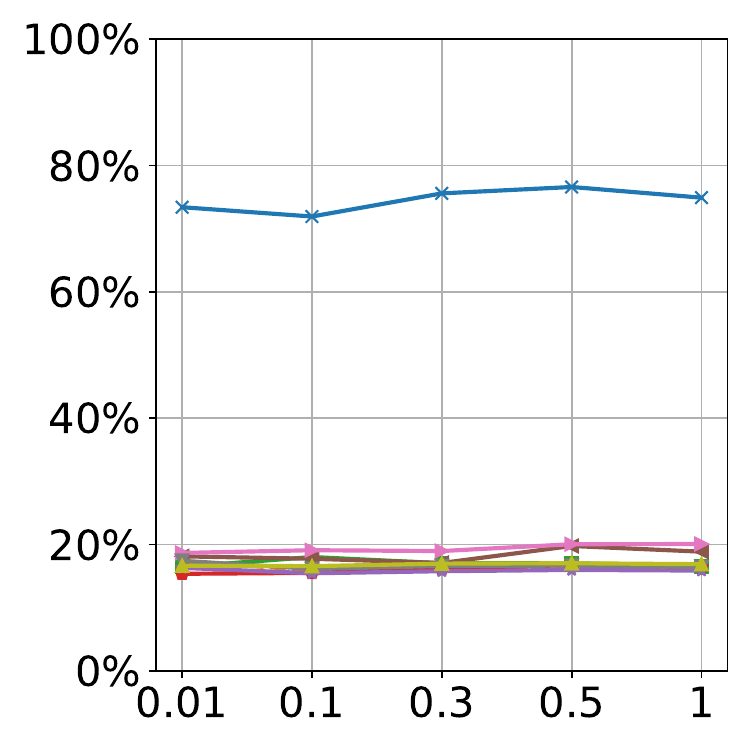}
        \caption{FLAME}
    \end{subfigure}
                \begin{subfigure}{0.121\linewidth}
        \centering
        \includegraphics[width=1\linewidth]{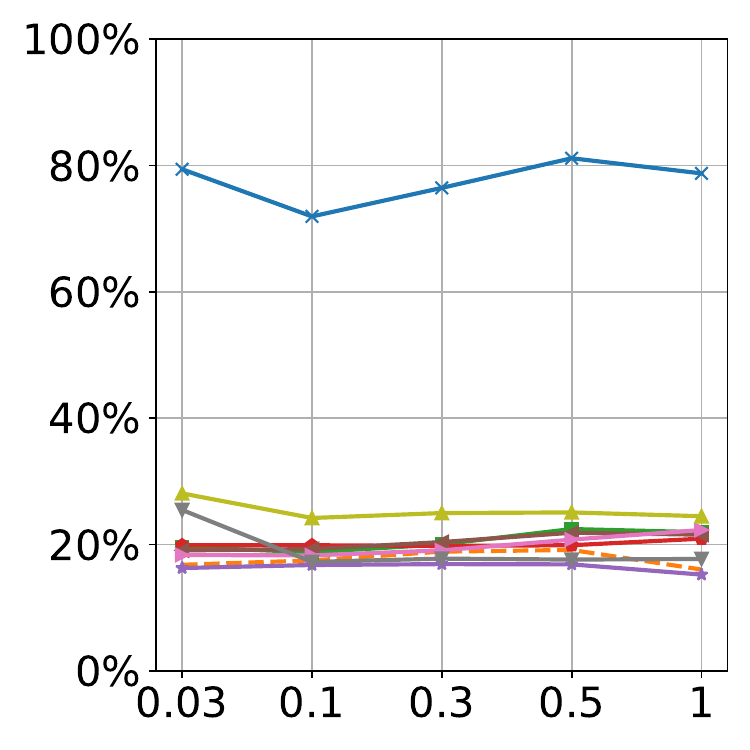}
        \caption{FLCert}
    \end{subfigure}
    \begin{subfigure}{0.121\linewidth}
        \centering
        \includegraphics[width=1\linewidth]{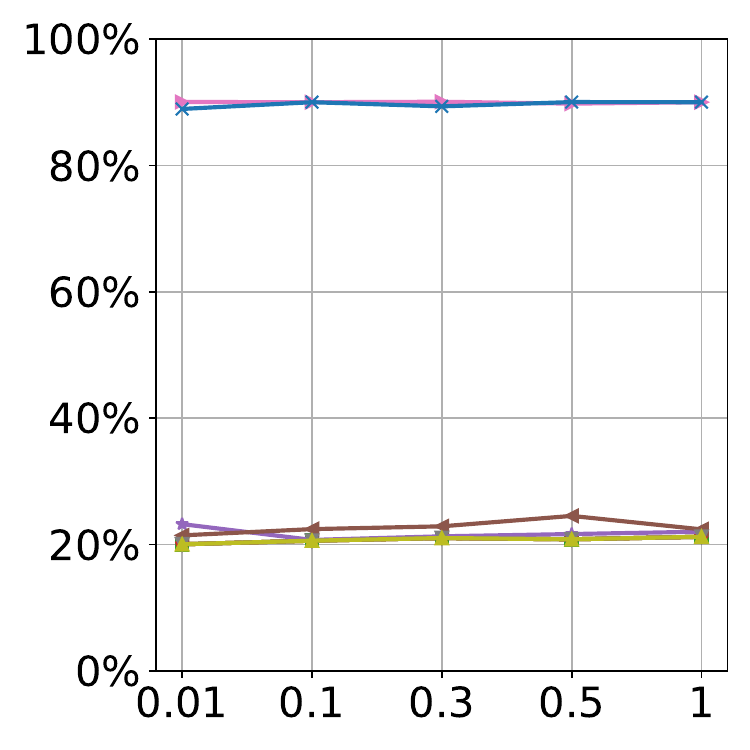}
        \subcaption{FLDetector}
    \end{subfigure}
    \caption{Testing error rate of the global model as a function of the participation rate under different defenses and attacks.}
    \label{fig:pr}
\end{figure*}

\subsection{Impact of the Participation Rate}
\label{asec:pr}
In each training round, the server typically selects a fraction (called \emph{participation rate}) of the clients to participate in training. We use participation rate of 0.1 by default in our experiments. 
Fig.~\ref{fig:pr} shows the testing error rates under different defenses and attacks when the participation rate ranges from 0.01 to 1, where FashionMNIST dataset is used. 
Note that for these experiments, we use FashionMNIST instead of CIFAR-10 dataset, due to limited memory of our GPU server. Moreover, we do not have results for Multi-Krum in Fig.~\ref{fig:pr}  when the participation rate is 1 due to the same reason on insufficient memory. 
For FLCert, the smallest participation rate is set to 0.03 instead of 0.01, because FLCert divides the clients into 10 groups,  and if a participation rate of 0.01 is used, only one client is selected to participate in each training  round. Consequently, attacks requiring genuine local models are not applicable.
We observe that  participation rate almost has no impact on the effectiveness of both existing attacks and our attack. 
Moreover, our attack consistently outperforms existing attacks across all considered participation rates.

\subsection{Impact of Different \ourmodel{} Variants}
\label{asec:variants}
\ourmodel{} crafts a malicious model update in each training round as a product of a given sign vector and a dynamically set magnitude vector, which is further decomposed into a product of an unit magnitude vector and a scaling factor. The unit magnitude vector and scaling factor play important roles in a malicious model update. 
Thus, we study different variants for designing them. We show results on CIFAR-10 dataset for simplicity.

\begin{figure}[t]
        \begin{subfigure}{0.49\linewidth}
        \includegraphics[width=1\linewidth]{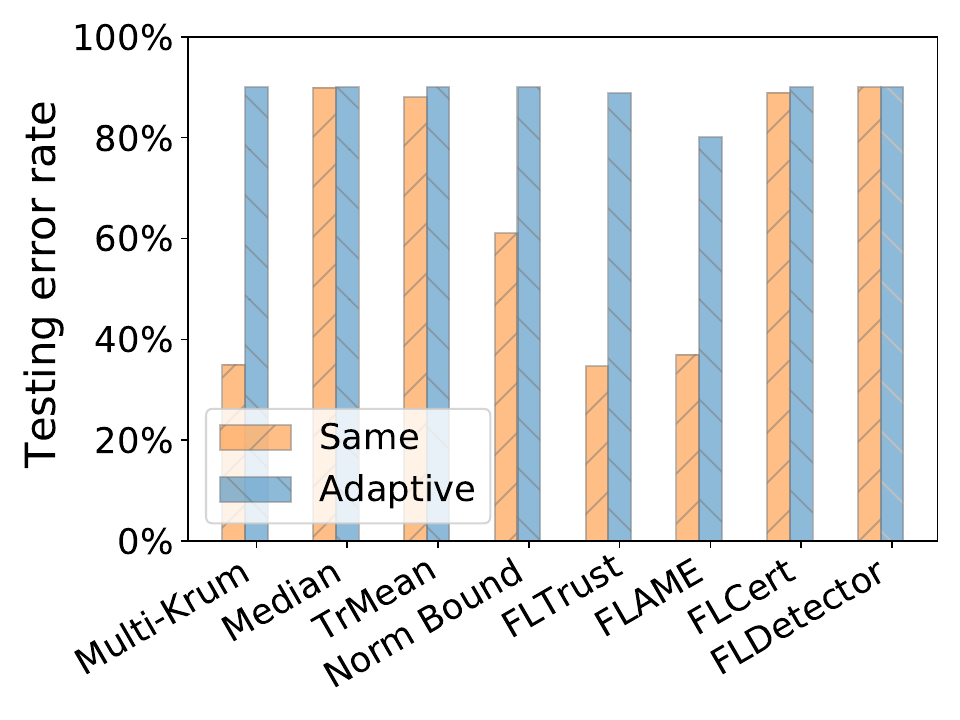}
        \subcaption{}
         \label{fig:unit}
    \end{subfigure}
        \begin{subfigure}{0.49\linewidth}
        \includegraphics[width=1\linewidth]{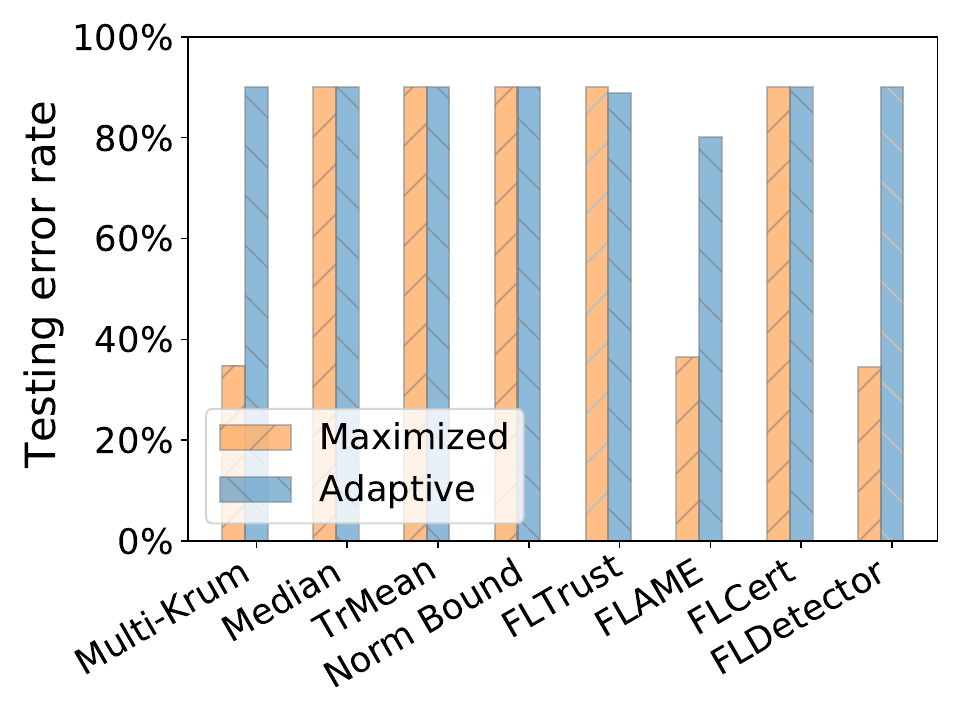}
        \subcaption{}
         \label{fig:scale}
    \end{subfigure}
    \caption{(a) Comparing the two variants of setting the unit magnitude vector in \ourmodel{}. (b) Comparing the two variants of setting the scaling factor in \ourmodel{}. }
\end{figure}

\myparatight{Variants of the unit magnitude vector} 
For the unit magnitude vector, we consider the following two variants.
\begin{itemize}
    \item \textbf{Same magnitude}: 
    In this variant, the attacker sets each dimension of the  unit magnitude vector $\boldsymbol{v}^t$ to the same value. In other words, we set each dimension to be $\frac{1}{\sqrt{d}}$, where $d$ is the number of dimensions/parameters.

    \item \textbf{Adaptive magnitude}: In this variant, the dimensions of the unit magnitude vector have different magnitudes and are dynamically set in each training round based on how the global model changes and the malicious model updates in the previous training round. Equation~\ref{unitvector} shows our  adaptive unit magnitude vector adopted by \ourmodel{}.  
    
\end{itemize}
 Fig.~\ref{fig:unit} shows the testing error rates under different defenses when \ourmodel{} uses the two variants to set the unit  magnitude vector, where the scaling factor is set using our adaptive method in Equation~\ref{scalingfactor}. Our results show that adaptive magnitude substantially outperforms same magnitude for some defenses (e.g., Multi-Krum and FLTrust) while the two variants achieve comparable effectiveness for other defenses. This is because  adaptive magnitude can target specific dimensions with larger effectiveness. 

\myparatight{Variants of the scaling factor}
We also consider two variants of setting the scaling factor as follows:
\begin{itemize}
    \item \textbf{Maximized scaling factor}: This variant sets a very large scaling factor with a goal to maximize the magnitude of the aggregated model update.  We use $100,000$ in our experiments.
    
    \item \textbf{Adaptive scaling factor}: This variant, adopted by \ourmodel{},  leverages how the global model changes in the previous training round to set the scaling factor (i.e., Equation~\ref{scalingfactor}) to avoid that the malicious model updates are filtered out by the server. 
\end{itemize}

Fig.~\ref{fig:scale} illustrates the testing error rates under different defenses when \ourmodel{} uses the two variants of setting the scaling factor, where adaptive magnitude is used to set the unit magnitude vector. We observe that, under some defenses (e.g., Median, TrMean, and Norm Bound), the variant of maximized scaling factor achieves comparable  testing error rates with adaptive scaling factor. This is because, for dimension-wise aggregation rules like Median and TrMean, although the malicious model updates with very large magnitudes are filtered out, they still deviate the aggregated model update; and although Norm Bound normalizes the norms of malicious model updates, the normalized malicious model updates still maintain consistency across training rounds. However, for other defenses like Multi-Krum and FLDetector that filter out entire model updates, adaptive scaling factor substantially outperforms maximized scaling factor. This is because these defenses filter out the entire malicious model updates with very large magnitudes. In contrast,  adaptive scaling factor leverages how the global model changes in the past  to achieve a balance between attack effectiveness  and undetectability. Since \ourmodel{} aims to be defense-agnostic, adaptive scaling factor is preferred. 

\subsection{Impact of the Sign Vector}
\label{app:sign}
\begin{table}[t]
    \centering
        \caption{Testing error rate of the global model under \ourmodel{} with various random sign vectors obtained from different random seeds on CIFAR-10.}
        \resizebox{\linewidth}{!}{
    \begin{tabular}{c|cccccc}
    \toprule
\midrule
         &seed=1&seed=2&seed=3&seed=4&seed=5&seed=6   \\
         \midrule
         Norm Bound & 90.00 & 90.00 & 89.95&90.00&88.94&90.00\\
         FLTrust & 90.00&88.15&89.09&90.00&90.00&90.00\\
         FLAME & 80.79&86.59&84.43&81.39&80.68&82.37\\
         \midrule
         \bottomrule
    \end{tabular}}
    \label{tab:sign}
\end{table}

We further analyze the impact of the random sign vectors.
We experiment with PoisonedFL on CIFAR-10 with $6$ random seeds ($1$-$6$) for various random sign vectors and defenses including TrMean, FLTrust, and FLAME, as shown in Table~\ref{tab:sign}. The result aligns with our intuition that these random sign vectors do not affect the attack performance and PoisonedFL can successfully break the defenses with these $6$ different random sign vectors. The reason is that in a continuous attack with the same random direction, the model inevitably updates substantially in a random direction in an uncontrolled way, and therefore leads to significantly degraded accuracy.

\section{Discussion and Limitations}
\label{sec:discussion}
\subsection{Countermeasures and Adaptive Attacks}
\label{app:counter}


\begin{figure*}[t]
    \begin{subfigure}{0.33\linewidth}
        \centering
        \includegraphics[width=0.49\linewidth]{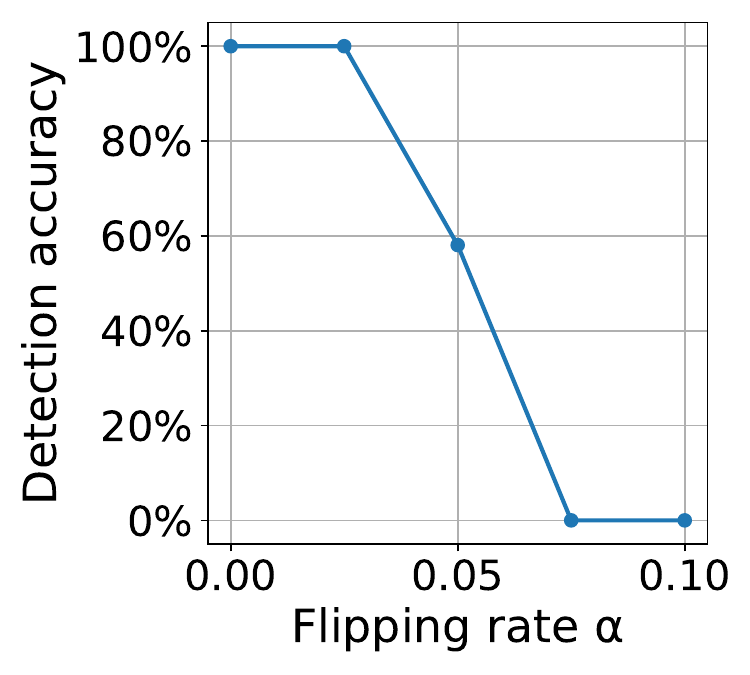}
        \includegraphics[width=0.49\linewidth]{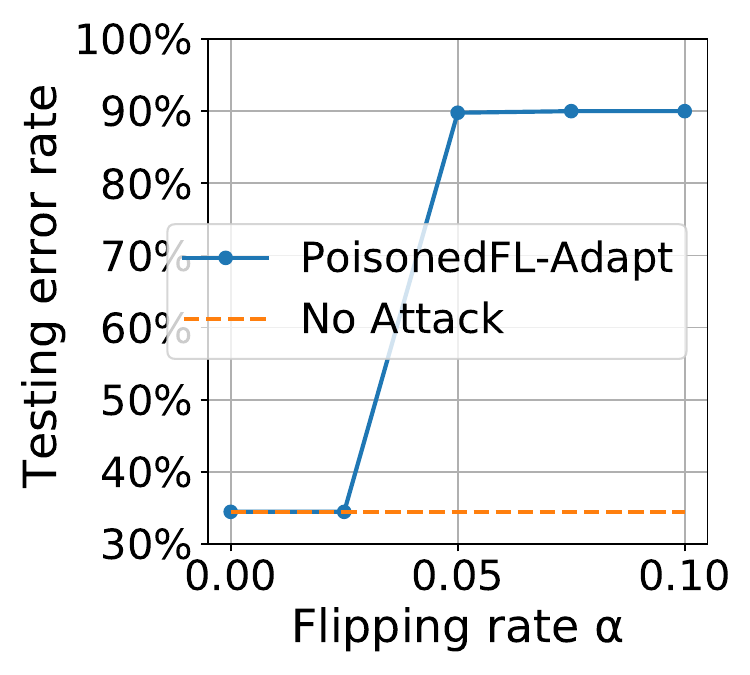}
        \subcaption{Median}
    \end{subfigure}
\begin{subfigure}{0.33\linewidth}
        \centering
        \includegraphics[width=0.49\linewidth]{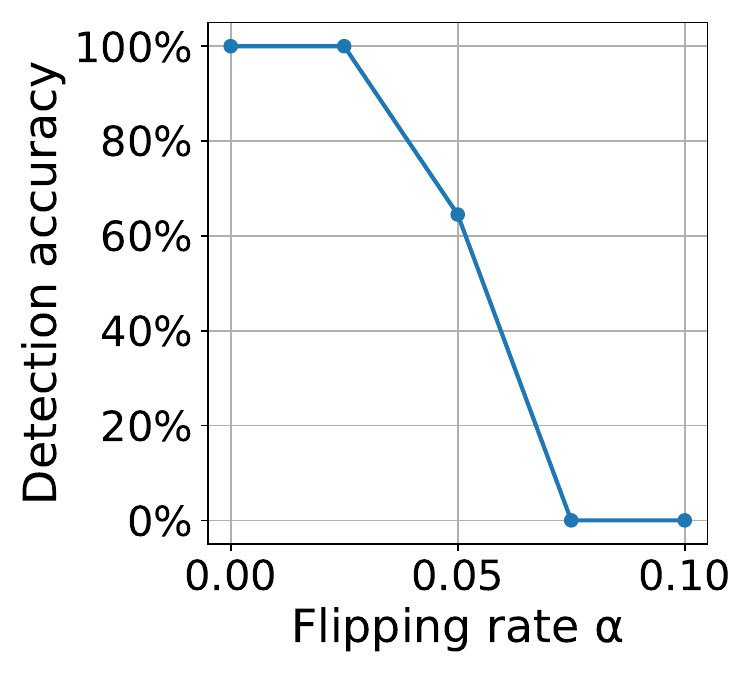}
        \includegraphics[width=0.49\linewidth]{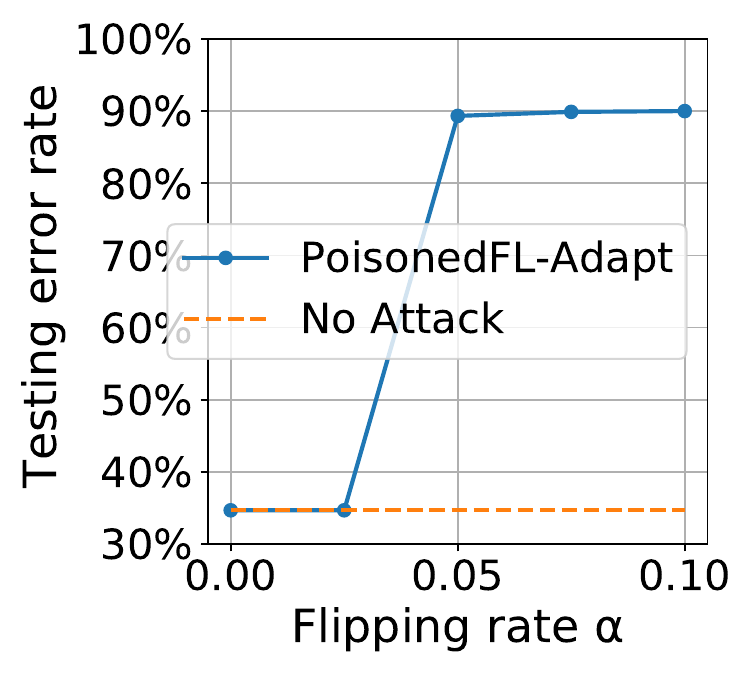}
        \subcaption{TrMean}
    \end{subfigure}
\begin{subfigure}{0.33\linewidth}
        \centering
        \includegraphics[width=0.49\linewidth]{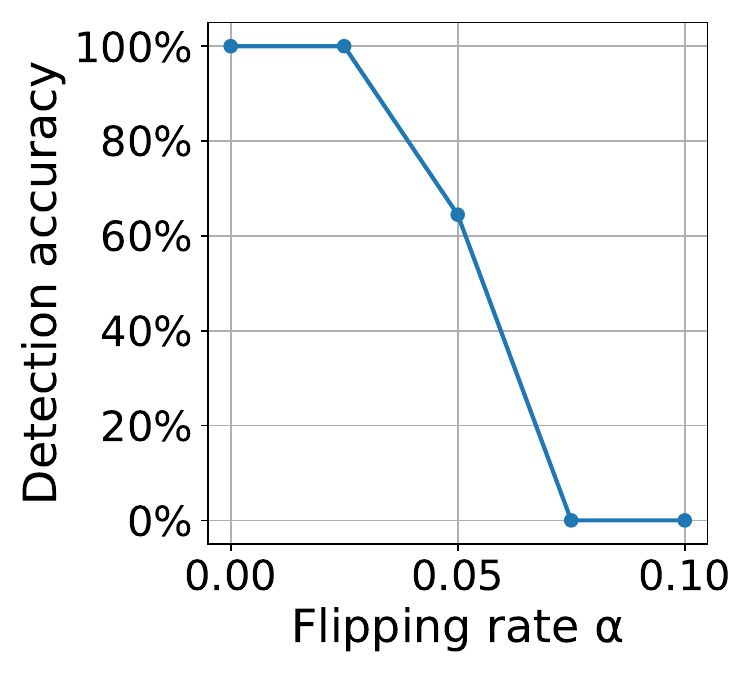}
        \includegraphics[width=0.49\linewidth]{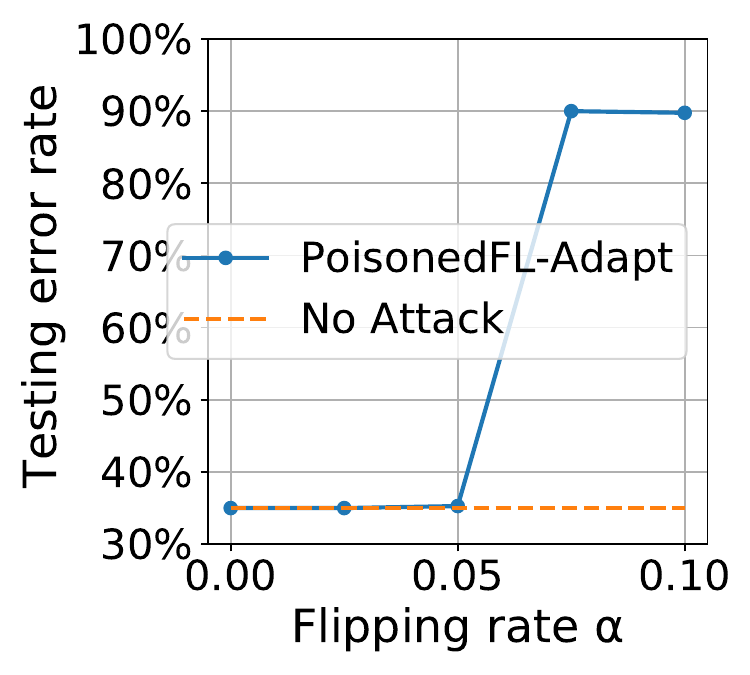}
        \subcaption{Norm Bound}
    \end{subfigure}
    \vspace{-6mm}
    \caption{Detection accuracy of GMM-Sign and testing error rate of the global model under \ourmodel{}-Adapt attack as the flipping rate $\alpha$ varies when the server uses different aggregation rules.}
    \label{fig:consistency}
    \vspace{-2mm}
\end{figure*}

We explore new defenses tailored to \ourmodel{}. Specifically, we study a new defense that normalizes the total aggregated model update since \ourmodel{} aims to increase its magnitude.  We also explore new tailored defenses to detect fake clients based on how \ourmodel{} crafts the malicious model updates. We show that the normalization-based defense has limited effectiveness at mitigating \ourmodel{} and we can still adapt \ourmodel{} to break the detection based defenses. 

In our tailored detection based defenses, the server extracts a feature for each client in each training round and then divides the clients into two clusters via fitting a two-component Gaussian Mixture Model (GMM)~\cite{permuter2006study} for the features.  We detect a particular cluster (e.g., with the smaller mean value) as fake clients. After detecting fake clients, the server removes them and re-trains a global model using the remaining clients. Note that if the distance between the mean values of the two clusters is smaller than the intra-cluster standard deviations of both clusters, no clients are detected as fake and the training continues since the difference between the two clusters may simply be caused by randomness. 

We extract features for clients based on how \ourmodel{} crafts malicious model updates. Depending on which information of the malicious model updates is used to extract features, we consider two variants of GMM, i.e., \emph{GMM-Magnitude} and \emph{GMM-Sign}, which extract features based on the magnitudes and signs of the model updates, respectively. 
Due to limited space, we discuss GMM-Magnitude in Appendix~\ref{asec:gmm-magnitude}.

\subsubsection{GMM-Sign}
\myparatight{Feature} GMM-Sign exploits the consistency of the malicious model updates to detect fake clients. Specifically, for a  client $i$ in training round $t$,  we define the feature $x_i^t$  as the  number of flipped dimensions of its model updates in two consecutive training rounds averaged in the recent $N$ training rounds. Formally, we have: 
\begin{equation}
    x_i^t = \sum_{t^\prime = t-N+1}^{t} \sum_{j=1}^{d} \mathbb{I}(\text{sign}(\bm{g}_{i}^{t'}[j]) \neq \text{sign}(\bm{g}_{i}^{t'-1}[j])), 
\end{equation}
where $d$ is the number of dimensions/parameters of a model update;  $\mathbb{I}$ is the indicator function that has a value of 1 if $\text{sign}(\bm{g}_{i}^{t'}[j]) \neq \text{sign}(\bm{g}_{i}^{t'-1}[j])$ and 0 otherwise; and $\bm{g}_{i}^{t'}[j]$ is the $j$-th dimension of $\bm{g}_{i}^{t'}$. For \ourmodel{}, we have $x_i^t = 0$ for all fake clients. 
After dividing the clients into two clusters, the cluster with a smaller mean value is detected as fake clients.

\myparatight{Adaptive attack}
In response to the defense, we adapt \ourmodel{}.  
Specifically, we  randomly flip $\alpha$ (called \emph{flipping rate}) fraction of the signs in each fake client's malicious model update and adjust the magnitudes of the corresponding dimensions to a small random value. 
Formally, we set the $d$-th dimension of the sign vector $\boldsymbol{s}^t_{i}$ and magnitude vector $\boldsymbol{k}^t_{i}$ for fake client $i$ in training round $t$  as follows: 
\begin{equation}
\begin{split}
    \boldsymbol{s}^t_{i}[j] &= \begin{cases} 
-\boldsymbol{s}{[j]} & \text{with probability } \alpha \\ 
\boldsymbol{s}{[j]} & \text{with probability } 1-\alpha
\end{cases},   \\
\boldsymbol{k}^t_{i}[j] &=  \begin{cases} 
\epsilon & \text{if } \boldsymbol{s}^t_{i}[j] = -\boldsymbol{s}{[j]}  \\ 
\boldsymbol{k}^t{[j]} & \text{otherwise }  
\end{cases},
\end{split}
\end{equation}
where $j\in[1,d]$, $\boldsymbol{s}$ is the sign vector picked randomly at the beginning of an attack, $\boldsymbol{k}^t$ is the magnitude vector for training round $t$ crafted by \ourmodel{}, and $\epsilon$ is a small value. 

\myparatight{Experimental results} Fig.~\ref{fig:consistency} shows the detection accuracy of GMM-Sign and the testing error rates as the flipping rate $\alpha$ varies when the server uses different aggregation rules, where the dataset is CIFAR-10 and $\epsilon=1e{-6}$.  We observe that GMM-Sign can accurately detect the fake clients when $\alpha$ is small. However, the detection accuracy reduces to 0 and \ourmodel{}-Adapt still makes the learnt global models random guessing when $\alpha$ is larger than a threshold, e.g., 0.08. 

\begin{figure*}[t]
    \begin{subfigure}{0.33\linewidth}
        \centering
        \includegraphics[width=0.49\linewidth]{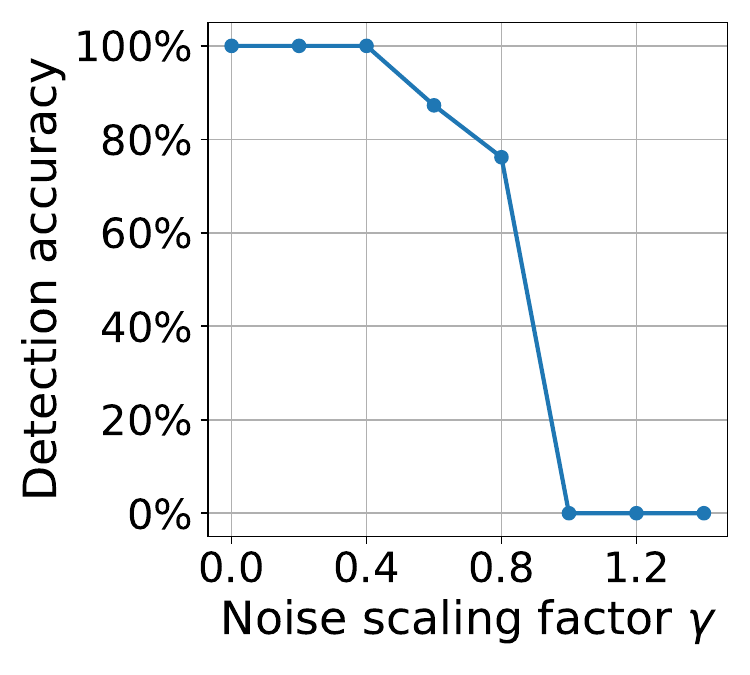}
        \includegraphics[width=0.49\linewidth]{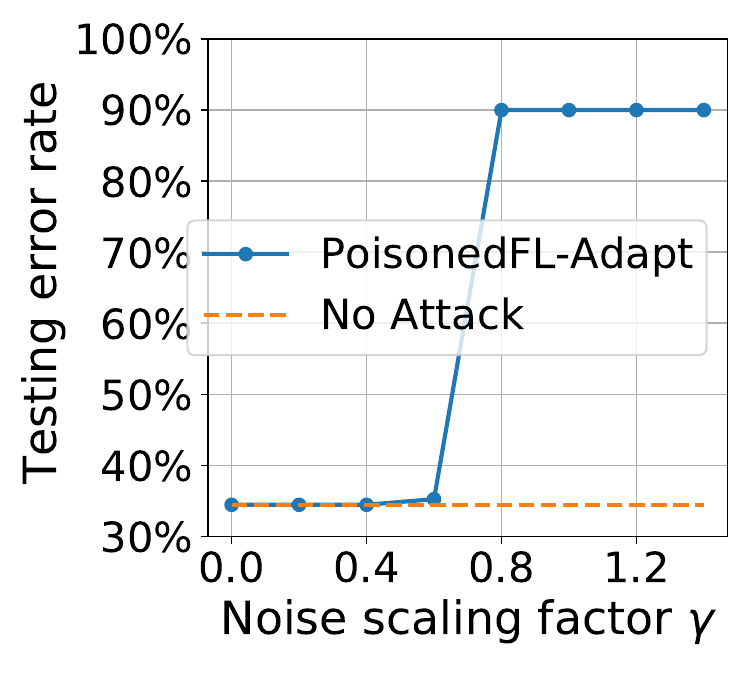}
        \subcaption{Median}
    \end{subfigure}
\begin{subfigure}{0.33\linewidth}
        \centering
        \includegraphics[width=0.49\linewidth]{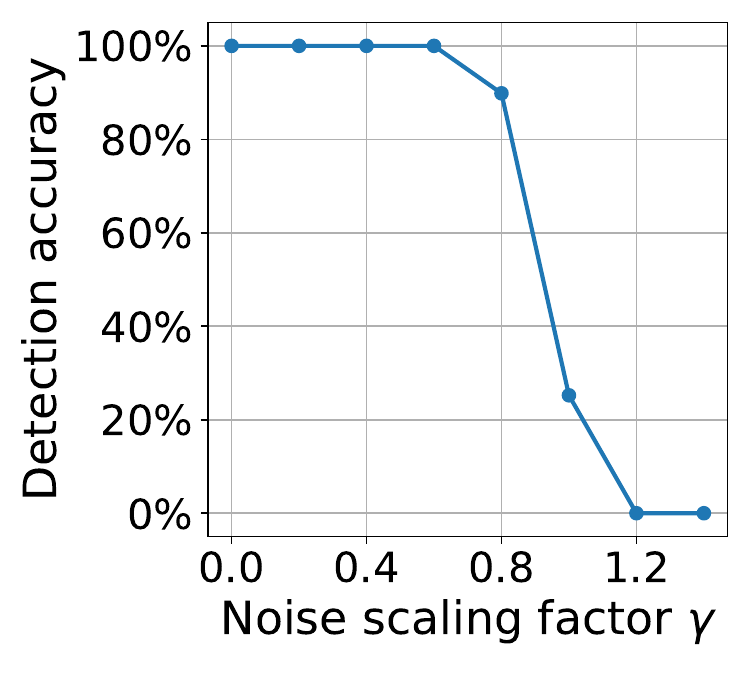}
        \includegraphics[width=0.49\linewidth]{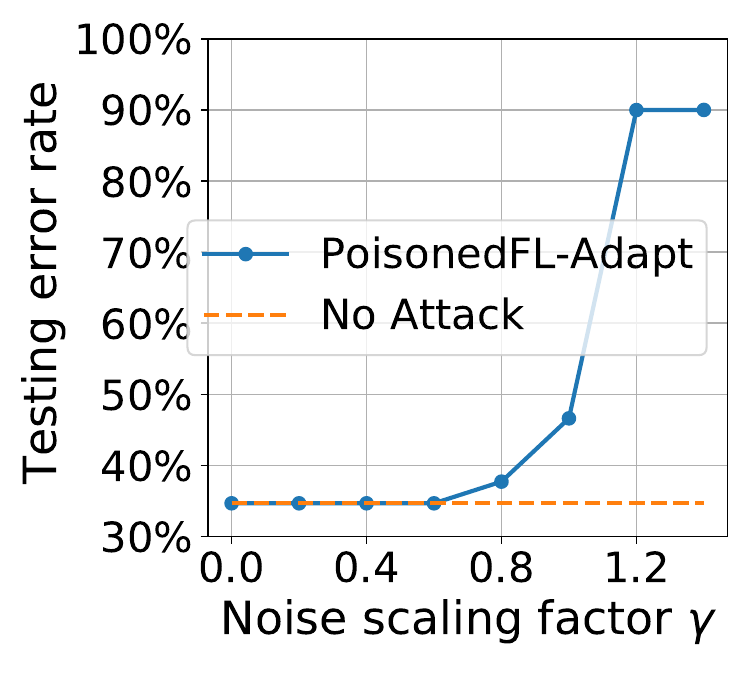}
        \subcaption{TrMean}
    \end{subfigure}
\begin{subfigure}{0.33\linewidth}
        \centering
        \includegraphics[width=0.49\linewidth]{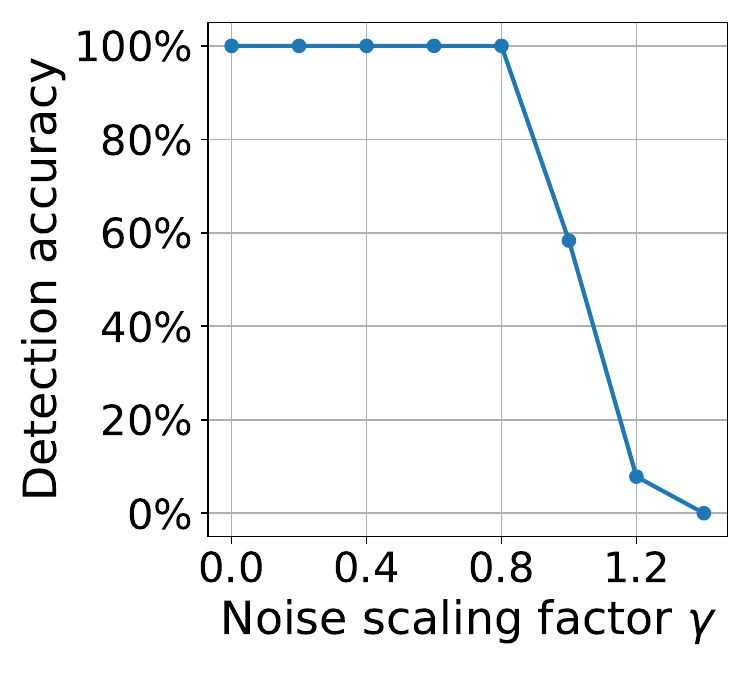}
        \includegraphics[width=0.49\linewidth]{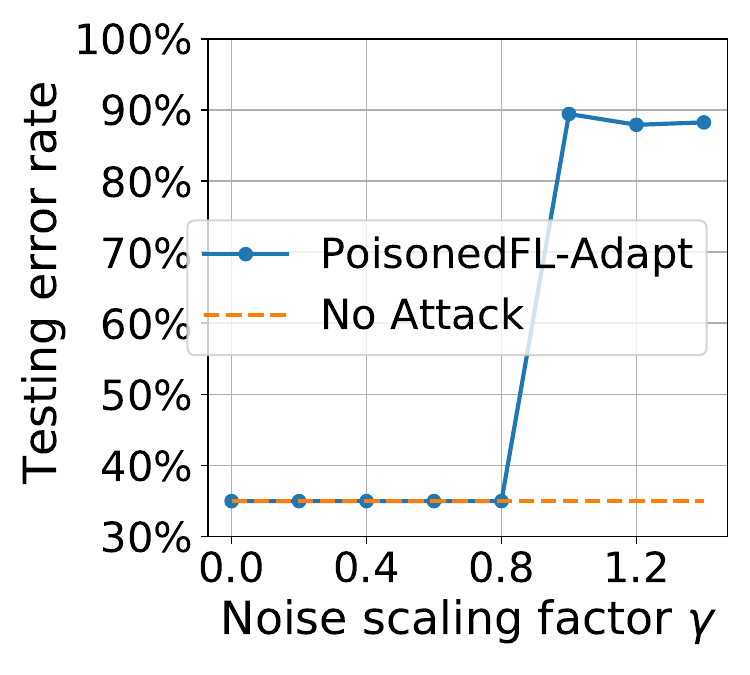}
        \subcaption{Norm Bound}
    \end{subfigure}
    \vspace{-6mm}
    \caption{Detection accuracy of GMM-Magnitude and testing error rate of the global model under \ourmodel{}-Adapt attack as the noise scaling factor $\gamma$ varies when the server uses different aggregation rules.} 
    \label{fig:collaboration}
\end{figure*}
\subsubsection{GMM-Magnitude}
\label{asec:gmm-magnitude}
\myparatight{Feature} \ourmodel{} uses the same malicious model update (i.e., $\bm{g}_{i}^t =\bm{k}^t \odot \bm{s}$) on all fake clients in training round $t$. Therefore, the server can extract  features for the clients based on the similarity between their model updates.  Specifically, for each client $i$ in training round $t$, we propose to calculate a feature $x_i^t$ as the sum of the average square distance between the client's model update  and its  closest $m-1$ model updates in the recent $N$ training rounds. Formally, we have:  
\begin{equation}
    x_i^t = \sum_{t^\prime = t-N+1}^{t}{{\sum_{j \in \mathbb{N}_i^{t'}}\frac{\left\|\bm{g}_i^{t^\prime}-\bm{g}_j^{t^\prime}\right\|^2}{m-1}}}, 
\end{equation}
where $\mathbb{N}_i^{t'}$ is the set of $m-1$ clients whose model updates are the closest to that of client $i$ in training round $t'$ and $m$ is the number of fake clients. Note that we give advantages to the defense by assuming that the number of fake clients is known to the server. In a training round, the server divides the clients into two clusters via GMM based on the features. The cluster with a smaller mean value is detected as fake clients. The mean value is exactly zero for \ourmodel{} since the malicious model updates are the same for the fake clients in each training round.

\myparatight{Adaptive attack} In response to such a defense, we can further adapt our \ourmodel{} by introducing random noise to the magnitude vector $\bm{k}^t$. Specifically, we can add a normalized noise to the magnitude vector for each fake client $i$ as follows: 
\begin{equation}
\boldsymbol{k}^t_i = \boldsymbol{k}^t 
 + \gamma \cdot  \frac{ ||\boldsymbol{k}^t||}{||\boldsymbol{\epsilon}_i|| }\cdot \boldsymbol{\epsilon}_i,
\end{equation}
where  $\boldsymbol{k}^t$ is the magnitude vector crafted by \ourmodel{}, $\boldsymbol{k}^t_i$ is the magnitude vector for fake client $i$ crafted by \ourmodel{}-Adapt, $\boldsymbol{\epsilon}_i \sim \mathcal{N}(\mathbf{0}, \mathbf{I})$ is the random noise sampled from the standard Gaussian distribution, $\frac{ ||\boldsymbol{k}^t||}{||\boldsymbol{\epsilon}_i|| }$ normalizes the random noise to have the same magnitude as $\boldsymbol{k}^t$, and $\gamma$ is a scaling factor for the random noise. The noise makes the malicious model updates of the fake clients less similar to each other and thus evade detection. 

\myparatight{Experimental results} Fig.~\ref{fig:collaboration} shows the detection accuracy of GMM-Magnitude and the testing error rates of the global models as the noise scaling factor $\gamma$ varies when the server uses different aggregation rules, where the dataset is CIFAR-10 and $N=20$.  
We observe that when $\gamma$ is small, i.e., a small amount of noise is added to each malicious model update, GMM-Magnitude can accurately detect the fake clients and thus our \ourmodel{}-Adapt is not effective. However,  when $\gamma$ is larger than a threshold (e.g., 1.2), GMM-Magnitude fails to detect the fake clients and \ourmodel{}-Adapt still makes the learnt global models nearly random guessing (i.e., testing error rates are around 90\%). This is because fake clients and genuine clients have distinguishable features when  $\gamma$ is small while their features are indistinguishable when $\gamma$ is large. \ourmodel{}-Adapt still substantially increases the testing error rates when $\gamma$ is large because the malicious model updates still have consistent sign vectors. 

\begin{figure}[t]
\centering
\includegraphics[width=0.8\linewidth]{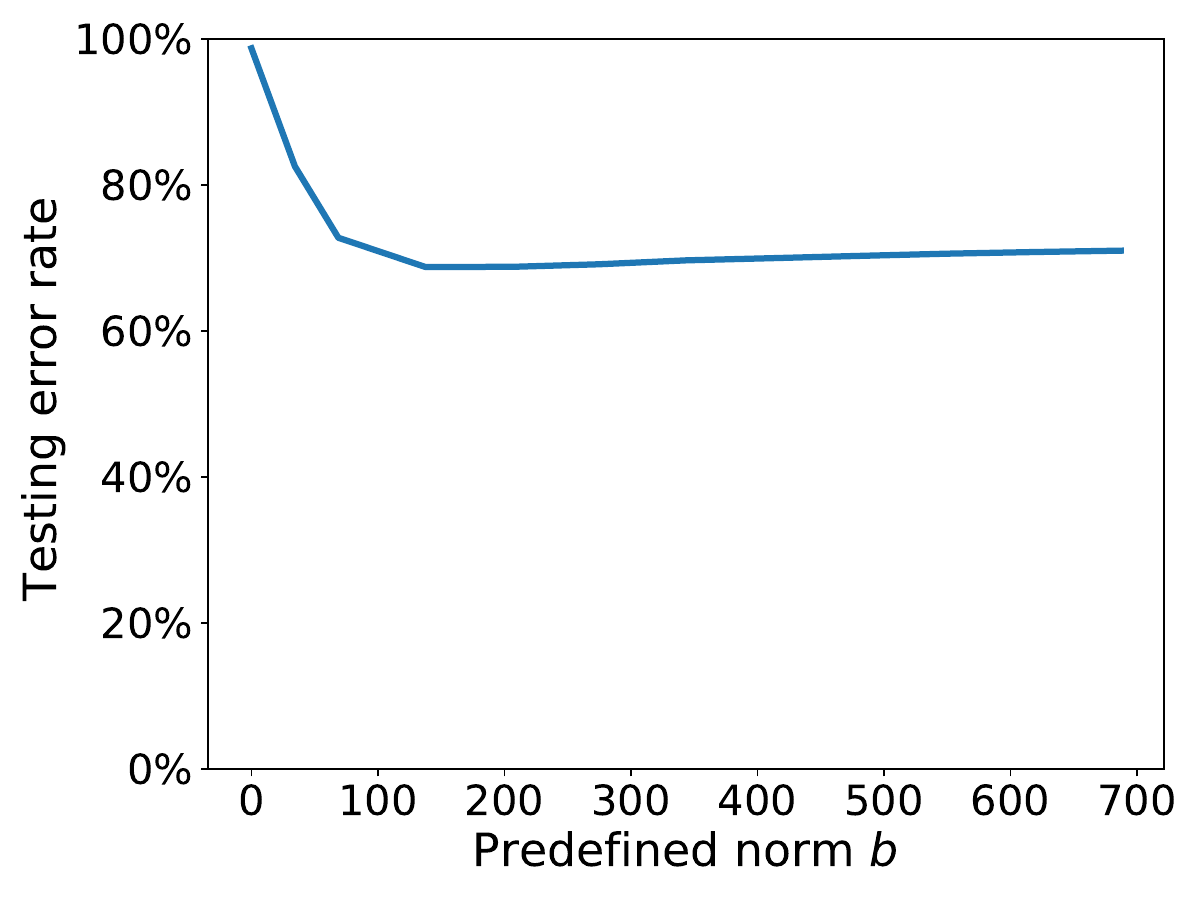}
    \caption{The testing error rate of a global model as a function of the predefined norm $b$ used to rescale the total aggregated model update. The dataset is Purchase and FL defense is Median.}
    \label{fig:b-performance}
\end{figure}

\subsubsection{Normalization-based Defense}
 \ourmodel{} aims to increase the magnitude of the total aggregated model update, i.e., $\left\| \bm{w}^T - \bm{w}^0\right\|$. Therefore, one tailored defense is to normalize the total aggregated model update. Specifically, the server can normalize $\bm{w}^T-\bm{w}^0$ to have a predefined $\ell_2$ norm $b$ and add the normalized total aggregated model update to the initial global model $\bm{w}^0$ to get a new global model, i.e., $\bm{w}^T \leftarrow \bm{w}^0 + b\frac{\bm{w}^T-\bm{w}^0}{\left\|\bm{w}^T-\bm{w}^0\right\|}$. Fig.~\ref{fig:b-performance} shows the testing error rate of the final global model under \ourmodel{} as the predefined norm $b$ varies. Our results show that  the testing error rate can be decreased by this defense but is still high. This is because the update direction of the total aggregated model update is randomly picked by the attacker, which normalization cannot change. 
\subsection{Using Synthetic Data} 
\label{app:sythetic}
In our previous experiments, when applying  attacks that require genuine clients' local models to craft malicious model updates on fake clients, we assume the attacker has access to the local models on all genuine clients. This assumption is strong and not realistic. Therefore, we also explore that an attacker uses the global models to reconstruct synthetic data, train local models using the synthetic data, and then craft malicious model updates based on them.

Reconstructing training data from a model is known as \emph{model inversion}~\cite{fredrikson2015model}. 
 Previous work~\cite{pi2023dynafed} shows that the global-model trajectory in multiple training rounds (especially early ones) of FL can be used to reconstruct synthetic data that is similar to the clients' genuine local training data. 
We follow the setting and generation strategy in \cite{pi2023dynafed}, which is the state-of-the-art reconstruction method in FL, to generate synthetic data for MNIST and FashionMNIST. We use  MNIST and FashionMNIST because they are simpler datasets and easier to reconstruct, giving advantages to these attacks. Specifically, we assume no attacks in the first 40 training rounds and use the corresponding clean global models to reconstruct  synthetic data. Table \ref{tab:fake_quality} shows the testing error rates of the classifiers trained on the synthetic data and evaluated on the genuine testing datasets of MNIST and FashionMNIST. The results show that the classifiers have decent testing error rates and thus the synthetic data mimic the genuine data well. These results are also consistent with prior work~\cite{pi2023dynafed}.

\begin{table}[t]
    \centering
        \caption{Testing error rate of a classifier trained on synthetic data and evaluated on genuine testing data.}
        \resizebox{0.7\linewidth}{!}{
    \begin{tabular}{c|cc}
    \toprule
\midrule
         &MNIST&FashionMNIST   \\
         \midrule
         Testing error rate & 9.94 & 24.74 \\
         \midrule
         \bottomrule
    \end{tabular}}
    \label{tab:fake_quality}
\end{table}

\begin{table}[t]
    \centering
        \caption{Testing error rates of attacks using synthetic data  on fake clients (i.e., ``Syn. Data + '') and  local models of all genuine clients (i.e., ``Genuine + ''), as well as our \ourmodel{} that does not require synthetic data nor genuine local models.}
    \resizebox{0.85\linewidth}{!}{
    \begin{tabular}{c|c|cc}
    \toprule
\midrule
         &&MNIST&FashionMNIST   \\
         \midrule
         \multirow{5}{*}{Median} 
                 &Syn. Data + Fang&  4.71& 23.30 \\
                &Genuine + Fang&  8.43& 27.82 \\
                    &Syn. Data + Min-Max& 4.79& 22.09\\
                    &Genuine + Min-Max& 8.52 & 23.30 \\
         &\ourmodel  &  \textbf{86.49} & \textbf{87.75}\\
         \midrule
         \multirow{5}{*}{TrMean}
                 &Syn. Data + Fang& 5.46& 25.48 \\
                  &Genuine + Fang&  5.66& 27.96 \\
                    &Syn. Data + Min-Max& 5.43& 19.77 \\
                    &Genuine + Min-Max&  5.66& 23.68 \\
         &\ourmodel &\textbf{88.73} & \textbf{85.36}  \\
         \midrule
                  \multirow{5}{*}{FLTrust}
                 &Syn. Data + Fang& 3.72 & 19.50 \\
                  &Genuine + Fang&  4.61& 18.78 \\
                    &Syn. Data + Min-Max& 3.68& 16.53 \\
                    &Genuine + Min-Max&  12.56& 21.32 \\
         &\ourmodel &\textbf{88.65} & \textbf{88.41}  \\
         \midrule
        \multirow{5}{*}{FLCert}
            &Syn. Data + Fang& 3.73& 20.80 \\
             &Genuine + Fang&  4.61& 18.78 \\
            &Syn. Data + Min-Max& 3.06 &16.27   \\
            &Genuine + Min-Max&  4.57& 19.28 \\
         &\ourmodel  &  \textbf{88.06} & \textbf{71.92} \\
         \midrule
         \bottomrule
    \end{tabular}
    }
    \label{tab:fake_lf}
     \vspace{-2mm}
\end{table}

We then use the synthetic data on the fake clients to perform model poisoning attacks in the remaining training rounds. Since each genuine client has 50 local training examples, we sample 50 synthetic training examples for each fake client. In each training round, the fake clients selected to participate in training train local models using their synthetic local training examples; and then we craft malicious model updates based on the local models for existing attacks.  
Table~\ref{tab:fake_lf} shows the testing error rates of the final global models under different attacks and defenses. We observe that an existing attack achieves higher testing error rates when having access to local models on the genuine clients in most cases, compared to using local models trained on synthetic data.  \ourmodel{} does not require genuine local models nor synthetic data but substantially outperforms these attacks.

\subsection{Cross-silo FL} 
\label{app:limitation}
\ourmodel{} primarily targets \emph{cross-device FL}~\cite{kairouz2021advances}, in which the clients are end-user devices such as smartphones. An attacker can inject fake clients into a cross-device FL system due to the open nature of the system. In particular, any client including fake ones can participate in the FL system. We acknowledge that it is harder to apply both \ourmodel{} and existing  attacks to  \emph{cross-silo FL}. Specifically, the clients are often a small number of verified institutions such as banks and hospitals in cross-silo FL. These institutions often go through certain verification processes before jointly training a model using FL. As a result, it is harder for an attacker to compromise genuine clients/institutions or inject fake clients/institutions in such a cross-silo FL system, making full-client control attacks like PoisonedFL—and all model poisoning
attacks—difficult to execute. We believe it is an interesting future work to identify a realistic threat model for cross-silo FL that takes into account the unique challenges and constraints.

\vspace{-2mm}


\end{document}